\documentstyle[psfig]{mn2e}

\title{Stellar populations in early-type Coma cluster galaxies --- I. The data}
\author[S. A. W. Moore et al.]
{Stephen~A.~W.~Moore,$^1$\thanks{E-mail: s.a.w.moore@durham.ac.uk} John~R.~Lucey,$^1$ Harald~Kuntschner,$^{1,2}$ Matthew~Colless$^3$\\
$^1$Extragalactic Astronomy Group, University of Durham, South Road, Durham DH1 3LE, UK.\\
$^2$European Southern Observatory, Karl-Schwarzschild-Str. 2, 85748 Garching, Germany.\\
$^3$Research School of Astronomy \& Astrophysics, The Australian National University, Weston Creek, ACT 2611, Australia.}

\date{Submitted 2002 March 18}
\pagerange{\pageref{firstpage}--\pageref{lastpage}}
\pubyear{2002}

\begin{document}

\maketitle 

\label{firstpage}

\begin{abstract}
We present a homogeneous and high signal-to-noise data set
(mean S/N of $\sim60$ per \AA)
of Lick/IDS stellar population line indices
 and central velocity dispersions 
for a sample of 132 bright ($b_j\leq18.0$)
galaxies within the central 1$^\circ$ ($\equiv$ 1.26\,$h^{-1}$\,Mpc) 
of the nearby rich Coma cluster (A1656).
Our observations include 73 per cent (100 out of 137) 
of the total early-type galaxy population ($b_j\leq18.0$).
Observations were made with the WHT 4.2 metre and 
the AUTOFIB2/WYFFOS multi-object spectroscopy instrument (resolution of $\sim2.2$~\AA\, FWHM)
using 2.7$''$ diameter fibres ($\equiv$ 0.94\,$h^{-1}$\,kpc).
The data in this paper have well characterised errors, 
calculated in a rigorous and statistical way.
Data are compared to previous studies and 
are demonstrated to be of high quality 
and well calibrated on to the Lick/IDS system.
Our data have median errors of $\sim0.1$~\AA\/ for atomic
line indices, $\sim0.008$~mag for molecular line indices, and 0.015\,dex for velocity dispersions.
This work provides a well-defined, high-quality baseline at $z\sim0$ for studies of 
medium to high redshift clusters. 
Subsequent papers will use this data set to probe the stellar populations
(which act as fossil records of galaxy formation and evolution) 
and the spectro-photometric relations of the
bright early-type galaxies within the core of the Coma cluster.

\end{abstract}


\section{Introduction}
\label{sec:intro}

Rich clusters provide a large sample of galaxies at a
common distance. This makes them ideal laboratories to study
global correlations between the dynamical, 
structural and stellar population properties of
galaxies in dense environments. 
One of the most important currently unsolved problems in observational cosmology though
is the formation process and subsequent evolutionary history of early-type
galaxies within such a rich cluster environment.
These early-type galaxies constitute the dominant population of galaxies within
rich cluster cores.
Studies using photometric and spectroscopic observations of bright ellipticals in clusters
have suggested that their luminosity and colour evolution is modest and consistent with
the passive evolution of stellar populations formed at
high redshifts, $z>2-4$ (Arag\'{o}n-Salamanca et al. 1993; Ellis et al. 1997;
Kodama et al. 1998; van Dokkum et al. 1998; Kelson et al. 2000).
However these studies have been unable to distinguish between  
dissipationless (or `monolithic') collapse
(e.g. Larson 1975) or hierarchical merging models (e.g. Baugh, Cole \& Frenk 1996; Kauffmann 1996)
of galaxy formation.
This relatively modest evolution contrasts with the claims of strong evolution
in the morphological mix in clusters, specifically the ratio of lenticular (or `S0')
to elliptical galaxies. Dressler et al. (1997) used the Hubble Space
Telescope to image 10 clusters at $z>0.3-0.6$ and observed a rapid
increase in the ratio of lenticular to elliptical galaxies 
from 10-20 per cent at $z\sim0.5$ to the 60 per cent seen today (see also Fasano et al. 2000).
Poggianti et al. (1999) used spectroscopic observations to analyse the 
Dressler et al. (1997) sample and suggested that the increase in the numbers
of lenticular galaxies (assuming that the number of ellipticals remains constant)
rather than simply being due to the slower formation of lenticulars from primordial origins
(i.e. from methods similar to ellipticals)
is in fact due to the morphological transformation of accreted field spiral galaxies 
into lenticulars. 
This implies that lenticulars and ellipticals are not constituent members 
a single morphological class of early-type galaxies, despite their
broad similarity and their closeness on the Hubble sequence, but instead
have contrasting evolutionary histories.
In clusters, the presence of a large population of red, apparently
passive galaxies (i.e. little evidence of recent star formation) with late-type
morphologies together with the absence of blue lenticular 
galaxies (i.e. galaxies with young stellar populations) 
was hypothesised by Poggianti et al. (1999) to be evidence that any morphological
transformation occurs on a longer time scale than the decline in the star
formation rates of the accreted field galaxies (see also Kodama \& Smail 2001).
Their hypothesis was that the accreting spiral galaxies first suffered a decline in
their star formation due to the stripping of their gas
as they encountered the dense intra-cluster medium within the cluster potential
(by e.g. galaxy-galaxy collisions, ram pressure
stripping, gas evaporation in the hot intra-cluster medium, or galactic winds methods,
see Spitzer \& Baade 1951; Gunn \& Gott 1972; Faber \& Gallagher 1976;
Cowie \& Songaila 1977; Burstein 1979; Dressler 1980b; Larson, Tinsley \& Caldwell 1980)
\emph{before} their morphological transformation into lenticular galaxies (this transformation could
be by a completely separate process, e.g. Moore, Lake \& Katz 1998).
If this hypothesis is correct and morphological transformation occurs on a time-scale of $\leq2-3$\,Gyr
after a galaxy's entry into a cluster there should still be evidence of the previous 
star formation activity in the 
stellar populations of the lenticular galaxies, with a wider spread in age for lenticulars
than for ellipticals.
Therefore a study of the 
stellar population ages of elliptical
and lenticular galaxies can be a powerful probe of the evolutionary history of rich clusters.

To date there have been many observational studies 
probing the evolutionary history of nearby cluster early-type galaxies
(e.g. Bower, Lucey \& Ellis 1992; Caldwell et al. 1993; Kuntschner \& Davies 1998;
Colless et al. 1999; J{\o}rgensen 1999; Castander et al. 2001; 
Poggianti et al. 2001;  Vazdekis et al. 2001).
However these studies have reached somewhat differing conclusions.
Some of the significant results,
particularly on the Coma cluster of galaxies (A1656), are discussed below in
more detail.

Caldwell et al. (1993) (see also Caldwell \& Rose 1997) obtained multi-fibre spectroscopy for 
125 early-type Coma cluster galaxies from two 45$'$ diameter fields: one
centred on the cluster core ($-21.4\la{B}\la-17.6$) and one centred 40$'$ south west (SW) of
the cluster centre ($-21.4\la{B}\la-16.7$). 
They found that for $B\la-18.5$, 11 out of 
the 28 galaxies (39 per cent) in the SW region are `abnormal', 
as opposed to only 3 out of 68 (4 per cent) in the central field. 
A subsequent line-strength analysis of the Caldwell spectra by
Terlevich et al. (1999) confirms the previous results and demonstrates
that the colour-magnitude-relation in Coma is driven primarily by a
luminosity-metallicity relation. These results imply an old, passively
evolving cluster core whilst the SW corner, possibly infalling to the
main, older core of galaxies, shows a spread of stellar population
ages.

J{\o}rgensen (1999) observed 71 Coma cluster early-type galaxies
($B\la-19.2$) within the central
$64'\times70'$ region. She combined these observations
with literature data to create a data set of 115 early-type galaxies with Mg$_2$, H$\beta_{\rm{G}}$
and $\langle\rm{Fe}\rangle$ Lick/IDS index measurements 
(though there were only 68 with all
of the indices measured). 
Using stellar population models and the Lick/IDS indices she 
found a low mean age and a sizeable spread in age (5.25~Gyr $\pm$ 0.166 dex)
for a sub-sample of 71 early-type galaxies 
($B\la-19.2$). She also observed 
a small spread in metallicity, [Fe/H] of $+0.08\,\pm\,0.194$.
Taken at face value this result seems to disagree with the analysis of the
Caldwell et al. (1993) spectra.

The SDSS team (Castander et al. 2001) 
observed a 3$^\circ$ field centred on the south western part of the Coma cluster.
They found 25 per cent of 
galaxies (49 out of 196 galaxies, 
$B\la-17.7$) 
showed signs of recent star formation activity, giving
them a young luminosity-weighted mean age. Their total population
of galaxies therefore has a large spread in age. This result is in broad agreement 
with the findings of Caldwell et al. (1993) and J{\o}rgensen (1999).
However since they used spectral morphological classification techniques it
is difficult to relate their findings directly to the early-type galaxy sub-population.

Poggianti et al. (2001)
observed two $32.5'\times50.8'$ fields towards the centre
and the south west region of the Coma cluster.
They found that for their sample of
of 52 early-type galaxies 
($-21.0\la{B}\la-18.0$), 
95 per cent (18 out of 19) of ellipticals were consistent
with ages older than 9~Gyr, whilst 41 per cent (13 out of 32) of lenticulars
had ages smaller than 5~Gyr (one lenticular was excluded from their
analysis because it had strong emission lines). 
The early-type galaxies show a large
metallicity spread (--0.7 to $+0.4$ in [Fe/H]). 
This more detailed study suggests that there are real differences
between the formation processes of the ellipticals and lenticulars.
However the individual measurements are not actually contradictory with
those of previous Coma cluster studies, but the study does imply that a better
understanding of the sample selection is needed. It seems, that if a
sample of early-type galaxies is incomplete then an over-representation
of ellipticals would lead to the conclusion that there is a small
spread in age, whereas a sample with a larger number of lenticulars
would conclude the reverse. 
It is therefore important to understand the selection effects and to deal with the 
elliptical and lenticular morphological classes separately.
The caveat to the findings of Poggianti et al. (2001) is that they are based
upon data from both the core of the cluster \emph{and}\/ the south west region. 
Caldwell et al. (1993) previously showed that these parts of the cluster
have \emph{different}\/ levels of star formation activity. This would affect 
conclusions based upon a conglomerate sample from the two regions 
(a sample with more galaxies from the SW region would have a larger 
age distribution than a sample with more from the core).

The different evolutionary histories of
elliptical galaxies have previously been reported by Kuntschner \&
Davies (1998, see also Kuntschner 2000) who conducted a study of the
small Fornax cluster (11 ellipticals and 11 lenticulars, $B\la-17.0$). 
They found a large spread in metallicity of
--0.25 to $+0.30$ in [Fe/H] for the ellipticals, while the data are
consistent with no spread in age suggesting an early
formation epoch at $\sim$8~Gyr. For the lenticulars, however, they
found a large spread in age (1~Gyr to older than $\sim12$~Gyr) 
and metallicity (--0.5$\leq$[Fe/H]$\leq$$+0.5$). Overall, the mean
metallicity of early-type galaxies increases with luminosity in the
Fornax cluster.

Emerging from the studies discussed is a picture of a different evolutionary
history of lenticular and elliptical galaxies within the cluster environment.
Additionally, the growth of clusters due to
accretion of small groups or mergers with other clusters can affect the
stellar populations of the individual galaxies (evidenced by the differences
between the SW region and the core of the Coma cluster).
It can be seen that it is essential in all observational studies of galaxy evolution to build up
homogeneous and complete, or at least representative, samples.
Somewhat surprisingly this has not been done for the Coma cluster up to this date.
This has been the cause of much disagreement and controversy in the study
of galaxy stellar population ages and metallicities.
A new study is therefore needed which 
does not suffer from these limitations and which can answer
the question of the 
formation processes and subsequent evolutionary histories of early-type
galaxies within a rich cluster environment. This is the aim
of this study.

In this study we measure spectroscopic 
line indices to derive accurate  luminosity-weighted mean ages and metallicities 
for the central stellar populations of bright early-type 
galaxies in Coma to probe their evolutionary history through an analysis
of the `noise' of galaxy formation. This study will also use 
several important early-type galaxy correlations to place further constraints
on their evolution. The correlations that will be investigated are: 
the colour--magnitude relation
(Faber 1973; Visvanathan \& Sandage 1977; Bower et al. 1992), 
the Mg$_2$ line strength versus velocity dispersion relation 
(Terlevich et al. 1981; Burstein et al. 1988; Bender, Burstein \& Faber 1993)
and the Fundamental Plane (Djorgovski \& Davis 1987; Dressler et al. 1987). 
These early-type galaxy relations provide a rich source of constraints for galaxy
formation scenarios, probing their underlying physical mechanisms (Bower et al. 1992; 
Bender, Burstein \& Faber 1992, 1993; Guzm\'{a}n, Lucey \& Bower 1993; 
Ciotti, Lanzoni \& Renzini 1996; Bernardi et al. 1998).

This paper is the first in a series on the rich Coma cluster (early results can be
found in Moore 2001 and Moore et al. 2001).
In this paper (Paper I) we present the sample selection, spectroscopic reduction
and final data catalogue.
Section \ref{sec:selection} describes the
sample selection, Section \ref{sec:astrometry} details the astrometry, 
Section \ref{sec:observations} the observations, 
Section \ref{sec:datared} the data reduction, Section \ref{sec:veldispcorr} the corrections
to the velocity dispersion measurements, 
Section \ref{sec:linestrengths} the measurement of stellar population absorption line strengths,
Section \ref{sec:comparison} the comparison with previous data, 
and finally Section \ref{sec:conclusions} presents the conclusions
on the quality of the data.
The companion papers (Moore et al. 2002a,b -- hereafter referred to as Papers II and III)
will use the data in this paper to measure stellar population mean ages and metallicities
to probe the evolutionary history of the Coma cluster (Paper II) and combine the data
with photometry to investigate in detail various spectro-photometric relations (Paper III).

\section{Sample selection}
\label{sec:selection}

Our aim was to construct a representative sample of the bright early-type
galaxy population in the central region of the rich Coma cluster
in order to measure central velocity dispersions and stellar population line strengths.
Furthermore, to minimize systematic errors we aimed to
provide a large overlap with previous studies; we also aimed to obtain many repeat 
observations with high S/N to characterise the random errors. 
The sample was selected using firstly the
Godwin, Metcalfe \& Peach (1983) catalogue (GMP), which contains
227 galaxies ($b_j\leq18.0$, 
equivalent to $B\la-17.7$\footnote{taking the mean heliocentric
redshift of the Coma cluster to be 6841\,km\,s$^{-1}$ (\emph{this study})
and assuming H$_0$ = 50 km\,s$^{-1}$\,Mpc$^{-1}$, $q_o=0$, no extinction and
that the cluster is at rest with respect to the Cosmic Microwave
Background (i.e. no cluster peculiar velocity) yields a 
distance modulus of 35.68\,mag.\label{fn:distancemod}})
within a 1$^\circ$ field centred close to the cD galaxy NGC 4874 (a small offset
is applied to improve the AUTOFIB2/WYFFOS setup).
Magnitudes ($b_j$) and colours 
($b-r\simeq{{\rm{B}}-{\rm{R}}}$) for the galaxies 
were also taken from GMP. 
A magnitude limit of $b_j\leq18.0$ (this is the magnitude limit implied
in all subsequent discussions herein)
was chosen since this study
aims to obtain high quality data of bright early-type galaxies and this limit
corresponds to $\sim1.7$\,mag below the peak of the Coma cluster luminosity
function (at $b_j\simeq16.3 \equiv B\simeq-19.4$, see Biviano et al. 1995) and therefore samples
the bulk of the early-type galaxy luminosity function with little contamination
from dwarf galaxies (see Binggeli, Sandage \& Tammann 1988 for a review); it also allowed high
S/N measurements ($\geq35$ per \AA) to be obtained within the observing program
time constraints (see Section \ref{sec:observations}).
In addition to the complete GMP 
data set we also used 816 redshifts 
in the Coma cluster region
(223 within the 1$^\circ$ field and with $b_j\leq18.0$) 
kindly provided by M.~Colless (see also Edwards et al. 2002). 
The morphological typing for
the galaxies was taken from Dressler (1980a) where available. Within
the central 1$^\circ$ field there are 210 confirmed cluster members. A
sub-sample of 158 galaxies have been classified by Dressler and 137
galaxies are of early-type morphology. The sample definition is
summarised in Table \ref{tab:selection_criteria}.

\begin{table}
\caption{Sample selection}
\label{tab:selection_criteria}
\begin{tabular}{lr}
\hline
\hline
total number of galaxies in field                       & 227 \\
number of galaxies with redshifts                       & 223 \\
galaxies with cluster membership confirmed by redshifts & 210 \\
confirmed cluster member galaxies with morphologies     & 158 \\
confirmed cluster member early-type galaxies            & 137 \\
\hline
\end{tabular} \\
\begin{minipage}{8.25cm}
  {\em Notes:}\/ Only galaxies with $b_j\leq18.0$ and within a 1$^\circ$
  field centred on the Coma cluster are considered.
\end{minipage}
\end{table}

Selection criteria are then applied to the sample of 210 Coma galaxies to prioritise their importance
within the AUTOFIB2/WYFFOS multi-fibre configuration program. This program uses a weighting scheme to maximise 
the scientific return of any observations, with priorities from 9 (most important) to 1 (least important),
and takes into account the limitations of the instrument (e.g. constraints on the minimum
distance between fibres).
The highest priority was given to galaxies with early-type morphologies and with 
previously measured velocity dispersions (the goal being to tie down the systematics of any measurements).
The next highest priority was given to galaxies with early-type morphologies but without
previous velocity dispersion measurements. Lower priorities are then given to those
galaxies with no morphological types in Dressler (1980a), with preference given to the
brighter galaxies. The lowest priority was given to late-type galaxies 
within the cluster.
This prioritised sample is then passed to the multi-fibre instrument configuration
program. To increase the completeness of the observations of this sample (affected by 
constraints on fibre closeness and by there being only 126 available fibres), three
different AUTOFIB2/WYFFOS field configurations are observed at the same position. The second field
has the same priorities for the configuration program as the first, except that the galaxies that were
observed in the first field have a lower priority (2 levels lower). Similarly
the third field also has reduced priorities for the configuration program 
for the galaxies observed in the previous two fields. This technique increases the
completeness and scientific return of the observations, whilst ensuring repeats between 
between each of the three observed fields.

\section{Astrometry}
\label{sec:astrometry}

To determine our astrometry 
three Schmidt plates were used:

~\\
\noindent{-- 10\,min exposure plate (OR17491) taken on 3/4/1997;} \\
\noindent{-- 30\,min exposure plate (OR18041) taken on 18/6/1998;} \\
\noindent{-- 85\,min exposure plate (OR9945) $\;$  taken on 25/2/1985.} \\

\begin{figure}
\psfig{file=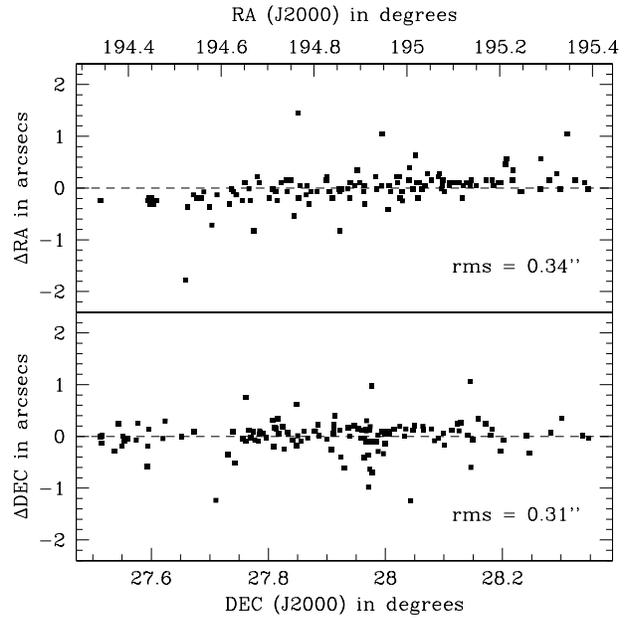,width=85mm}
\caption{Comparison of our astrometry with that of Colless (see also Edwards et al. 2002).}
\label{fig:astrometry}
\end{figure}

\noindent
The shorter exposure plates were specifically requested to measure accurate
astrometry for the bright Coma galaxies.
The plates were taken at the UK Schmidt Telescope 
using 3\,mm glass with emulsion IIIaF and filter OG590. These Schmidt plates
were scanned in using the SuperCOSMOS scanner at the Royal
Observatory Edinburgh (Hambly et al. 2001).
The data was then analysed and positions
of all the programme objects determined by matching field star positions to the 
USNOA2
catalogue (Monet at al. 1997) and creating an astrometry solution for the plate. 
Table \ref{tab:astrometry} lists the astrometry of the objects observed
in this study together with the different names associated with each
galaxy.
Comparison (Fig. \ref{fig:astrometry}) with the
Coma cluster astrometry of M.~Colless (Edwards et al. 2002)
 confirms that our astrometry is accurate to 
$\la0.3''$.
This is sufficient for multi-fibre spectroscopy to be undertaken 
(cf. the 2.7$''$ diameter of the WYFFOS fibres).

\begin{table*}
\begin{minipage}{160mm}
\caption{Coma cluster astrometry for the 135 galaxies observed.}
\label{tab:astrometry}
\begin{tabular}{lllllllccrrrrrr}
\hline \hline 
n1 & n2 & n3 & n4 & n5 & \multicolumn{2}{l}{type} & $b_j$ & $b-r$ & \multicolumn{3}{c}{RA (J2000)} & \multicolumn{3}{c}{DEC (J2000)} \\
\hline
d112     & \ \      & gmp4945  & \ \      & b40     & E        & E+A & 16.64    & 1.78     & 12 & 57 & 21.731 & +27 & 52 & 49.75 \\
d75      & \ \      & gmp4679  & \ \      & b91     & S0       & \ \ & 16.13    & 1.91     & 12 & 57 & 46.139 & +27 & 45 & 25.51 \\
d201     & \ \      & gmp4666  & \ \      & \ \     & S0       & \ \ & 17.35    & 1.80     & 12 & 57 & 46.697 & +28 & 8 & 26.77 \\
d93      & \ \      & gmp4664  & \ \      & b39     & S0       & \ \ & 16.26    & 2.06     & 12 & 57 & 47.296 & +27 & 50 & 0.03 \\
d74      & \ \      & gmp4656  & \ \      & b84     & E        & \ \ & 17.62    & 1.82     & 12 & 57 & 47.863 & +27 & 46 & 10.03 \\
d210     & \ \      & gmp4648  & \ \      & \ \     & E        & \ \ & 15.97    & 1.88     & 12 & 57 & 48.658 & +28 & 10 & 49.48 \\
d110     & \ \      & gmp4626  & \ \      & \ \     & S0/E     & \ \ & 16.60    & 1.93     & 12 & 57 & 50.627 & +27 & 52 & 46.34 \\
d220     & ngc4848  & gmp4471  & \ \      & \ \     & Scd      & \ \ & 14.50    & 1.56     & 12 & 58 & 5.598 & +28 & 14 & 33.31 \\
\ \      & \ \      & gmp4469  & \ \      & b79     & \ \      & \ \ & 17.69    & 1.88     & 12 & 58 & 6.820 & +27 & 34 & 37.09 \\
d29      & \ \      & gmp4447  & \ \      & b78     & E        & \ \ & 17.81    & 1.98     & 12 & 58 & 9.688 & +27 & 32 & 57.86 \\
\ \      & \ \      & gmp4420  & \ \      & b75     & \ \      & \ \ & 17.60    & 1.86     & 12 & 58 & 11.426 & +27 & 56 & 23.85 \\
d209     & \ \      & gmp4391  & \ \      & \ \     & S0       & \ \ & 16.04    & 1.77     & 12 & 58 & 13.792 & +28 & 10 & 57.20 \\
d200     & \ \      & gmp4379  & \ \      & a35     & S0       & \ \ & 16.08    & 1.82     & 12 & 58 & 15.032 & +28 & 7 & 33.25 \\
\ \      & \ \      & gmp4348  & \ \      & \ \     & \ \      & \ \ & 17.77    & 1.30     & 12 & 58 & 18.203 & +27 & 50 & 54.46 \\
d73      & \ \      & gmp4341  & rb183    & b24     & E        & E+A & 17.33    & 1.84     & 12 & 58 & 19.186 & +27 & 45 & 43.65 \\
d199     & ngc4851  & gmp4313  & \ \      & \ \     & S0       & \ \ & 16.00    & 1.95     & 12 & 58 & 21.722 & +28 & 8 & 56.18 \\
d137     & ngc4850  & gmp4315  & \ \      & a8      & E/S0     & \ \ & 15.39    & 1.87     & 12 & 58 & 21.828 & +27 & 58 & 4.05 \\
d44      & \ \      & gmp4255  & \ \      & b64     & S0       & E+A & 16.57    & 1.77     & 12 & 58 & 28.386 & +27 & 33 & 33.31 \\
d225     & \ \      & gmp4235  & \ \      & \ \     & S0       & \ \ & 16.80    & 1.53     & 12 & 58 & 29.503 & +28 & 18 & 4.60 \\
d161     & \ \      & gmp4230  & rb241    & \ \     & E        & \ \ & 15.19    & 1.87     & 12 & 58 & 30.202 & +28 & 0 & 53.20 \\
d59      & \ \      & gmp4209  & rb188    & \ \     & E        & \ \ & 16.90    & 1.85     & 12 & 58 & 31.596 & +27 & 40 & 24.73 \\
d182     & \ \      & gmp4200  & rb243    & a15     & S0       & \ \ & 16.84    & 1.72     & 12 & 58 & 31.908 & +28 & 2 & 58.66 \\
d43      & ngc4853  & gmp4156  & \ \      & b42     & S0p      & E+A & 14.38    & 1.66     & 12 & 58 & 35.193 & +27 & 35 & 47.00 \\
d197     & ic3943   & gmp4130  & \ \      & \ \     & S0/a     & \ \ & 15.55    & 1.97     & 12 & 58 & 36.343 & +28 & 6 & 49.46 \\
d28      & \ \      & gmp4117  & \ \      & b83     & E/S0     & \ \ & 16.67    & 1.99     & 12 & 58 & 38.405 & +27 & 32 & 39.09 \\
\ \      & \ \      & gmp4103  & rb245    & \ \     & \ \      & \ \ & 17.74    & 1.76     & 12 & 58 & 38.931 & +27 & 57 & 14.11 \\
\ \      & \ \      & gmp4083  & rb198    & a9/b3   & SA0      & \ \ & 17.82    & 1.91     & 12 & 58 & 40.780 & +27 & 49 & 37.41 \\
\ \      & \ \      & gmp4060  & rb199    & \ \     & \ \      & \ \ & 17.57    & 1.31     & 12 & 58 & 42.641 & +27 & 45 & 38.71 \\
d224     & \ \      & gmp4043  & \ \      & \ \     & S0       & \ \ & 17.19    & 1.77     & 12 & 58 & 43.903 & +28 & 16 & 57.62 \\
d91      & ic3946   & gmp3997  & \ \      & a57/b77 & S0       & \ \ & 15.28    & 1.95     & 12 & 58 & 48.723 & +27 & 48 & 37.72 \\
d181     & \ \      & gmp3972  & rb252    & a2      & S0       & \ \ & 16.52    & 1.87     & 12 & 58 & 50.767 & +28 & 5 & 2.47 \\
d72      & ic3947   & gmp3958  & \ \      & a74/b61 & E        & \ \ & 15.94    & 1.91     & 12 & 58 & 52.102 & +27 & 47 & 6.45 \\
d90      & \ \      & gmp3943  & rb209    & a69     & S0       & \ \ & 16.93    & 1.88     & 12 & 58 & 53.020 & +27 & 48 & 48.51 \\
d136     & \ \      & gmp3914  & rb257    & \ \     & E        & \ \ & 16.57    & 1.81     & 12 & 58 & 55.254 & +27 & 57 & 53.02 \\
d71      & \ \      & gmp3882  & rb214    & a96/b44 & S0       & \ \ & 16.97    & 1.85     & 12 & 58 & 57.638 & +27 & 47 & 7.81 \\
d42      & \ \      & gmp3879  & \ \      & b55     & S0       & \ \ & 16.31    & 1.86     & 12 & 58 & 58.103 & +27 & 35 & 41.06 \\
d135     & \ \      & gmp3851  & rb260    & \ \     & E        & \ \ & 16.98    & 1.86     & 12 & 59 & 0.068 & +27 & 58 & 3.19 \\
\ \      & \ \      & gmp3829  & \ \      & \ \     & \ \      & \ \ & 17.44    & 1.85     & 12 & 59 & 1.590 & +27 & 32 & 12.87 \\
d194     & ngc4860  & gmp3792  & \ \      & \ \     & E        & \ \ & 14.69    & 1.93     & 12 & 59 & 3.902 & +28 & 7 & 25.29 \\
d134     & \ \      & gmp3794  & rb261    & \ \     & E        & \ \ & 17.37    & 1.98     & 12 & 59 & 4.143 & +27 & 57 & 33.07 \\
d108     & \ \      & gmp3782  & rb262    & a76     & S0       & \ \ & 16.55    & 1.85     & 12 & 59 & 4.639 & +27 & 54 & 39.69 \\
d109     & ic3960   & gmp3733  & \ \      & \ \     & S0       & \ \ & 15.85    & 1.89     & 12 & 59 & 7.948 & +27 & 51 & 17.95 \\
d69      & ic3959   & gmp3730  & \ \      & a19/b86 & E        & \ \ & 15.27    & 1.94     & 12 & 59 & 8.211 & +27 & 47 & 3.10 \\
\ \      & \ \      & gmp3706  & rb223    & \ \     & \ \      & \ \ & 17.61    & 1.85     & 12 & 59 & 9.626 & +27 & 52 & 2.71 \\
d53      & \ \      & gmp3697  & rb224    & a50/b93 & E        & \ \ & 16.59    & 1.87     & 12 & 59 & 10.302 & +27 & 37 & 11.70 \\
d159     & ngc4864  & gmp3664  & \ \      & a58     & E        & \ \ & 14.70    & \ \      & 12 & 59 & 13.176 & +27 & 58 & 36.55 \\
d68      & ic3963   & gmp3660  & \ \      & \ \     & S0       & \ \ & 15.76    & 1.87     & 12 & 59 & 13.493 & +27 & 46 & 28.73 \\
d133     & ngc4867  & gmp3639  & \ \      & a82     & E        & \ \ & 15.44    & 1.83     & 12 & 59 & 15.227 & +27 & 58 & 14.88 \\
\ \      & \ \      & gmp3588  & \ \      & b43     & \ \      & \ \ & 17.76    & 1.72     & 12 & 59 & 18.453 & +27 & 30 & 48.74 \\
\ \      & \ \      & gmp3585  & \ \      & \ \     & \ \      & \ \ & 17.29    & \ \      & 12 & 59 & 18.541 & +27 & 35 & 36.67 \\
d107     & \ \      & gmp3557  & rb6      & \ \     & E        & \ \ & 16.35    & 1.81     & 12 & 59 & 20.162 & +27 & 53 & 9.56 \\
d158     & \ \      & gmp3534  & rb7      & \ \     & S0       & \ \ & 17.20    & 1.77     & 12 & 59 & 21.393 & +27 & 58 & 24.96 \\
d105     & ngc4869  & gmp3510  & \ \      & \ \     & E        & \ \ & 14.97    & 2.06     & 12 & 59 & 23.356 & +27 & 54 & 41.89 \\
d67      & \ \      & gmp3493  & rb230    & \ \     & S0       & \ \ & 16.50    & 1.94     & 12 & 59 & 24.924 & +27 & 44 & 19.93 \\
d132     & \ \      & gmp3487  & rb13     & \ \     & S0       & \ \ & 16.63    & 1.88     & 12 & 59 & 25.320 & +27 & 58 & 4.73 \\
d157     & \ \      & gmp3484  & rb14     & \ \     & S0       & \ \ & 16.26    & 1.81     & 12 & 59 & 25.479 & +27 & 58 & 23.72 \\
d156     & \ \      & gmp3471  & rb18     & a56     & E/S0     & \ \ & 16.45    & \ \      & 12 & 59 & 26.585 & +27 & 59 & 54.69 \\
d88      & ic3976   & gmp3423  & \ \      & a21     & S0       & \ \ & 15.80    & 1.95     & 12 & 59 & 29.393 & +27 & 51 & 0.56 \\
d87      & \ \      & gmp3403  & rb234    & \ \     & E        & \ \ & 16.87    & 1.79     & 12 & 59 & 30.632 & +27 & 47 & 29.31 \\
d103     & ic3973   & gmp3400  & \ \      & a68     & S0/a     & \ \ & 15.32    & 1.88     & 12 & 59 & 30.823 & +27 & 53 & 3.27 \\
d155     & ngc4873  & gmp3367  & \ \      & a20     & S0       & \ \ & 15.15    & 1.91     & 12 & 59 & 32.781 & +27 & 59 & 1.16 \\
\hline
\end{tabular}
\end{minipage}
\end{table*}
\begin{table*}
\begin{minipage}{160mm}
\contcaption{}
\begin{tabular}{lllllllccrrrrrr}
\hline \hline 
n1 & n2 & n3 & n4 & n5 & \multicolumn{2}{l}{type} & $b_j$ & $b-r$ & \multicolumn{3}{c}{RA (J2000)} & \multicolumn{3}{c}{DEC (J2000)} \\
\hline
d130     & ngc4872  & gmp3352  & \ \      & a47     & E/S0     & \ \ & 14.79    & 1.78     & 12 & 59 & 34.110 & +27 & 56 & 48.85 \\
d129     & ngc4874  & gmp3329  & \ \      & \ \     & cD       & \ \ & 12.78    & \ \      & 12 & 59 & 35.694 & +27 & 57 & 33.62 \\
\ \      & \ \      & gmp3298  & \ \      & \ \     & \ \      & \ \ & 17.26    & 1.79     & 12 & 59 & 37.838 & +27 & 46 & 36.68 \\
d104     & ngc4875  & gmp3296  & \ \      & a54     & S0       & \ \ & 15.88    & 1.96     & 12 & 59 & 37.904 & +27 & 54 & 26.40 \\
d154     & \ \      & gmp3291  & rb38     & a7      & S0       & \ \ & 16.41    & 1.78     & 12 & 59 & 38.304 & +27 & 59 & 14.08 \\
d153     & \ \      & gmp3213  & rb45     & \ \     & E        & \ \ & 16.14    & 1.83     & 12 & 59 & 43.730 & +27 & 59 & 40.84 \\
d124     & ngc4876  & gmp3201  & \ \      & a66     & E        & \ \ & 15.51    & 1.91     & 12 & 59 & 44.393 & +27 & 54 & 44.97 \\
d152     & ic3998   & gmp3170  & \ \      & a59     & SB0      & \ \ & 15.70    & 1.90     & 12 & 59 & 46.770 & +27 & 58 & 26.13 \\
d57      & \ \      & gmp3165  & \ \      & a4      & S0/a     & \ \ & 15.15    & 1.78     & 12 & 59 & 47.138 & +27 & 42 & 37.32 \\
\ \      & \ \      & gmp3129  & rb153    & \ \     & \ \      & \ \ & 17.94    & 1.71     & 12 & 59 & 50.271 & +28 & 8 & 40.61 \\
\ \      & \ \      & gmp3126  & rb60     & \ \     & \ \      & \ \ & 17.55    & 1.82     & 12 & 59 & 51.000 & +27 & 49 & 58.78 \\
\ \      & \ \      & gmp3113  & rb58     & \ \     & \ \      & \ \ & 17.82    & 1.81     & 12 & 59 & 51.750 & +28 & 5 & 54.80 \\
d85      & \ \      & gmp3092  & \ \      & \ \     & E        & \ \ & 17.55    & 1.59     & 12 & 59 & 54.870 & +27 & 47 & 45.63 \\
d193     & \ \      & gmp3084  & rb155    & a16     & E        & \ \ & 16.43    & 1.82     & 12 & 59 & 55.095 & +28 & 7 & 42.21 \\
d175     & ngc4883  & gmp3073  & \ \      & a97     & S0       & \ \ & 15.43    & 1.89     & 12 & 59 & 56.012 & +28 & 2 & 5.09 \\
d123     & \ \      & gmp3068  & rb64     & \ \     & SB0      & \ \ & 16.47    & 1.93     & 12 & 59 & 56.685 & +27 & 55 & 48.45 \\
\ \      & \ \      & gmp3058  & rb66     & \ \     & \ \      & \ \ & 17.71    & 1.78     & 12 & 59 & 57.600 & +28 & 3 & 54.47 \\
d217     & ngc4881  & gmp3055  & \ \      & \ \     & E        & \ \ & 14.73    & 1.87     & 12 & 59 & 57.738 & +28 & 14 & 48.02 \\
\ \      & \ \      & gmp3017  & rb71     & \ \     & \ \      & \ \ & 17.91    & 1.65     & 13 & 0 & 0.936 & +27 & 56 & 43.95 \\
\ \      & \ \      & gmp3012  & \ \      & \ \     & \ \      & \ \ & 17.49    & 1.83     & 13 & 0 & 1.530 & +27 & 43 & 50.39 \\
d216     & \ \      & gmp2989  & rb160    & a65     & Sa       & E+A & 17.05    & \ \      & 13 & 0 & 2.998 & +28 & 14 & 25.16 \\
d151     & ngc4886  & gmp2975  & \ \      & a95     & E        & \ \ & 14.83    & 1.76     & 13 & 0 & 4.448 & +27 & 59 & 15.45 \\
\ \      & \ \      & gmp2960  & rb74     & \ \     & SA0      & \ \ & 16.78    & 1.74     & 13 & 0 & 5.396 & +28 & 1 & 28.24 \\
d84      & \ \      & gmp2956  & \ \      & a51     & S0       & \ \ & 16.20    & 1.98     & 13 & 0 & 5.503 & +27 & 48 & 27.87 \\
\ \      & \ \      & gmp2942  & \ \      & \ \     & \ \      & \ \ & 16.34    & 1.73     & 13 & 0 & 6.263 & +27 & 41 & 7.01 \\
d65      & \ \      & gmp2945  & \ \      & a11     & S0       & \ \ & 16.15    & 1.77     & 13 & 0 & 6.285 & +27 & 46 & 32.93 \\
d150     & ic4011   & gmp2940  & \ \      & a86     & E        & \ \ & 16.08    & 1.82     & 13 & 0 & 6.383 & +28 & 0 & 14.94 \\
d174     & ic4012   & gmp2922  & \ \      & \ \     & E        & \ \ & 15.93    & 1.86     & 13 & 0 & 7.997 & +28 & 4 & 42.89 \\
d148     & ngc4889  & gmp2921  & \ \      & \ \     & cD       & \ \ & 12.62    & 1.91     & 13 & 0 & 8.125 & +27 & 58 & 37.22 \\
d207     & \ \      & gmp2912  & rb167    & a45     & E        & \ \ & 16.07    & 1.80     & 13 & 0 & 9.109 & +28 & 10 & 13.49 \\
d40      & \ \      & gmp2894  & \ \      & \ \     & S0       & \ \ & 17.15    & 1.84     & 13 & 0 & 10.413 & +27 & 35 & 42.20 \\
d64      & \ \      & gmp2866  & \ \      & a94     & E        & \ \ & 16.90    & 1.79     & 13 & 0 & 12.629 & +27 & 46 & 54.75 \\
d122     & ngc4894  & gmp2815  & \ \      & a12     & S0       & \ \ & 15.87    & 1.74     & 13 & 0 & 16.510 & +27 & 58 & 3.16 \\
d171     & \ \      & gmp2805  & rb91     & a17     & S0       & \ \ & 16.57    & 1.78     & 13 & 0 & 17.024 & +28 & 3 & 50.24 \\
d206     & ngc4895  & gmp2795  & \ \      & a24     & S0       & \ \ & 14.38    & \ \      & 13 & 0 & 17.915 & +28 & 12 & 8.57 \\
\ \      & \ \      & gmp2783  & \ \      & \ \     & \ \      & \ \ & 17.37    & 1.83     & 13 & 0 & 18.569 & +27 & 48 & 56.09 \\
\ \      & \ \      & gmp2778  & rb94     & \ \     & SB0/a    & \ \ & 16.69    & 1.81     & 13 & 0 & 18.767 & +27 & 56 & 13.52 \\
d39      & \ \      & gmp2776  & \ \      & \ \     & S0/E     & \ \ & 16.17    & 1.89     & 13 & 0 & 19.101 & +27 & 33 & 13.37 \\
d170     & ic4026   & gmp2727  & \ \      & a23     & SB0      & \ \ & 15.73    & 1.77     & 13 & 0 & 22.123 & +28 & 2 & 49.26 \\
\ \      & \ \      & gmp2721  & \ \      & \ \     & \ \      & \ \ & 17.50    & 1.82     & 13 & 0 & 22.376 & +27 & 37 & 24.85 \\
\ \      & \ \      & gmp2688  & \ \      & \ \     & \ \      & \ \ & 17.71    & 1.87     & 13 & 0 & 25.165 & +27 & 33 & 8.25 \\
d27      & \ \      & gmp2670  & \ \      & \ \     & E        & \ \ & 16.45    & 1.88     & 13 & 0 & 26.833 & +27 & 30 & 56.26 \\
d147     & \ \      & gmp2651  & rb100    & a93     & S0       & \ \ & 16.19    & 1.85     & 13 & 0 & 28.376 & +27 & 58 & 20.77 \\
d26      & \ \      & gmp2640  & \ \      & \ \     & S0p      & \ \ & 16.18    & \ \      & 13 & 0 & 29.210 & +27 & 30 & 53.72 \\
d232     & ngc4896  & gmp2629  & \ \      & \ \     & S0       & \ \ & 15.06    & 2.01     & 13 & 0 & 30.762 & +28 & 20 & 47.12 \\
d63      & \ \      & gmp2615  & \ \      & \ \     & S0/a     & \ \ & 16.97    & 1.90     & 13 & 0 & 32.508 & +27 & 45 & 58.27 \\
d83      & \ \      & gmp2603  & \ \      & \ \     & S0       & \ \ & 17.36    & 1.80     & 13 & 0 & 33.357 & +27 & 49 & 27.44 \\
d192     & \ \      & gmp2584  & \ \      & a61     & S0       & \ \ & 16.14    & 1.79     & 13 & 0 & 35.572 & +28 & 8 & 46.15 \\
d38      & \ \      & gmp2582  & \ \      & \ \     & Sbc      & \ \ & 16.20    & 1.74     & 13 & 0 & 35.709 & +27 & 34 & 27.27 \\
d118     & ngc4906  & gmp2541  & \ \      & a62     & E        & \ \ & 15.44    & 1.98     & 13 & 0 & 39.753 & +27 & 55 & 26.45 \\
d145     & ic4041   & gmp2535  & \ \      & a42     & S0       & \ \ & 15.93    & 1.90     & 13 & 0 & 40.830 & +27 & 59 & 47.81 \\
d144     & ic4042   & gmp2516  & \ \      & \ \     & S0/a     & \ \ & 15.34    & 1.86     & 13 & 0 & 42.761 & +27 & 58 & 16.87 \\
d116     & \ \      & gmp2510  & rb113    & a64     & SB0      & \ \ & 16.13    & 1.90     & 13 & 0 & 42.825 & +27 & 57 & 47.44 \\
d231     & \ \      & gmp2495  & \ \      & \ \     & S0       & \ \ & 15.78    & 2.09     & 13 & 0 & 44.226 & +28 & 20 & 14.26 \\
d191     & \ \      & gmp2489  & rb116    & \ \     & S0       & \ \ & 16.69    & 1.77     & 13 & 0 & 44.629 & +28 & 6 & 2.38 \\
d117     & \ \      & gmp2457  & rb119    & a83     & S0/a     & \ \ & 16.56    & 1.88     & 13 & 0 & 47.383 & +27 & 55 & 19.76 \\
d168     & ic4045   & gmp2440  & \ \      & a6      & E        & \ \ & 15.17    & 1.85     & 13 & 0 & 48.631 & +28 & 5 & 26.92 \\
d205     & ngc4907  & gmp2441  & \ \      & \ \     & Sb       & \ \ & 14.65    & 1.74     & 13 & 0 & 48.804 & +28 & 9 & 30.30 \\
\ \      & \ \      & gmp2421  & \ \      & a81     & \ \      & \ \ & 17.98    & 1.90     & 13 & 0 & 51.124 & +27 & 44 & 34.43 \\
d167     & ngc4908  & gmp2417  & \ \      & \ \     & S0/E     & \ \ & 14.91    & 1.87     & 13 & 0 & 51.525 & +28 & 2 & 35.10 \\
d62      & \ \      & gmp2393  & \ \      & a25     & S0       & \ \ & 16.51    & 1.90     & 13 & 0 & 54.217 & +27 & 47 & 2.60 \\
\hline
\end{tabular}
\end{minipage}
\end{table*}
\begin{table*}
\begin{minipage}{160mm}
\contcaption{}
\begin{tabular}{lllllllccrrrrrr}
\hline \hline 
n1 & n2 & n3 & n4 & n5 & \multicolumn{2}{l}{type} & $b_j$ & $b-r$ & \multicolumn{3}{c}{RA (J2000)} & \multicolumn{3}{c}{DEC (J2000)} \\
\hline
d143     & ic4051   & gmp2390  & \ \      & \ \     & E        & \ \ & 14.47    & 1.82     & 13 & 0 & 54.457 & +28 & 0 & 27.59 \\
\ \      & \ \      & gmp2385  & rb122    & \ \     & \ \      & \ \ & 17.62    & 1.82     & 13 & 0 & 54.769 & +27 & 50 & 31.47 \\
d50      & \ \      & gmp2355  & \ \      & \ \     & SBa      & \ \ & 16.56    & 1.81     & 13 & 0 & 58.371 & +27 & 39 & 7.64 \\
d98      & \ \      & gmp2347  & rb124    & a78     & S0/a     & \ \ & 15.85    & 1.91     & 13 & 0 & 59.262 & +27 & 53 & 59.59 \\
d81      & \ \      & gmp2252  & \ \      & \ \     & E        & \ \ & 16.10    & 1.85     & 13 & 1 & 9.215 & +27 & 49 & 6.00 \\
\ \      & \ \      & gmp2251  & rb128    & \ \     & \ \      & \ \ & 17.35    & 1.79     & 13 & 1 & 9.435 & +28 & 1 & 59.25 \\
\ \      & \ \      & gmp2201  & rb129    & a43     & unE      & \ \ & 16.86    & 1.85     & 13 & 1 & 13.616 & +27 & 54 & 51.64 \\
d79      & ngc4919  & gmp2157  & \ \      & a88     & S0       & \ \ & 15.06    & 1.92     & 13 & 1 & 17.595 & +27 & 48 & 32.95 \\
\ \      & \ \      & gmp2141  & rb131    & \ \     & \ \      & \ \ & 17.78    & 1.44     & 13 & 1 & 19.317 & +27 & 51 & 37.94 \\
d204     & \ \      & gmp2091  & \ \      & \ \     & E        & \ \ & 15.99    & 1.75     & 13 & 1 & 22.767 & +28 & 11 & 45.86 \\
d142     & \ \      & gmp2048  & rb133    & a49     & E        & \ \ & 17.06    & 1.94     & 13 & 1 & 27.147 & +27 & 59 & 57.20 \\
d78      & ngc4923  & gmp2000  & \ \      & a36     & E        & \ \ & 14.78    & 1.93     & 13 & 1 & 31.794 & +27 & 50 & 51.37 \\
\ \      & \ \      & gmp1986  & \ \      & \ \     & \ \      & \ \ & 17.91    & 1.78     & 13 & 1 & 33.817 & +27 & 54 & 40.39 \\
\hline
\multicolumn{15}{l}{{\em Notes:}\/ RA and DEC are given in J2000 coordinates. Columns 1--5
give the different names associated with the} \\
\multicolumn{15}{l}{galaxy according to the following key:}\\
`n1'    & \multicolumn{14}{l}{names from Dressler (1980a)} \\
`n2'    & \multicolumn{14}{l}{names from New General Catalogue or Index Catalogue (Dreyer 1888,1908)}\\
`n3'    & \multicolumn{14}{l}{names from Godwin, Metcalfe \& Peach (1983)} \\
`n4'    & \multicolumn{14}{l}{names from Rood \& Baum (1967)}\\
`n5'    & \multicolumn{14}{l}{names from Caldwell et al. (1993). a = Table 1(a), b = Table 1(b).
64 out of 125 galaxies in common}\\
\multicolumn{15}{l}{Columns 6--8 are described below:} \\
`type'  & \multicolumn{14}{l}{morphological types from Dressler (1980a). `E+A' typing from Caldwell et al. (1993)}\\
`$b_j$' & \multicolumn{14}{l}{magnitudes from Godwin, Metcalfe \& Peach (1983), accurate to $\pm{0.15}$}\\
`$b-r$' & \multicolumn{14}{l}{colours from Godwin, Metcalfe \& Peach (1983), accurate to $\pm{0.15}$. Note that $b-r\simeq{{\rm{B}}-{\rm{R}}}$}\\
\end{tabular}
\end{minipage}
\end{table*}

\section{Observations}
\label{sec:observations}

Observations were carried out over 3 half-nights between 13-18 April 1999
with the wide-field (1$^\circ$ $\equiv$ 1.26\,$h^{-1}$\,Mpc 
at Coma with H$_0$ = 100 $h$ km\,s$^{-1}$\,Mpc$^{-1}$)
multi-object spectroscopy instrument AUTOFIB2/WYFFOS and
the H1800V grating on the 
William Herschel 4.2\,metre telescope (WHT).
Table \ref{tab:sumobs} summarises the observation parameters.

\begin{table}
\caption{The principal parameters of the observations.}
\label{tab:sumobs}
\begin{tabular}{lr}
\hline
\hline
Telescope                   & WHT 4.2m at ING, La Palma\\
Instrument                  & WYFFOS + AUTOFIB2 \\
Field diameter              & 1$^\circ$ $\equiv$ 1.26\,$h^{-1}$\,Mpc at Coma \\
Unvignetted field           & 40 arcmin \\
Number of fibres            & 126 \\
Fibre diameter              & $126\times153\mu$m \\
\ \                         & (2.7$''$ $\equiv$ 0.94\,$h^{-1}$\,kpc at Coma) \\
Positioning accuracy        & better than 10\,$\mu$m (i.e. 0.18$''$ rms) \\
CCD                         & thinned Tektronix (TEK6), 1k $\times$ 1k \\
Pixel size                  & $24 \times 24\,\mu{\rm{m}} \equiv 0.93 \times 0.93\,\AA$ \\
Grating used                & H1800V \\
Resolution                  & $\sim$2.2--4.0~\AA\, FWHM \\
Gain                        & 1.7e$^-$ per ADU \\
Readout noise               & 5.6e$^-$ \\
Wavelength range$^\dag$    & 4600--5600~\AA \\
Number of nights            & 3 half nights \\
Typical exposure time     & $6\times1650$\,secs \\
$\;\;\;\;$ per configuration & \ \ \\
Date of observations        & 13--18 April 1999 \\
Observers                   & Stephen Moore, John Lucey \\
Field centre (J2000)        & 12$^h$59$^m$32.9$^s$, +27$^\circ$55$'$49$''$ \\
\hline
\end{tabular} \\
\begin{minipage}{8.25cm}
{\em Notes:}\/ $^\dag$the wavelength range varies by $\sim4$ per cent due to the 
stepping of the fibres on the CCD
\end{minipage}
\end{table}

Exposure times of typically 6$\times$1650\,secs per configuration 
were sufficient to obtain high S/N ($\geq35$ per \AA) on
our programme objects ($b_j\leq18.0$). 
Each individual exposure on the brighter galaxies ($b_j\leq16.0$) was also 
long enough to achieve the same S/N.
Therefore a large 
number of repeats both during a night and night-to-night were gathered
enabling a detailed treatment of the random and day-to-day systematic errors
of our velocity dispersion and line-strength measurements.

\begin{figure}
\psfig{file=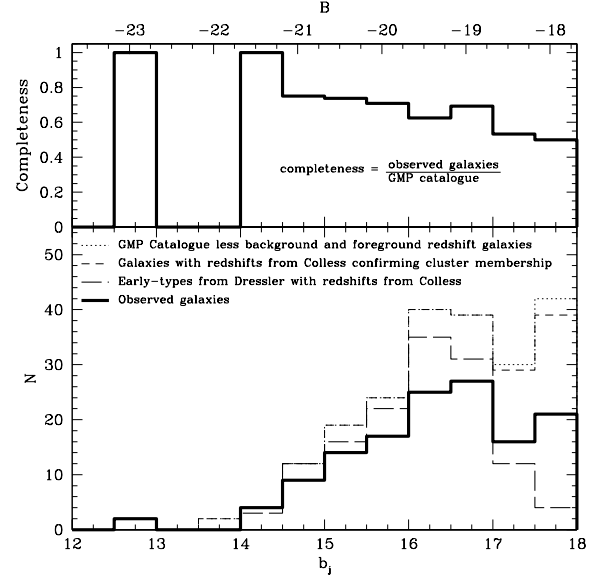,width=80mm} 
\caption{Luminosity function of observed galaxies overlaid on top of the
luminosity functions of the sample defined in Section \ref{sec:selection}.
The completeness function of the observed galaxies with respect to
the galaxies with cluster membership confirmed by redshifts is shown
at the top of the figure. Both apparent and absolute magnitudes are plotted
along the x-axes; this assumes a distance modulus of 35.68\,mag (see footnote \ref{fn:distancemod}).}
\label{fig:luminosityfunction}
\end{figure}

\begin{figure}
\psfig{file=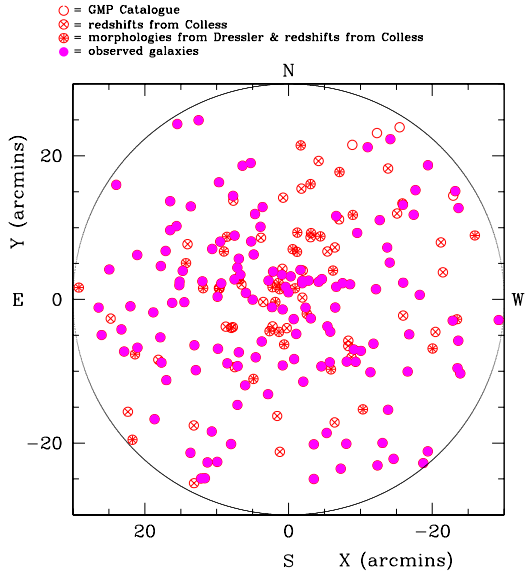,width=80mm}
\caption{Coma galaxies observed (solid points) overlaid on the
total sample defined in Section \ref{sec:selection}, i.e. galaxies
with confirmed cluster membership and with $b_j\leq18.0$.}
\label{fig:observed}
\end{figure}

The completeness of the observations versus 
the sample defined in Section \ref{sec:selection} is summarised in 
Table \ref{tab:completeness}, Fig. \ref{fig:luminosityfunction} and Fig. \ref{fig:observed}.
These completeness calculations assume that the GMP catalogue is 100 per cent complete
within the central 1$^\circ$ of the Coma cluster core down to the faint limit
of this study;
this is justified
as at this magnitude limit missing blue compact dwarfs and potential problems
of stellar contamination are at a minimum.
High S/N spectroscopic data of a 
homogeneous nature were collected for 73 per cent 
of the known Coma cluster early-type galaxies brighter than $b_j=18.0$. 

\begin{table}
\caption{Completeness of observed sample of bright
early-type galaxies ($b_j\leq18.0$) within a 1$^\circ$ field centred
on the Coma cluster all with cluster membership confirmed by redshifts.}
\label{tab:completeness}
\begin{centering}
\begin{tabular}{lr}
\hline
\hline
early-type galaxies                & 100 / 137 = 73 per cent \\
galaxies with no morphologies      &  30 / 55  = 55 per cent \\
late-type galaxies                 &   5 / 18  = 28 per cent \\
total number of galaxies observed  & 135 / 210 = 64 per cent \\
\hline
\end{tabular} \\
\end{centering}
\end{table}

~\\
~\\

\section{Data reduction}
\label{sec:datared}

\subsection{Basic data reduction}
\label{sec:basicdatared}

The first step in the raw reduction was performed with the \verb+IRAF+ WYFFOS
data reduction software (Pollacco et al. 1999). This software performs the following basic tasks:
bias subtraction; aperture identification; scattered light 
correction; flat fielding; throughput correction; fibre extraction;
wavelength calibration (using an Argon I lamp); and sky subtraction.
Cosmic ray and night sky line removal were then performed using our own software.
The wavelength calibration residuals had a median rms of 0.050 to 0.083~\AA\, for 
the three field configurations.

Individual spectra from a particular night were  summed together to 
produce a spectrum of higher S/N. Where galaxies
were observed on multiple nights 
it was decided not to co-add the spectra but rather
treat each night as a separate measurement. This was because
each spectra had sufficient S/N to yield high quality
measurements without co-adding.

\subsection{Redshift and velocity dispersion measurement}
\label{sec:redshift_veldisp}

Velocity dispersions, $\sigma$ of the central 2.7$''$ region of a galaxy 
(determined by the fibre diameter, equivalent to 0.94\,$h^{-1}$\,kpc) 
are measured using
the well-known Fourier Quotient method of Sargent et al. (1977). 
Recession velocities, $cz$, are obtained simultaneously with the velocity
dispersions, as a result of the Fourier Quotient fit.

In the Fourier Quotient method a galaxy spectrum is approximated by the 
convolution of a representative stellar spectrum with 
an appropriate broadening function. This function is then calculated in 
Fourier parameter space.
Prior to transforming to Fourier space,
 continuum fits are
subtracted (using a 5th order
polynomial) from both the stellar template spectrum and the galaxy spectrum, and
modulated by a cosine bell function  to fix the ends of the spectrum to zero. 
The latter step is necessary to avoid unphysical signals appearing at all 
frequencies in the Fourier transforms.
These spectra then require filtering in Fourier space to remove signals arising
from noise, inadequate continuum removal and the application of the cosine bell.  
A cut is made at high frequencies, to suppress channel-to-channel 
noise. The resulting $\sigma$ values are fairly insensitive
to the exact value, $k_{\rm{high}}$, chosen for the high frequency
cut.  $k_{\rm{high}} = 200$ has been used
throughout.  At low frequencies, a filter must be applied to
remove residual continuum features and the effects of the cosine-bell
modulation function described above. For the case of the low frequency cut,
results are found to exhibit a clear trend: velocity dispersions are measured 
to be smaller when $k_{\rm{low}}$ is larger.
The cut-off frequency must therefore be chosen with care.  
It is required that the low frequency filter should remove the signal arising from
the cosine bell modulation, whilst preserving intrinsic features in
spectra of velocity dispersion $\leq500$\,km\,s$^{-1}$. For the spectra herein, 
these constraints leave a range of $k_{\rm{low}} = 6 - 9$ over 
which an average velocity dispersion is calculated.

In order for the recovered width to represent only the intrinsic velocity 
broadening of the galaxy spectrum, it is necessary to ensure that the stellar 
spectrum has been subject to the same instrumental resolution effects as
the galaxy spectrum. This is done by observing standard stars throughout 
an observing run 
and comparing these spectra to the observed galaxy spectra.
In this study, five radial velocity standard stars (stellar types G8{\sc{iii}} to K3{\sc{iii}})
were observed down a number of different fibres.
Since these observed standard stars have small, but non-zero radial velocities it
is necessary to first zero-redshift them.
This is performed by applying the Fourier Quotient method to calculate the
radial velocity of the standard stars against a high resolution (0.5~\AA\, per pixel),
very high S/N ($\sim250-1000$ per \AA) 
spectrum that has been precisely zero-redshifted through
the identification of spectral features and then shifting them
to their laboratory rest frame. Four spectra were kindly supplied
by Claire Halliday (\emph{private communication}, hereafter referred 
to as the Halliday spectra). These allow the standard stars
to be zero-redshifted to an accuracy of $\pm6$\,km\,s$^{-1}$.

Redshifts and velocity dispersions were measured both from each galaxy exposure
and from the combined galaxy exposures for each night the galaxy was observed.
The redshifts are corrected to heliocentric redshifts and the results
averaged (weighting by their S/N) to yield a final value
for a galaxy.

\subsection{Spectral resolution variation}
\label{sec:fieldvariation}

Using any multi-fibre spectroscopy instrument introduces intra-fibre and fibre-to-fibre 
variations in spectral resolution and throughput that necessarily have to be removed before 
an accurate
analysis can be undertaken.
These variations are due to the optical 
performance of the telescope plus instrumentation setup both across the 
field and along the slit where the fibres are fed into the spectrograph.
These variations are found to be significant in the WHT/WYFFOS system and
affect velocity dispersion and line-strength measurements (though to a 
lesser extent, see Section \ref{sec:veldispcorr}). 
In this section we describe how these variations are quantified.

The variation in
spectral resolution
was mapped by analysing the calibration
spectra taken on each night\footnote{It would have been preferable to use twilight sky flats to characterise
the performance of the system, but these observations could not be obtained}.
The arc lamps were
located at the Cassegrain focus of the WHT and shine their 
light on to the tertiary mirror from which it is then relayed to
the fibres. These allowed accurate mapping of the vignetting caused
by the IDS grating H1800V convolved with a function representing
the vignetting of the telescope optics from that point onwards. This
approach is capable of removing the majority of the overall vignetting, since any
effects superimposed by the telescope optics prior to the tertiary mirror
are significantly smaller than the dominant effects subsequent to the 
tertiary mirror.

\begin{figure*}
\begin{minipage}{170mm}
\begin{tabular}{cc}
\psfig{file=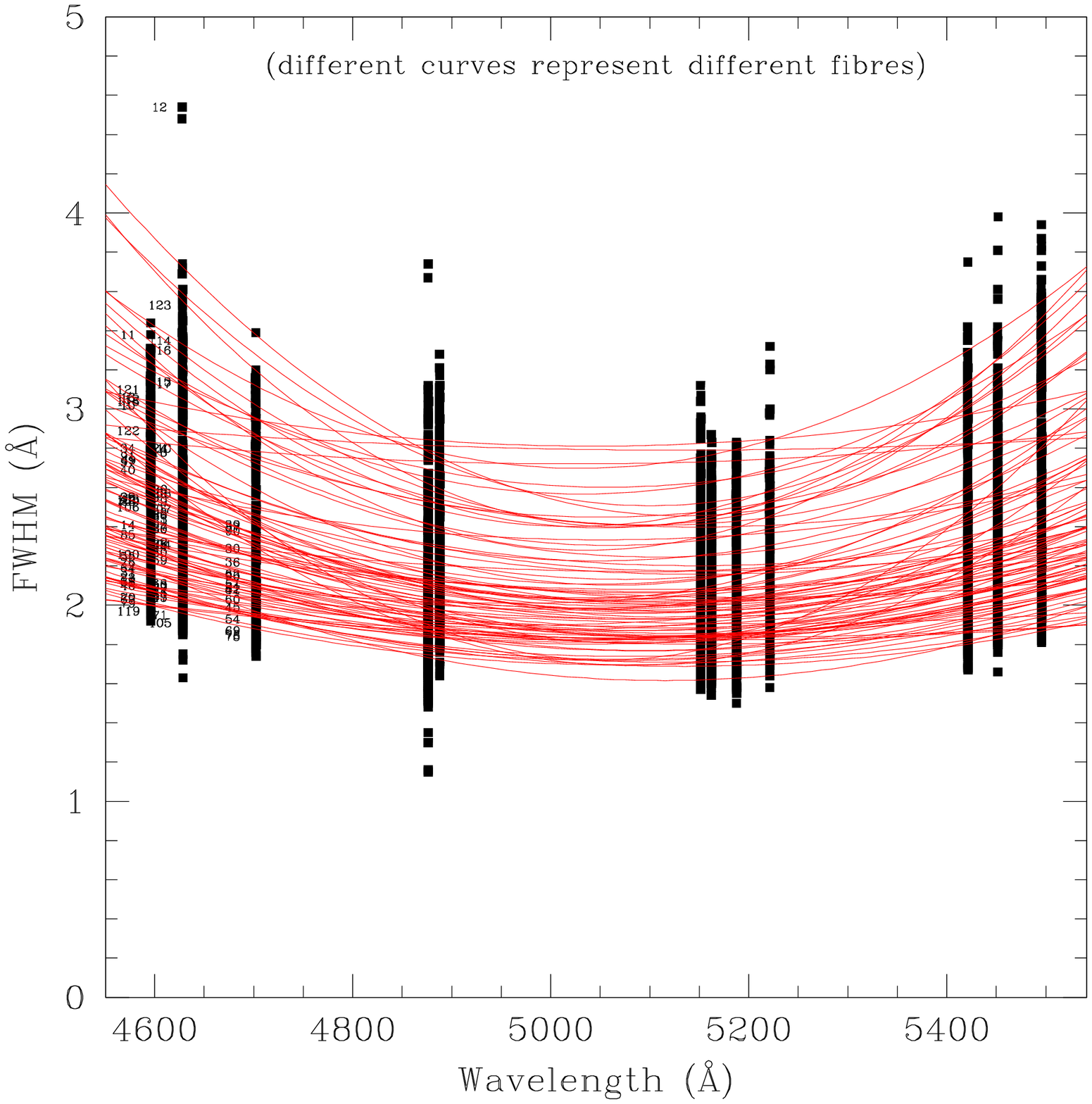,width=80mm} &
\psfig{file=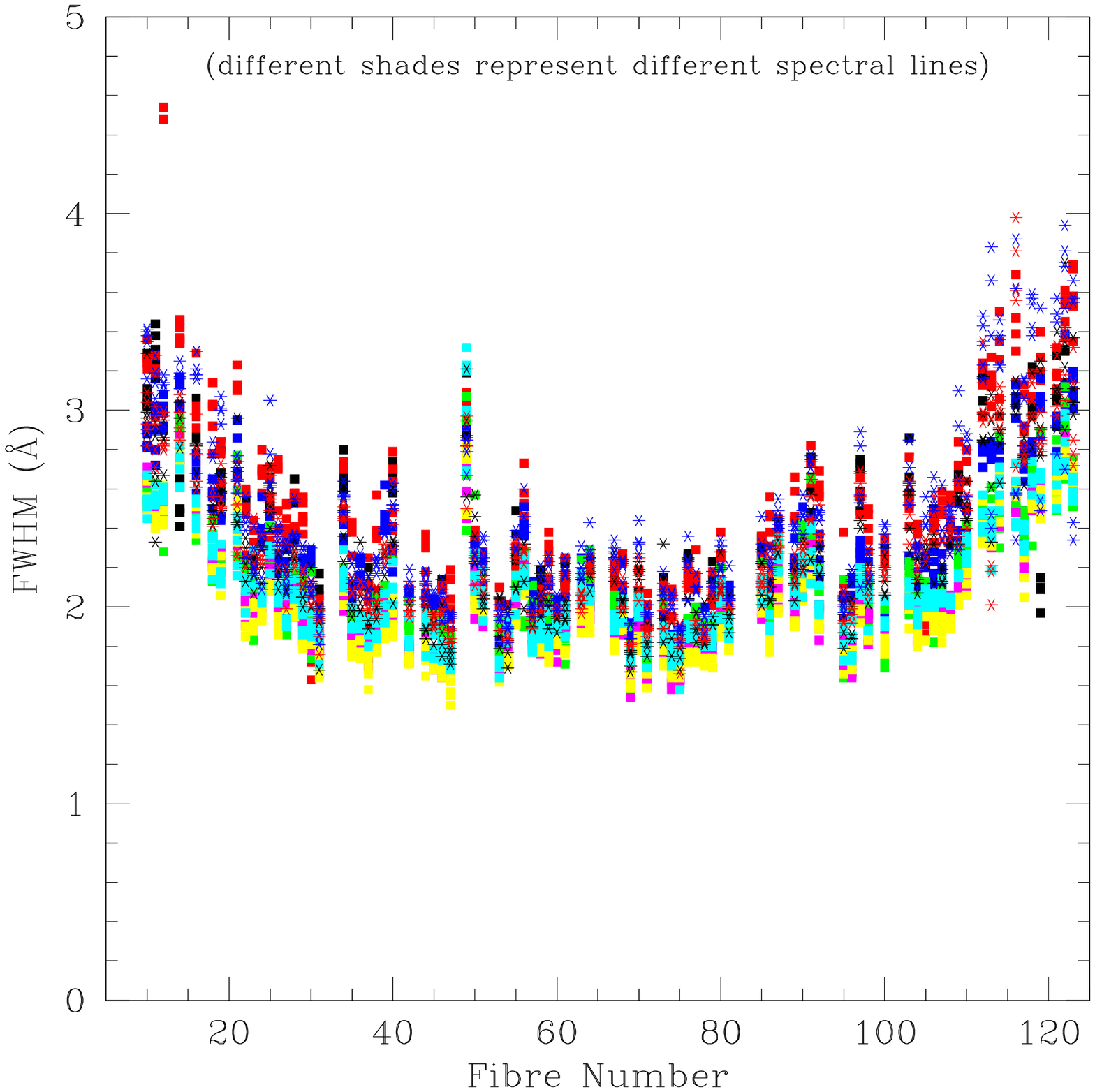,width=80mm} \\
(a) & (b) \\
\end{tabular}
\caption{Dependence of Argon I emission line width on wavelength and fibre 
number.
Fig. (a) shows the spectral resolution dependence on wavelength, whilst Fig. (b)
shows the dependence on fibre number. 
Each point corresponds to a spectral line width measurement. Fig. (a) has a 
2nd order polynomial fit
to the spectral resolution variation with wavelength for each fibre superimposed.
In Fig. (b) the points are shaded to indicate the different line width measurements
for a fibre.
See text for a more detailed explanation. }
\label{fig:fwhm_variations}
\end{minipage}
\end{figure*}

The results of this mapping can be seen in Fig. \ref{fig:fwhm_variations}.
A clear and sizable variation in spectral resolution is seen for each of the
WYFFOS fibres. 
This effect is inherent in all telescopes due to the 
natural vignetting caused by their optical performance. However
it is exacerbated in the WHT/WYFFOS system by the mis-match between the \emph{smaller}\/ IDS grating H1800V used
and the \emph{larger}\/ WYFFOS beam; this mis-match further vignettes the spectra observed in fibres at the 
top and bottom ends of the slit.
The widths of the Argon I calibration spectrum can be 
seen to vary from $\sim4$~\AA\,
at the end of the slit and at the edge of the field to $\sim1.7$~\AA\,
at the centre of the slit and at the centre of the field.
This variation is a function both of fibre number (reflecting 
the position of the fibre on the slit and within the observed field) and of
wavelength. 
These variations in line width will 
affect any velocity dispersion measurements (with a maximum error of 12\,km\,s$^{-1}$
for $\sigma=100$\,km\,s$^{-1}$), 
but will not change the redshift measurements (there is no systematic shift introduced by 
this problem) and will have a minor effect on line strengths
(since the spectrum are broadened before measurement, see Section \ref{sec:lickids}).

In principle it is possible 
that the variable throughput of the fibres might play a role here, with 
a dependence of spectral resolution on the strength of an arc line. 
However after further investigation this was ruled out.
There could also be a further complication in that the tracking of the field centre 
on the Coma cluster during an exposure is not perfect, causing a drift in field 
position (and hence fibre position) relative to the target being observed. This drift, both
during an exposure and between exposures, could introduce a time dependence into the 
mapping of the fibre characteristics. This effect was investigated by 
comparing the first and last spectra observed on a night and was found 
to be negligible ($\sim0.1$~\AA) compared to the other effects discussed.

\section{Velocity dispersion corrections}
\label{sec:veldispcorr}

\subsection{Modelling the effect of intra--fibre and fibre--fibre variations}
\label{sec:modellingfieldvariation}

The first stage in removing the effect of any intra--fibre and fibre--fibre variations
is to fit a function to the spectral resolution variation. 
We fit 2nd order polynomials to the variation with wavelength 
for each fibre and each field configuration (see Fig. \ref{fig:fwhm_variations}a).
These functions can then be convolved with an ideal template spectrum and the result used to cross-correlate 
against that of an observed galaxy to find its dispersion with any intrinsic 
fibre variation removed. The problem with this method is in having a 
template spectrum of very high resolution that is not itself suffering from any 
internal or instrumental variation.

To counter this problem we have devised the following method. 
We use the Halliday spectra 
which have been precisely zero-redshifted through the identification 
and then subsequent shifting of spectral features 
to their laboratory rest frame. These are the same spectra that were previously 
used to removed any redshift from observed standard stars in Section \ref{sec:redshift_veldisp}.
These spectra will still suffer from some intrinsic variation (due to e.g. the telescope/instrument 
setup they were obtained with), but this is unimportant 
in the proposed method. These spectra are convolved with a particular fibre model 
in linear wavelength space, resulting 
in a `template spectrum'. The new template spectra are then cross-correlated against 
a mock galaxy created by convolving the original high-resolution spectra 
with a fibre model (not necessarily 
the same fibre) and broadening it by a fixed amount in logarithmic wavelength space 
(to simulate the Doppler broadening caused 
by a galaxy). A correction can then be calculated for each fibre configuration and each 
galaxy dispersion case:

\begin{equation}
\begin{array}{c} 
\rm{dispersion} \\
\rm{correction}
\end{array}
\quad = \quad
\begin{array}{c} 
\rm{measured} \\
\rm{dispersion}
\end{array}
\quad - \quad
\begin{array}{c} 
\rm{true} \\
\rm{dispersion}
\end{array}
\label{eq:dispcorr}
\end{equation}

\begin{figure}
\psfig{file=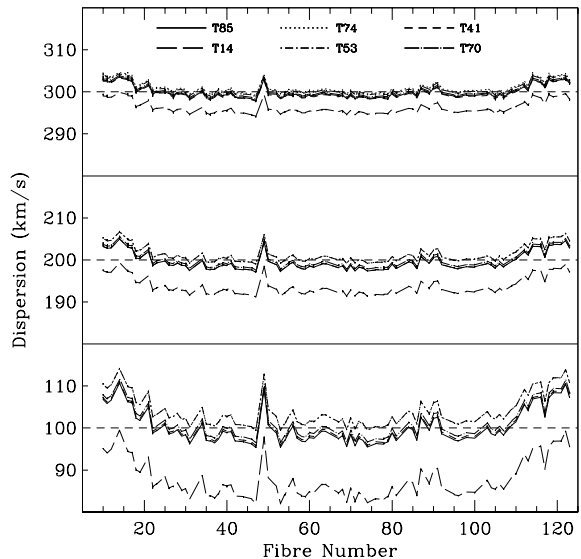,width=80mm}
\caption{Velocity dispersion measured versus galaxy fibre number 
for template stars observed with different fibres (template fibre
numbers are given by the key at the top of the figure). 
Three plots are shown for the Halliday 
spectra broadened to 100, 200 and 300\,km\,s$^{-1}$.
The dashed line separate from the others corresponds to velocity
dispersions measured versus a template star observed with fibre 14.
This fibre suffers a large amount of spectral variation which 
greatly affects velocity dispersion measurements.
See text.}
\label{fig:fibres1}
\end{figure}

\begin{figure}
\psfig{file=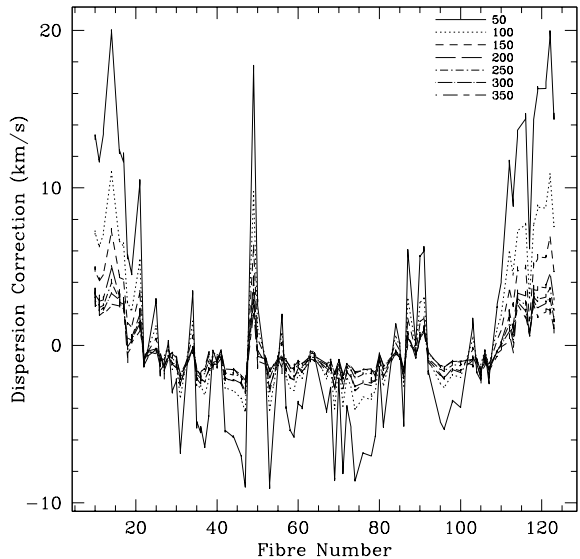,width=80mm}
\caption{Velocity dispersion correction versus galaxy fibre number 
for one template star observed with a particular fibre
cross-correlated against a mock galaxy observed with different 
fibres and with various broadening factors (the different
broadenings used are given by the key in the top right of the figure,
these are 50--350\,km\,s$^{-1}$). See text.}
\label{fig:fibres2}
\end{figure}

\begin{figure}
\psfig{file=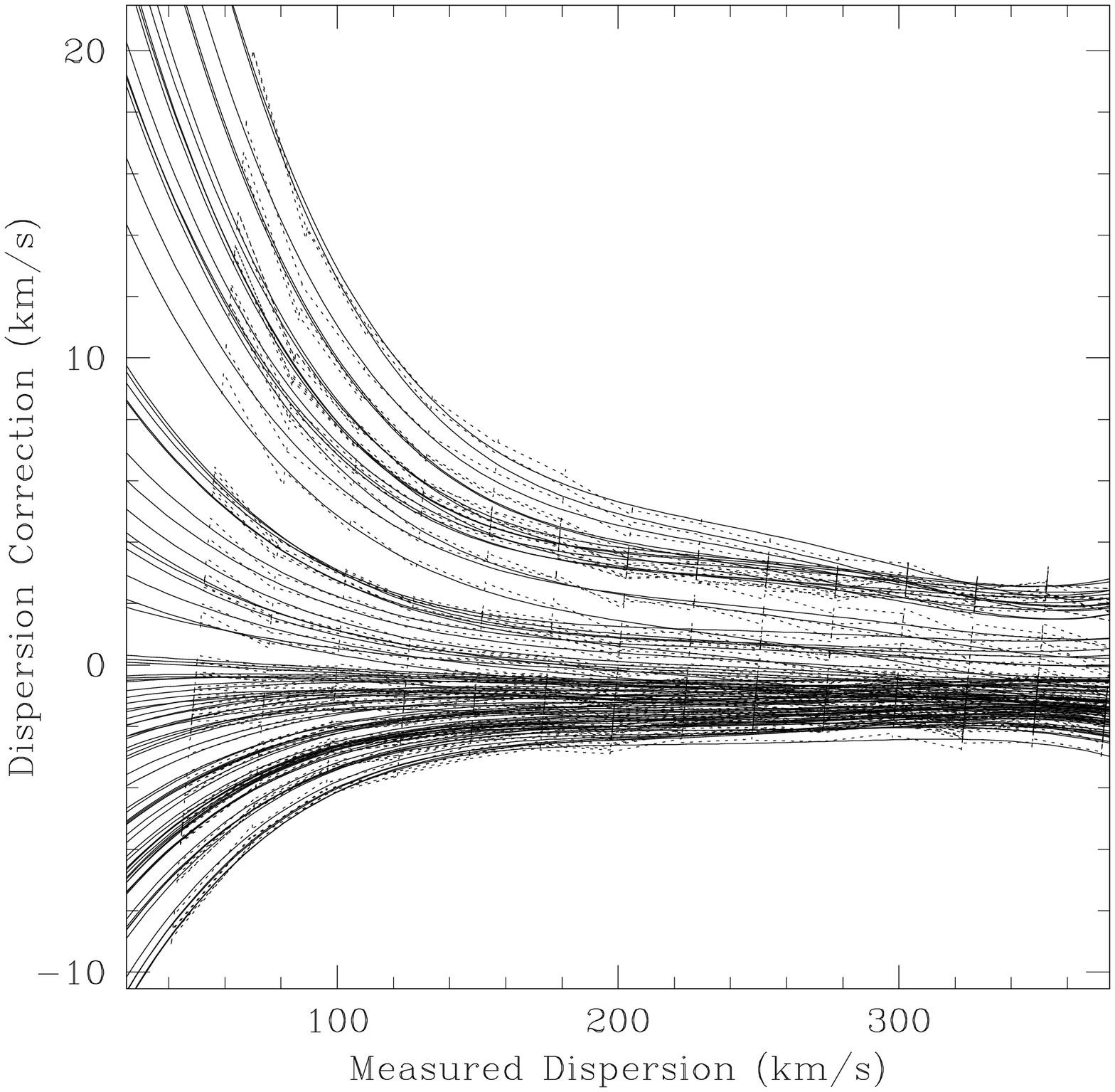,width=80mm}
\caption{Velocity dispersion correction versus measured velocity 
dispersion for each fibre in a particular field setup cross-correlated against a standard star 
observed with a particular fibre (in this case fibre number 41). 
The curves are constructed by taking a 
Halliday spectrum, 
convolving it with a previously computed broadening function for a given
fibre number and then broadening it by a given velocity dispersion from 
25 to 350\,km\,s$^{-1}$. This resultant mock galaxy spectra is then 
cross-correlated against a similar template spectra which has been convolved with a 
broadening function for a different fibre (matching a fibre number with
which a standard star was observed during the run). The computed 
velocity dispersion is compared to the true velocity dispersion and a correction
computed. A curve is then fit to the variation of the velocity 
dispersion correction with measured velocity dispersion. This curve (overlaid 
on the plot) can then subsequently used to correct an actual galaxy velocity 
dispersion measurement. 
Ideally all lines should be straight and coincident with a zero correction. 
See text.}
\label{fig:fibres3}
\end{figure}

\noindent
These corrections are then used to modify the real calculated dispersions 
which are calculated using template spectra observed on the night to cross-correlate 
against the galaxy spectra. This is done by subtracting the calculated correction from 
the measured value. In this way the `true' (or best estimate) dispersion 
is derived, with any modifications due to intra--fibre and fibre--fibre variations removed.
Some results of this method can be seen in Figs. \ref{fig:fibres1} to \ref{fig:fibres3}.

Figs. \ref{fig:fibres1} and \ref{fig:fibres2} show the size of the 
velocity dispersion correction required
versus galaxy fibre number for a range of template stars with different
broadening factors (simulating the typical velocity dispersions
of the observed galaxies) and different spectral resolution variation functions
(simulating the variations introduced through observing a galaxy with different fibres). 
Ideally each plot should be a straight line, however this is not the case 
because of the WHT/WYFFOS spectral resolution variations. A correction
therefore needs to be applied to the velocity dispersion measurements
of the observed early-type galaxies. This correction is a function
of the fibre number that the galaxy was observed with and of the fibre number
that the star used for the velocity dispersion measurement was observed with.
The size of the correction is also a function of the velocity dispersion
of the galaxy observed (Fig. \ref{fig:fibres3}), with larger velocity
dispersions requiring a smaller correction.
These corrections are typically small, but where an unfavourable pairing between 
galaxy fibre and standard star fibre occurs the velocity dispersion correction can be as large 
as 20\,km\,s$^{-1}$ for low velocity dispersion galaxies ($\sigma\simeq50$\,km\,s$^{-1}$). 
However for the typical velocity dispersions of the 
galaxies that we observed in this project ($\sigma\sim100$\,km\,s$^{-1}$ and greater)
the corrections are not large, but are significant.
As expected though, when a galaxy is cross-correlated against a standard star observed
in the same field configuration and with the same fibre the velocity dispersion 
correction is zero.

This modelling of the effect of intra-fibre and fibre-fibre variations 
results in accurate
velocity dispersions for the galaxies, which are subsequently required during the Lick/IDS
stellar population index measurement process (see Section \ref{sec:veldispcorrections}).
Bootstrap tests on the accuracy of this method using the 
high-resolution, very high S/N 
stellar template spectra have shown that the random errors for the 
velocity dispersion corrections are $\sim1-{2}$\,km\,s$^{-1}$, demonstrating the 
success of this approach.

\subsection{Redshift and velocity dispersion data}
{\label{sec:final_kinematics}}

Table \ref{tab:kinematics} lists
the heliocentric redshift (135 galaxies) and central velocity dispersion data (132 galaxies) 
for the galaxies observed in the Coma cluster in this study.
The dispersions have been corrected for the field variations (Section \ref{sec:fieldvariation}).
These data are average values (weighted by their S/N) of all the 
measurements from the multiple exposures. 
Blank entries indicate a measurement was not possible.

The redshift errors are calculated by combining the error in the
wavelength calibration in quadrature with the error resulting from
the cross-correlation technique plus the template mis-matching
error (calculated through cross-correlating the galaxy spectrum
against \emph{different}\/ stellar spectra) and an additional 
error component factor (calculated from the variance between multiple
exposures on a galaxy cross-correlated against a single stellar spectrum).
The median heliocentric redshift error is 12\,km\,s$^{-1}$.

The velocity dispersion errors are calculated by combining the 
error resulting from the cross-correlation technique
in quadrature with the error resulting from
template mis-matching (calculated through cross-correlating the galaxy spectrum
against \emph{different}\/ stellar spectra) plus an additional 
error component factor (calculated from the variance between multiple
exposures on a galaxy cross-correlated against a single stellar spectrum).
The median velocity dispersion error is 0.015\,dex.

\begin{table}
\caption{Heliocentric redshifts and central velocity dispersions. }
\label{tab:kinematics}
\begin{centering}
\begin{tabular}{llccr}
\hline \hline 
name     & type  & S/N & \multicolumn{1}{c}{cz$_\odot$ (km\,s$^{-1}$)} & \multicolumn{1}{c}{$\sigma$ (km\,s$^{-1}$)} \\ 
\hline
d26      & S0p      &  53.5 &   7396 \/ $\pm$ \/ 12 &     71.5 \/ $\pm$ \/   9.4 \\
d27      & E        &  41.3 &   7762 \/ $\pm$ \/ 12 &    107.4 \/ $\pm$ \/   3.6 \\
d28      & E/S0     &  57.9 &   5974 \/ $\pm$ \/ 12 &    103.5 \/ $\pm$ \/   4.5 \\
d29      & E        &  33.9 &   6973 \/ $\pm$ \/ 16 &     63.1 \/ $\pm$ \/   8.6 \\
d38      & Sbc      &  38.8 &   5084 \/ $\pm$ \/ 12 &     71.3 \/ $\pm$     14.4 \\
d39      & S0/E     &  76.1 &   5897 \/ $\pm$ \/ 12 &    120.4 \/ $\pm$ \/   3.4 \\
d40      & S0       &  47.0 &   5597 \/ $\pm$ \/ 12 &     72.9 \/ $\pm$ \/   6.2 \\
d42      & S0       &  80.7 &   6016 \/ $\pm$ \/ 12 &    147.1 \/ $\pm$ \/   7.0 \\
d44      & S0       &  55.7 &   7533 \/ $\pm$ \/ 12 &     55.4 \/ $\pm$     11.5 \\
d50      & SBa      &  38.4 &   5211 \/ $\pm$ \/ 11 &     54.0 \/ $\pm$ \/   6.3 \\
d53      & E        &  80.2 &   5742 \/ $\pm$ \/ 12 &    128.4 \/ $\pm$ \/   5.4 \\
d57      & S0/a     &  97.4 &   8384 \/ $\pm$ \/ 12 &    142.5 \/ $\pm$ \/   4.7 \\
d59      & E        &  66.0 &   6947 \/ $\pm$ \/ 12 &    129.9 \/ $\pm$ \/   5.0 \\
d62      & S0       &  51.9 &   8359 \/ $\pm$ \/ 16 &    126.2 \/ $\pm$     10.9 \\
d63      & S0/a     &  34.8 &   6675 \/ $\pm$ \/ 12 &     87.3 \/ $\pm$ \/   4.8 \\
d64      & E        &  50.5 &   7010 \/ $\pm$ \/ 12 &     80.9 \/ $\pm$ \/   5.6 \\
d65      & S0       &  65.1 &   6191 \/ $\pm$ \/ 12 &    116.3 \/ $\pm$ \/   3.2 \\
d67      & S0       &  52.3 &   6039 \/ $\pm$ \/ 12 &    150.8 \/ $\pm$ \/   2.0 \\
d71      & S0       &  42.3 &   6919 \/ $\pm$ \/ 12 &     63.9 \/ $\pm$ \/   7.7 \\
d73      & E        &  49.2 &   5440 \/ $\pm$ \/ 12 &     73.5 \/ $\pm$ \/   5.5 \\
d74      & E        &  27.9 &   5793 \/ $\pm$ \/ 11 &     41.1 \/ $\pm$     10.9 \\
d75      & S0       &  48.2 &   6132 \/ $\pm$ \/ 13 &     79.6 \/ $\pm$ \/   5.8 \\
d81      & E        &  48.7 &   5928 \/ $\pm$ \/ 12 &    143.3 \/ $\pm$ \/   2.3 \\
d83      & S0       &  31.3 &   8184 \/ $\pm$ \/ 12 &     37.5 \/ $\pm$ \/   9.8 \\
d84      & S0       &  46.8 &   6553 \/ $\pm$ \/ 11 &    120.6 \/ $\pm$ \/   3.5 \\
d85      & E        &  42.4 &   8251 \/ $\pm$ \/ 12 &     65.0 \/ $\pm$ \/   5.8 \\
d87      & E        &  63.2 &   7770 \/ $\pm$ \/ 12 &     94.0 \/ $\pm$ \/   4.7 \\
d90      & S0       &  52.0 &   5522 \/ $\pm$ \/ 12 &     88.5 \/ $\pm$ \/   4.1 \\
d93      & S0       &  78.4 &   6063 \/ $\pm$ \/ 12 &    136.3 \/ $\pm$ \/   4.9 \\
d98      & S0/a     &  77.7 &   6868 \/ $\pm$ \/ 12 &    130.0 \/ $\pm$ \/   5.4 \\
d107     & E        &  39.3 &   6491 \/ $\pm$ \/ 12 &     87.7 \/ $\pm$ \/   3.7 \\
d108     & S0       &  66.8 &   6424 \/ $\pm$ \/ 12 &    115.9 \/ $\pm$ \/   3.2 \\
d110     & S0/E     &  60.3 &   6948 \/ $\pm$ \/ 12 &    114.4 \/ $\pm$ \/   3.2 \\
d112     & E        &  50.8 &   7433 \/ $\pm$ \/ 13 &     58.3 \/ $\pm$ \/   6.5 \\
d116     & SB0      &  75.7 &   8437 \/ $\pm$ \/ 12 &    123.2 \/ $\pm$ \/   4.2 \\
d117     & S0/a     &  38.2 &   8561 \/ $\pm$ \/ 12 &     93.1 \/ $\pm$ \/   4.8 \\
d123     & SB0      &  50.0 &   7712 \/ $\pm$ \/ 12 &    100.6 \/ $\pm$ \/   3.3 \\
d132     & S0       &  46.7 &   7698 \/ $\pm$ \/ 12 &     96.2 \/ $\pm$ \/   3.5 \\
d134     & E        &  63.7 &   7009 \/ $\pm$ \/ 12 &    126.7 \/ $\pm$ \/   2.2 \\
d135     & E        &  36.8 &   8323 \/ $\pm$ \/ 12 &    100.2 \/ $\pm$ \/   3.9 \\
d136     & E        &  82.0 &   5682 \/ $\pm$ \/ 11 &    168.8 \/ $\pm$ \/   2.3 \\
d142     & E        &  79.0 &   7652 \/ $\pm$ \/ 12 &    161.4 \/ $\pm$ \/   2.3 \\
d147     & S0       &  58.9 &   7713 \/ $\pm$ \/ 12 &    107.7 \/ $\pm$ \/   3.9 \\
d153     & E        &  52.7 &   6684 \/ $\pm$ \/ 11 &    127.9 \/ $\pm$ \/   2.7 \\
d154     & S0       &  51.1 &   6833 \/ $\pm$ \/ 11 &     57.1 \/ $\pm$ \/   5.0 \\
d156     & E/S0     &  51.8 &   6671 \/ $\pm$ \/ 12 &     84.8 \/ $\pm$ \/   7.9 \\
d157     & S0       &  74.8 &   6107 \/ $\pm$ \/ 12 &    131.5 \/ $\pm$ \/   2.4 \\
d158     & S0       &  28.9 &   6058 \/ $\pm$ \/ 12 &     64.8 \/ $\pm$ \/   6.1 \\
d161     & E        &  86.9 &   7146 \/ $\pm$ \/ 12 &    190.3 \/ $\pm$ \/   4.9 \\
d171     & S0       &  81.0 &   6135 \/ $\pm$ \/ 12 &    127.5 \/ $\pm$ \/   2.9 \\
d181     & S0       &  63.0 &   6090 \/ $\pm$ \/ 12 &    120.3 \/ $\pm$ \/   4.5 \\
d182     & S0       &  44.0 &   5702 \/ $\pm$ \/ 12 &    120.2 \/ $\pm$ \/   2.3 \\
d191     & S0       &  44.4 &   6592 \/ $\pm$ \/ 12 &     90.9 \/ $\pm$ \/   5.2 \\
d192     & S0       &  56.4 &   5435 \/ $\pm$ \/ 12 &     87.5 \/ $\pm$ \/   5.5 \\
d193     & E        &  72.4 &   7567 \/ $\pm$ \/ 12 &    117.6 \/ $\pm$ \/   3.4 \\
d200     & S0       & 104.0 &   7466 \/ $\pm$ \/ 12 &    189.3 \/ $\pm$ \/   4.5 \\
d201     & S0       &  36.5 &   6409 \/ $\pm$ \/ 12 &     59.6 \/ $\pm$ \/   9.4 \\
d204     & E        &  53.1 &   7578 \/ $\pm$ \/ 12 &    126.1 \/ $\pm$ \/   4.0 \\
d207     & E        &  78.1 &   6743 \/ $\pm$ \/ 12 &    146.9 \/ $\pm$ \/   2.8 \\
d209     & S0       &  48.5 &   7182 \/ $\pm$ \/ 12 &     80.7 \/ $\pm$ \/   5.2 \\
d210     & E        &  66.6 &   7252 \/ $\pm$ \/ 12 &    144.6 \/ $\pm$ \/   3.8 \\
d216     & Sa       &  43.5 &   7684 \/ $\pm$ \/ 12 &     71.5 \/ $\pm$     13.0 \\
d224     & S0       &  42.2 &   7597 \/ $\pm$ \/ 12 &     59.5 \/ $\pm$ \/   6.2 \\
\hline
\end{tabular} \\
\end{centering}
\end{table}
\begin{table}
\contcaption{}
\begin{centering}
\begin{tabular}{llccr}
\hline \hline 
name     & type  & S/N & \multicolumn{1}{c}{cz$_\odot$ (km\,s$^{-1}$)} & \multicolumn{1}{c}{$\sigma$ (km\,s$^{-1}$)} \\ 
\hline
d225     & S0       &  38.1 &   5879 \/ $\pm$ \/ 14 &     71.7 \/ $\pm$ \/   6.7 \\
d231     & S0       &  62.9 &   7878 \/ $\pm$ \/ 13 &    127.8 \/ $\pm$ \/   5.0 \\
ic3943   & S0/a     &  97.8 &   6789 \/ $\pm$ \/ 12 &    168.6 \/ $\pm$ \/   1.9 \\
ic3946   & S0       &  73.8 &   5927 \/ $\pm$ \/ 12 &    199.6 \/ $\pm$ \/   2.6 \\
ic3947   & E        &  93.6 &   5675 \/ $\pm$ \/ 12 &    158.8 \/ $\pm$ \/   2.1 \\
ic3959   & E        &  95.1 &   7059 \/ $\pm$ \/ 12 &    215.9 \/ $\pm$ \/   6.0 \\
ic3960   & S0       &  95.5 &   6592 \/ $\pm$ \/ 12 &    174.3 \/ $\pm$ \/   2.9 \\
ic3963   & S0       &  74.7 &   6839 \/ $\pm$ \/ 12 &    122.4 \/ $\pm$ \/   3.9 \\
ic3973   & S0/a     &  78.3 &   4716 \/ $\pm$ \/ 12 &    228.0 \/ $\pm$ \/   3.1 \\
ic3976   & S0       & 105.8 &   6814 \/ $\pm$ \/ 14 &    255.2 \/ $\pm$ \/   6.4 \\
ic3998   & SB0      &  75.5 &   9420 \/ $\pm$ \/ 12 &    136.9 \/ $\pm$ \/   4.9 \\
ic4011   & E        &  52.5 &   7253 \/ $\pm$ \/ 11 &    123.2 \/ $\pm$ \/   3.6 \\
ic4012   & E        &  90.7 &   7251 \/ $\pm$ \/ 12 &    180.7 \/ $\pm$ \/   3.7 \\
ic4026   & SB0      &  86.3 &   8168 \/ $\pm$ \/ 12 &    132.2 \/ $\pm$ \/   3.0 \\
ic4041   & S0       &  76.6 &   7088 \/ $\pm$ \/ 12 &    132.5 \/ $\pm$ \/   2.3 \\
ic4042   & S0/a     &  67.8 &   6371 \/ $\pm$ \/ 12 &    170.6 \/ $\pm$ \/   3.3 \\
ic4045   & E        & 107.9 &   6992 \/ $\pm$ \/ 22 &    217.6 \/ $\pm$ \/   3.6 \\
ic4051   & E        &  56.1 &   4994 \/ $\pm$ \/ 12 &    228.8 \/ $\pm$ \/   2.5 \\
ngc4848  & Scd      &  46.7 &   7199 \/ $\pm$ \/ 16 &    106.8 \/ $\pm$ \/   7.4 \\
ngc4850  & E/S0     & 105.6 &   6027 \/ $\pm$ \/ 12 &    189.8 \/ $\pm$ \/   2.5 \\
ngc4851  & S0       &  50.0 &   7861 \/ $\pm$ \/ 12 &    126.8 \/ $\pm$ \/   3.3 \\
ngc4853  & S0p      &  88.5 &   7676 \/ $\pm$ \/ 12 &    140.8 \/ $\pm$ \/   4.4 \\
ngc4860  & E        &  76.6 &   7926 \/ $\pm$ \/ 12 &    277.3 \/ $\pm$ \/   7.2 \\
ngc4864  & E        & 103.4 &   6828 \/ $\pm$ \/ 12 &    187.6 \/ $\pm$ \/   3.2 \\
ngc4867  & E        & 117.3 &   4817 \/ $\pm$ \/ 12 &    208.5 \/ $\pm$ \/   2.0 \\
ngc4869  & E        & 101.9 &   6844 \/ $\pm$ \/ 12 &    203.1 \/ $\pm$ \/   4.4 \\
ngc4872  & E/S0     &  80.1 &   7198 \/ $\pm$ \/ 12 &    217.8 \/ $\pm$ \/   3.4 \\
ngc4873  & S0       & 100.8 &   5818 \/ $\pm$ \/ 12 &    176.9 \/ $\pm$ \/   1.8 \\
ngc4874  & cD       &  64.4 &   7180 \/ $\pm$ \/ 12 &    274.5 \/ $\pm$ \/   3.3 \\
ngc4875  & S0       &  88.7 &   8014 \/ $\pm$ \/ 13 &    180.1 \/ $\pm$ \/   4.3 \\
ngc4876  & E        &  82.0 &   6710 \/ $\pm$ \/ 12 &    164.1 \/ $\pm$ \/   3.1 \\
ngc4881  & E        &  94.7 &   6730 \/ $\pm$ \/ 12 &    193.9 \/ $\pm$ \/   4.9 \\
ngc4883  & S0       &  85.3 &   8161 \/ $\pm$ \/ 12 &    166.1 \/ $\pm$ \/   2.7 \\
ngc4886  & E        &  41.7 &   6377 \/ $\pm$ \/ 12 &    153.8 \/ $\pm$ \/   2.8 \\
ngc4889  & cD       & 141.6 &   6495 \/ $\pm$ \/ 13 &    397.5 \/ $\pm$     10.1 \\
ngc4894  & S0       &  55.0 &   4640 \/ $\pm$ \/ 12 &     85.6 \/ $\pm$ \/   3.8 \\
ngc4895  & S0       & 106.9 &   8458 \/ $\pm$ \/ 15 &    239.8 \/ $\pm$ \/   5.0 \\
ngc4896  & S0       &  67.7 &   5988 \/ $\pm$ \/ 18 &    164.0 \/ $\pm$ \/   2.6 \\
ngc4906  & E        &  91.4 &   7505 \/ $\pm$ \/ 12 &    175.0 \/ $\pm$ \/   4.4 \\
ngc4907  & Sb       &  56.8 &   5812 \/ $\pm$ \/ 12 &    148.2 \/ $\pm$ \/   2.6 \\
ngc4908  & S0/E     &  72.5 &   8710 \/ $\pm$ \/ 12 &    193.9 \/ $\pm$ \/   4.3 \\
ngc4919  & S0       & 121.0 &   7294 \/ $\pm$ \/ 12 &    191.5 \/ $\pm$ \/   3.1 \\
ngc4923  & E        & 109.0 &   5487 \/ $\pm$ \/ 12 &    198.3 \/ $\pm$ \/   3.5 \\
rb58     & \ \      &  22.6 &   7634 \/ $\pm$ \/ 12 &     50.1 \/ $\pm$ \/   6.7 \\
rb60     & \ \      &  34.7 &   7895 \/ $\pm$ \/ 12 &     57.1 \/ $\pm$ \/   6.8 \\
rb66     & \ \      &  30.7 &   5822 \/ $\pm$ \/ 11 &     43.0 \/ $\pm$ \/   6.4 \\
rb71     & \ \      &  35.4 &   6839 \/ $\pm$ \/ 12 &              \ \           \\
rb74     & SA0      &  32.2 &   5899 \/ $\pm$ \/ 11 &     63.8 \/ $\pm$ \/   4.8 \\
rb94     & SB0/a    &  28.7 &   5283 \/ $\pm$ \/ 12 &     57.6 \/ $\pm$ \/   6.4 \\
rb122    & \ \      &  33.4 &   7082 \/ $\pm$ \/ 11 &     77.3 \/ $\pm$ \/   6.3 \\
rb128    & \ \      &  36.0 &   7013 \/ $\pm$ \/ 12 &    150.3 \/ $\pm$ \/   2.4 \\
rb129    & unE      &  58.2 &   5852 \/ $\pm$ \/ 12 &     89.9 \/ $\pm$ \/   4.6 \\
rb131    & \ \      &  20.5 &   8209 \/ $\pm$ \/ 12 &     45.7 \/ $\pm$     11.3 \\
rb153    & \ \      &  22.9 &   6780 \/ $\pm$ \/ 12 &     51.6 \/ $\pm$ \/   6.9 \\
rb198    & SA0      &  31.1 &   6177 \/ $\pm$ \/ 12 &     54.8 \/ $\pm$ \/   5.7 \\
rb199    & \ \      &  21.2 &   8476 \/ $\pm$ \/ 46 &              \ \           \\
rb223    & \ \      &  64.0 &   6916 \/ $\pm$ \/ 12 &     94.4 \/ $\pm$ \/   3.6 \\
rb245    & \ \      &  25.1 &   6009 \/ $\pm$ \/ 11 &     47.6 \/ $\pm$ \/   6.1 \\
gmp1986  & \ \      &  13.3 &   6591 \/ $\pm$ \/ 12 &     22.8 \/ $\pm$     17.8 \\
gmp2421  & \ \      &  28.0 &   8132 \/ $\pm$ \/ 13 &     30.0 \/ $\pm$     38.6 \\
gmp2688  & \ \      &  30.1 &   7261 \/ $\pm$ \/ 12 &     58.8 \/ $\pm$ \/   4.7 \\
gmp2721  & \ \      &  28.7 &   7580 \/ $\pm$ \/ 11 &     55.6 \/ $\pm$ \/   5.4 \\
gmp2783  & \ \      &  22.6 &   5360 \/ $\pm$ \/ 12 &     39.8 \/ $\pm$     11.3 \\
\hline
\end{tabular} \\
\end{centering}
\end{table}
\begin{table}
\contcaption{}
\begin{centering}
\begin{tabular}{llccr}
\hline \hline 
name     & type  & S/N & \multicolumn{1}{c}{cz$_\odot$ (km\,s$^{-1}$)} & \multicolumn{1}{c}{$\sigma$ (km\,s$^{-1}$)} \\ 
\hline
gmp2942  & \ \      &  47.1 &   7542 \/ $\pm$ \/ 12 &    149.9 \/ $\pm$ \/   2.8 \\
gmp3012  & \ \      &  25.8 &   8041 \/ $\pm$ \/ 12 &     60.4 \/ $\pm$ \/   8.2 \\
gmp3298  & \ \      &  28.5 &   6786 \/ $\pm$ \/ 12 &     51.3 \/ $\pm$ \/   8.3 \\
gmp3585  & \ \      &  29.5 &   5178 \/ $\pm$ \/ 22 &     52.8 \/ $\pm$     23.1 \\
gmp3588  & \ \      &  24.1 &   6033 \/ $\pm$ \/ 13 &     55.5 \/ $\pm$ \/   7.2 \\
gmp3829  & \ \      &  18.6 &   8577 \/ $\pm$ \/ 12 &     48.4 \/ $\pm$ \/   5.0 \\
gmp4348  & \ \      &  29.2 &   7581 \/ $\pm$ \/ 12 &     56.3 \/ $\pm$     18.8 \\
gmp4420  & \ \      &  40.6 &   8520 \/ $\pm$ \/ 13 &     59.6 \/ $\pm$     12.0 \\
gmp4469  & \ \      &  15.8 &   7467 \/ $\pm$ \/ 12 &              \ \           \\
\hline
\end{tabular} \\
\end{centering}
\begin{minipage}{8.25cm}
{\em Notes:}\/ These results are average values (weighted by their S/N) 
of all of the measurements from the 
multiple exposures. Blank entries indicate a measurement was not possible.
There are a total of 135 galaxies in this data table.
Morphological types are taken from Dressler (1980a).
S/N is measured at the centre of index Fe5270.
\end{minipage}
\end{table}

\section{Stellar population absorption line strengths}
\label{sec:linestrengths}

One of the main goals of this study was to measure the luminosity weighted 
mean ages and metallicities of the dominant stellar populations within
the core of bright early-type galaxies. 
To measure these ages and metallicities, we used the Lick/IDS
system of line strength measurement and then compared the data
to models (e.g. Worthey 1994).
The principal line indices used were
H$\beta_{\rm{G}}$ (predominantly age dependent) and $[$MgFe$]$ (predominantly metallicity
dependent). 
This section details the measurement of these and other stellar population
absorption line strength indices in the wavelength range $\sim$4600--5600~\AA.

\subsection{Flux calibration}
\label{sec:fluxing}

It is first necessary to remove the overall instrument response function (IRF) from 
galaxy spectra prior to line strength measurement. Spectra are 
affected by: the response of the spectrograph optics; the response of the CCD; the
response of the grating; atmospheric conditions; and airmass. These
effects are removed by observing flux standard stars (see e.g. Massey et al. 1988). 
Flux calibration
of the galaxy spectra was performed by comparing the observed stars to their
standard spectrum and computing the IRF which
was then used to transform the galaxy spectra continuum. Since 
the conditions these observations were performed in were not photometric,
it is only possible to correct the galaxy spectra to some arbitrary flux units.
However this does not affect the shape of the spectral continuum nor the
line strength measurements. To improve the calculation of the
IRF a number of flux standard stars were observed during the run. 
The IRFs calculated from each star were then compared and 
a mean IRF derived using a technique that minimized their 
maximum absolute deviations (MAD).

It was not possible to measure flux standard stars down each
fibre and for each field configuration 
because of the prohibitively long observing time required.
This means that the computed IRF function is only an \emph{overall}\/
IRF and does not include fibre-to-fibre continuum shape
variations\footnote{Note that the fibre-to-fibre throughput variations
are already removed in the basic data reduction process}.
This fibre-to-fibre IRF variation is however small and would only
affect line strength measurements at the level of $<0.1$~\AA.

\begin{figure}
\psfig{file=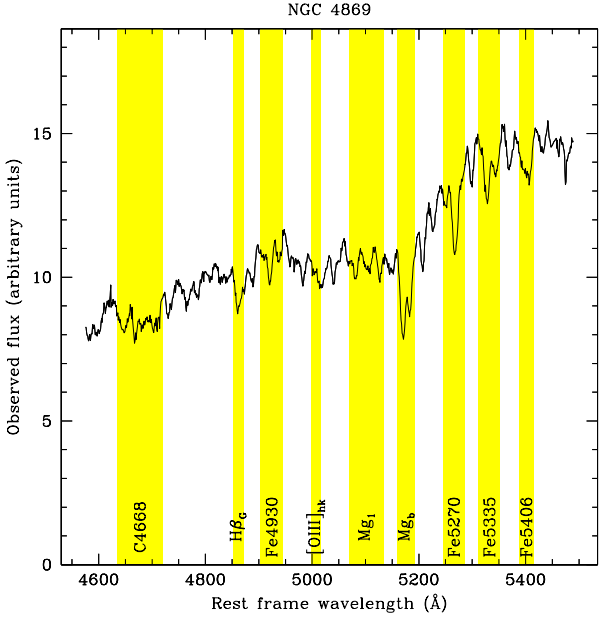,width=85mm}
\caption{Fluxed spectrum of galaxy NGC 4869 shifted to its rest
frame wavelength. The Lick/IDS features present in the spectrum
are shaded.}
\label{fig:ngc4869}
\end{figure}

A typical flux calibrated galaxy spectrum is shown in Fig. \ref{fig:ngc4869}.

\subsection{Lick/IDS system}
\label{sec:lickids}

\begin{figure}
\begin{centering}
\begin{tabular}{c}
\psfig{file=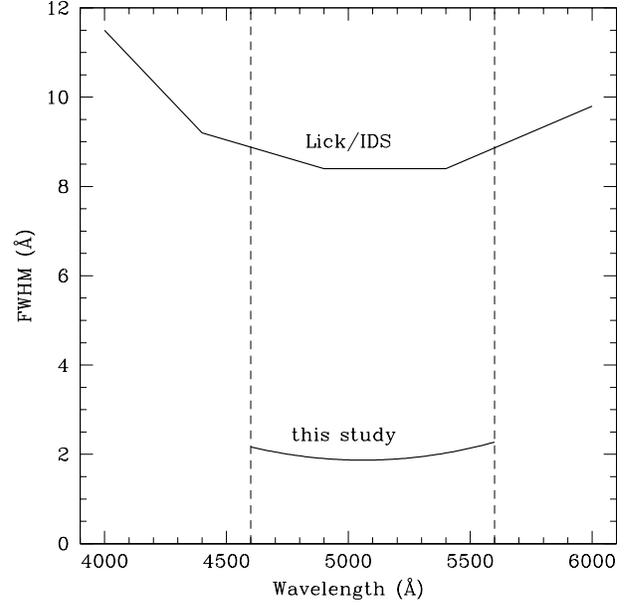,width=85mm} \\
\end{tabular}
\caption{Comparison of the variation in FWHM with wavelength for this study 
and for the Lick/IDS system.
The mean resolution of this study is $\sim$2.2~\AA\, (see Fig. \ref{fig:fwhm_variations}),
whereas the mean resolution 
of the Lick/IDS system is 9~\AA. This Lick/IDS resolution varies to values 
30 per cent higher at the ends of the spectra. To measure line strengths on the 
Lick/IDS system it is necessary to broaden any higher resolution spectra to 
the same resolution (Worthey \& Ottaviani 1997).}
\label{fig:lickplot}
\end{centering}
\end{figure}

Absorption line strengths were measured using 
the Lick/IDS system of indices, where a central feature
bandpass is flanked on either side by pseudo-continuum 
bandpasses (see Trager et al. 1998 for details). 
The Lick/IDS system is based upon spectra with a mean
resolution of 9~\AA\, (Fig. \ref{fig:lickplot}).
The spectra in this study need to be broadened to
the same resolution to measure indices on the Lick/IDS system.
This is done by calculating a transformation function
using the known resolution function of a galaxy spectrum (Section \ref{sec:fieldvariation})
and the Lick/IDS resolution function (see Fig. \ref{fig:lickplot} or Worthey \& Ottaviani 1997):

\begin{equation}
{\rm{FWHM}}_{trans}(\lambda)^2 = {\rm{FWHM}}_{Lick}(\lambda)^2 - {\rm{FWHM}}_{gal}(\lambda)^2
\end{equation}

\noindent
The spectrum is then broadened in linear wavelength space using a sliding Gaussian smoothing function derived
from this transformation function.

\subsection{Line strength measurement}
\label{sec:linestrengthmeasure}

Indices were measured by first zero-redshifting galaxy spectra 
to the laboratory rest frame 
using the previously measured 
heliocentric redshifts corrected back to the geocentric rest frame. 
Then the mean height in each of the two pseudo-continuum regions was
determined in either side of the feature bandpass, and a straight line 
drawn through the mid-point of each one. The difference in flux between 
this line and the observed spectrum within the feature bandpass determines 
the index. For narrow features, the indices are expressed in angstroms (\AA)
of equivalent width (EW); for broad molecular bands, in magnitudes. 
Specifically, the average pseudo-continuum flux level is:

\begin{equation}
F_P =  \int_{\lambda_1}^{\lambda_2} \frac{F_\lambda}{(\lambda_2-\lambda_1)} d\lambda
\label{eq:fluxlevel}
\end{equation}

\noindent
where $\lambda_1$ and $\lambda_2$ are the wavelength limits of the 
pseudo-continuum sideband. If $F_{C\lambda}$ represents the straight 
line connecting the midpoints of the blue and red pseudo-continuum levels, 
an equivalent width is then:

\begin{equation}
\rm{EW} =  \int_{\lambda_1}^{\lambda_2} \left( 1 - \frac{F_{I\lambda}}{F_{C\lambda}} \right) d\lambda
\label{eq:equivwidth}
\end{equation}

\noindent
where $F_{I\lambda}$ is the observed flux per unit wavelength and 
$\lambda_1$ and $\lambda_2$ are the wavelength limits of the feature passband.
Similarly, an index measured in magnitudes is:

\begin{equation}
\rm{mag} = -2.5 \log_{10} \left[ \left( \frac{1}{\lambda_2 - \lambda_1} \right)
\int_{\lambda_1}^{\lambda_2} \frac{F_{I\lambda}}{F_{C\lambda}} d\lambda \right]
\label{eq:indexmag}
\end{equation}

These definitions, after Trager et al. (1998), differ slightly from those 
used in Burstein et al. (1984) and Faber et al. (1985) for the original 
11 IDS indices. In the original scheme, the continuum was taken to be 
a horizontal line over the feature bandpass at the level $F_{C\lambda}$ 
taken at the midpoint of the bandpass. This flat rather than sloping 
continuum would induce erroneous small, systematic shifts in the feature strengths.

\begin{figure}
\begin{centering}
\begin{tabular}{c}
\psfig{file=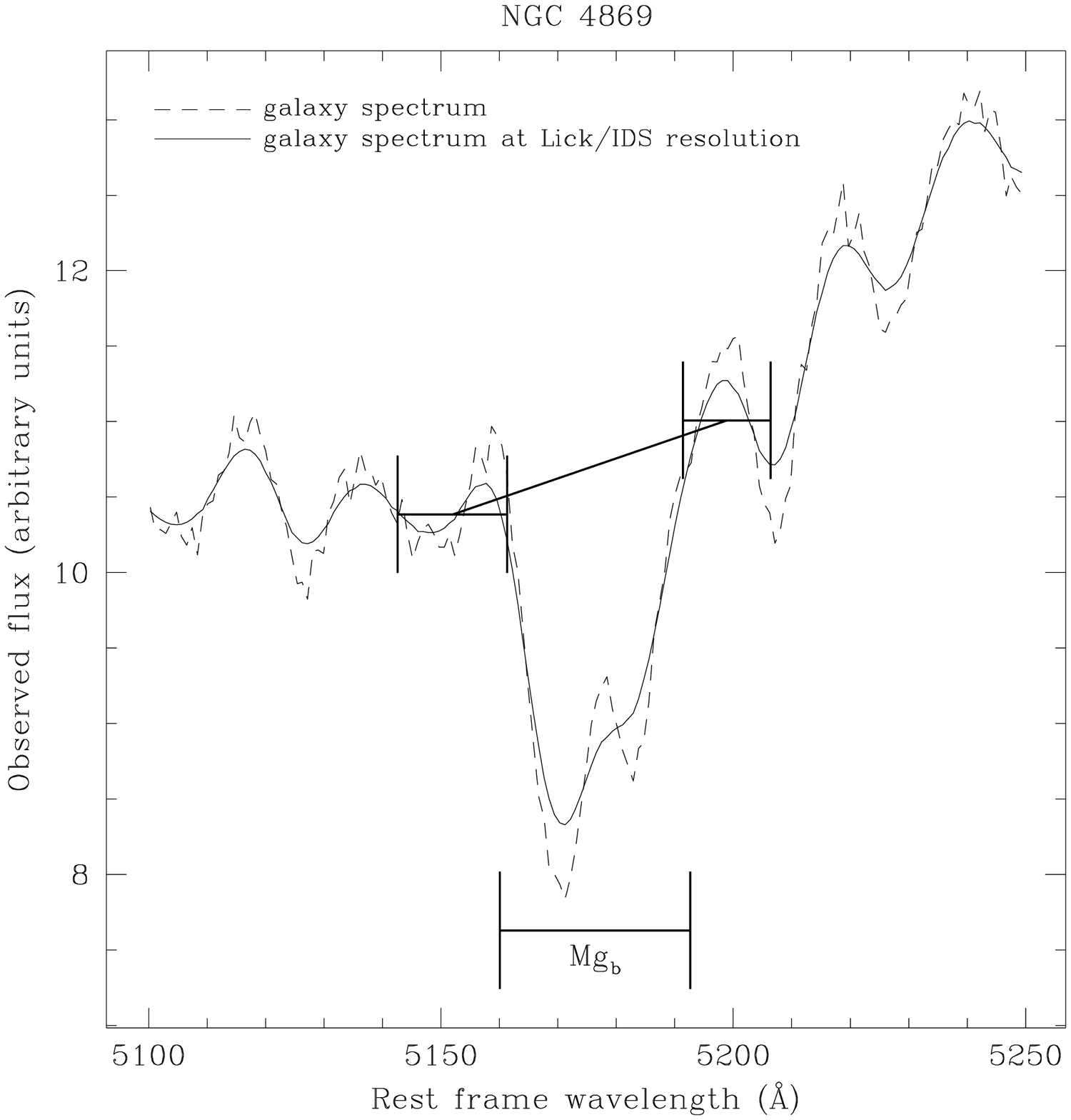,width=85mm} \\
\end{tabular}
\caption{Measurement of Mg$_{\rm{b}}$ index for galaxy NGC 4869
(an elliptical galaxy with $b_j=14.97$ and $\sigma=203$\,km\,s$^{-1}$). 
The fluxed spectrum at 
rest wavelength is overlaid with the spectrum 
transformed to the Lick/IDS system 
spectral resolution (solid line).
The two pseudo-continuum bandpasses are marked either 
side of the Mg$_{\rm{b}}$ feature (also marked); the mean level at the 
mid-point of the two pseudo-continuum bandpasses are joined by a straight line and
the flux in the index feature relative to this line computed.}
\label{fig:ngc4869_mgb}
\end{centering}
\end{figure}

An example of the measurement of the Mg$_{\rm{b}}$ index for galaxy NGC 4869,
an elliptical galaxy with $b_j=14.97$ and $\sigma=203$\,km\,s$^{-1}$, is
shown in Fig. \ref{fig:ngc4869_mgb}.

\subsection{Indices measured}
\label{sec:indicesmeasured}

Table \ref{tab:lickbands} presents the stellar population
absorption line indices measured in 
this work. Column (5) is after the work of Tripicco \& Bell (1995), 
who modelled the Lick/IDS system using synthetic stellar spectra. They 
found that many of the Lick/IDS indices do not in fact measure the 
abundances of the elements for which they are named.
The following composite indices were also measured in our study:

\begin{equation}
\langle\rm{Fe}\rangle = \frac{\rm{Fe5270} + \rm{Fe5335}}{2}
\label{eq:meanfe}
\end{equation}

\begin{equation}
[\rm{MgFe}] = \sqrt{ \rm{Mg}_{\rm{b}} \, \times \, \langle\rm{Fe}\rangle }
\label{eq:mgfe}
\end{equation}

\begin{table*}
\begin{minipage}{160mm}
\caption{Stellar population analysis spectral line index definitions.}
\label{tab:lickbands}
\begin{tabular}{lcccll}
\hline
\hline
Name            & Index Bandpass (\AA) & Pseudocontinua (\AA) & Units & Measures & Source \\
(1)             & (2)                  & (3)                  & (4)   & (5)      & (6)    \\
\hline
C4668$^\dag$    & 4634.000--4720.250 & 4611.500--4630.250 & \AA & C,(O),(Si)     & Lick \\
\ \             & \ \                & 4742.750--4756.500 & \ \ & \ \            & \ \  \\
H$\beta$        & 4847.875--4876.625 & 4827.875--4847.875 & \AA & H$\beta$,(Mg)  & Lick \\
\ \             & \ \                & 4876.625--4891.625 & \ \ & \ \            & \ \  \\
Fe5015          & 4977.750--5054.000 & 4946.500--4977.750 & \AA & (Mg),Ti,Fe     & Lick \\
\ \             & \ \                & 5054.000--5065.250 & \ \ & \ \            & \ \  \\
Mg$_1$          & 5069.125--5134.125 & 4895.125--4957.625 & mag & C,Mg,(O),(Fe)  & Lick \\
\ \             & \ \                & 5301.125--5366.125 & \ \ & \ \            & \ \  \\
Mg$_2$          & 5154.125--5196.625 & 4895.125--4957.625 & mag & Mg,C,(Fe),(O)  & Lick \\
\ \             & \ \                & 5301.125--5366.125 & \ \ & \ \            & \ \  \\
Mg$_{\rm{b}}$   & 5160.125--5192.625 & 5142.625--5161.375 & \AA & Mg,(C),(Cr)    & Lick \\
\ \             & \ \                & 5191.375--5206.375 & \ \ & \ \            & \ \  \\
Fe5270          & 5245.650--5285.650 & 5233.150--5248.150 & \AA & Fe,C,(Mg)      & Lick \\
\ \             & \ \                & 5285.650--5318.150 & \ \ & \ \            & \ \  \\
Fe5335          & 5312.125--5352.125 & 5304.625--5315.875 & \AA & Fe,(C),(Mg),Cr & Lick \\
\ \             & \ \                & 5353.375--5363.375 & \ \ & \ \            & \ \  \\
Fe5406          & 5387.500--5415.000 & 5376.250--5387.500 & \AA & Fe             & Lick \\
\ \             & \ \                & 5415.000--5425.000 & \ \ & \ \            & \ \  \\
H$\beta_{\rm{G}}$  & 4851.320--4871.320 & 4815.000--4845.000 & \AA & H$\beta$,(Mg) & Gonz\'{a}lez (1993), p116 \\
\ \             & \ \                & 4880.000--4930.000 & \ \ & \ \            & J{\o}rgensen (1997)  \\
Fe4930          & 4903.000--4945.500 & 4894.500--4907.000 & \AA & Fe I,Ba II,Fe II & Gonz\'{a}lez (1993), p34 \\
\ \             & \ \                & 4943.750--4954.500 & \ \ & \ \            & \ \  \\
$[$O{\sc{iii}}$]$$_1$ & 4948.920--4978.920 & 4885.000--4935.000 & \AA & $[$O{\sc{iii}}$]$ & Gonz\'{a}lez (1993), p116 \\
\ \             & \ \                & 5030.000--5070.000 & \ \ & \ \            & \ \  \\
$[$O{\sc{iii}}$]$$_2$ & 4996.850--5016.850 & 4885.000--4935.000 & \AA & $[$O{\sc{iii}}$]$ & Gonz\'{a}lez (1993), p116 \\
\ \             & \ \                & 5030.000--5070.000 & \ \ & \ \            & \ \  \\
$[$O{\sc{iii}}$]$$_{\rm{hk}}$ & 4998.000--5015.000 & 4978.000--4998.000 & \AA & $[$O{\sc{iii}}$]$ & Kuntschner (2000) \\
\ \             & \ \                & 5015.000--5030.000 & \ \ & \ \            & \ \  \\
\hline
\multicolumn{6}{l}{{\em Notes:}\/ $^\dag$Worthey (1994) called this index Fe4668. 
In publications after 1995 this index is called C$_2$4668} \\
\multicolumn{6}{l}{since it turned out to depend more on 
carbon than on iron.} \\
\end{tabular}
\end{minipage}
\end{table*}

\subsection{Signal-to-noise}
\label{sec:signal-to-noise}

\begin{figure}
\begin{tabular}{cc}
\psfig{file=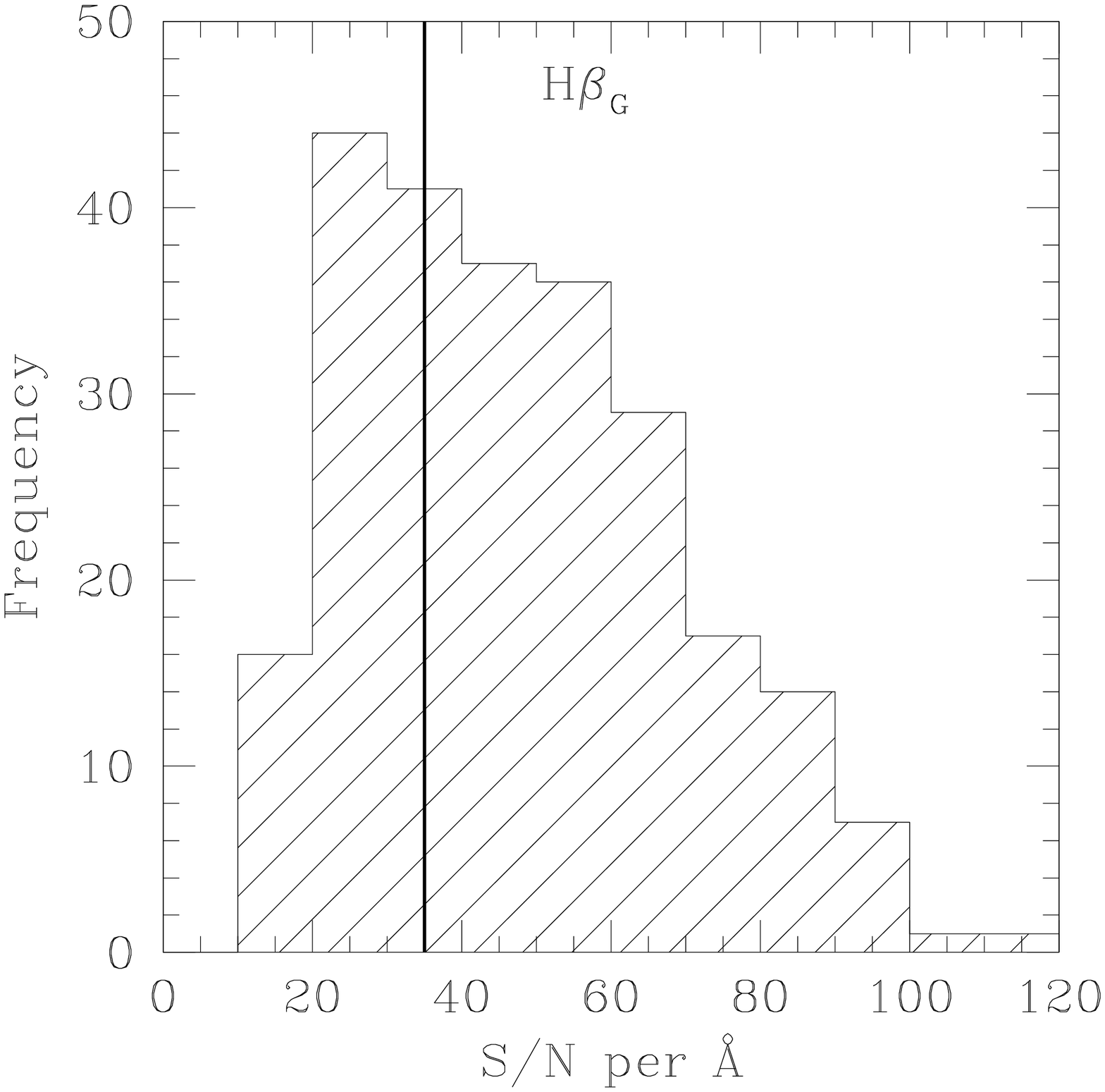,width=40mm} & 
\psfig{file=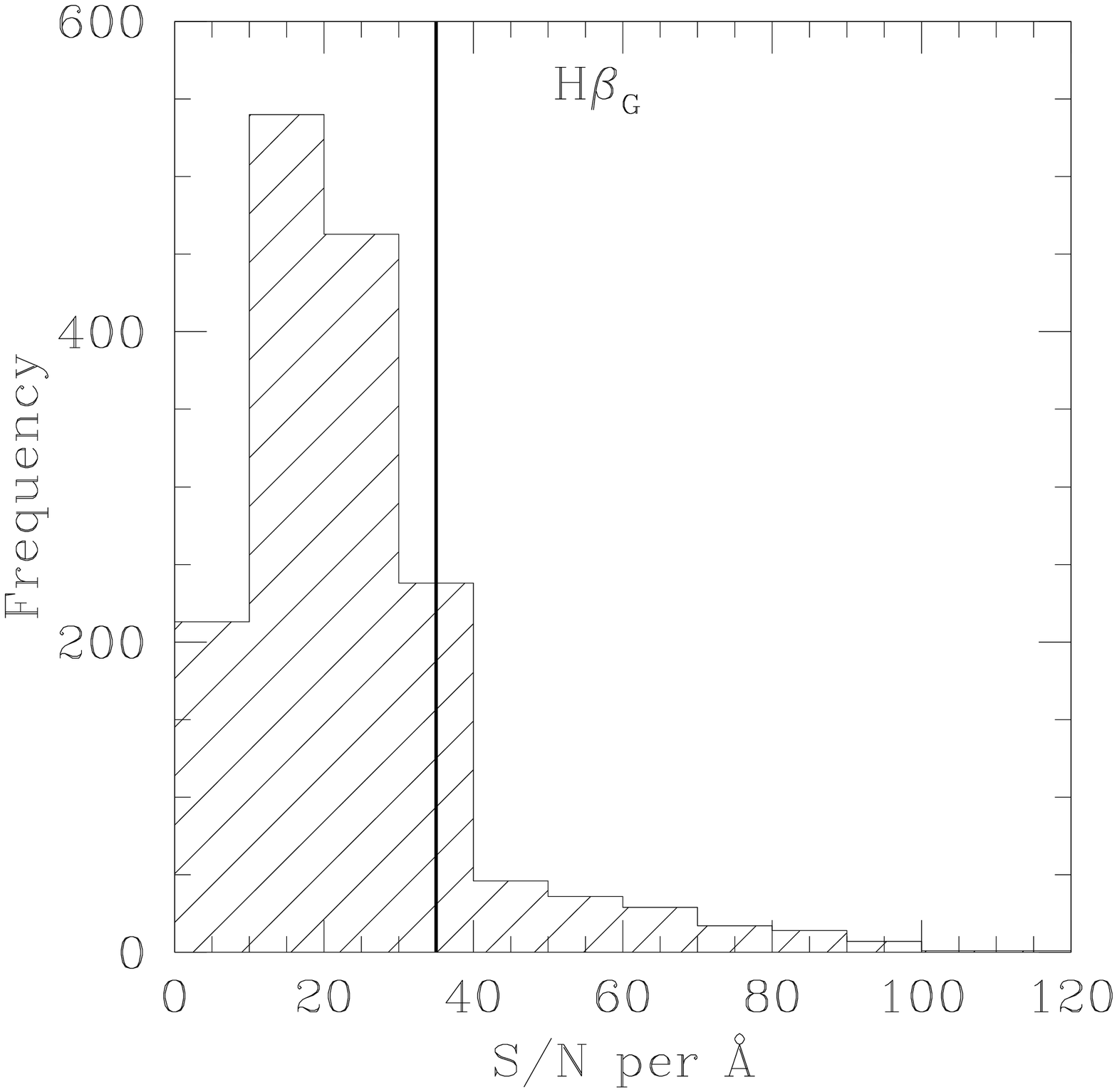,width=40mm} \\
(a) combined exposures only & (b) individual and combined \\
\ \ & exposures \\
\end{tabular}
\caption{A histogram of the S/N per \AA\, of the 
stellar population absorption line index H$\beta_{\rm{G}}$
measured from each galaxy spectrum. The bold vertical line indicates
a S/N of 35 per \AA.}
\label{fig:snplots}
\end{figure}

Our goal was to measure high signal-to-noise (S/N) 
line strength indices
to probe the age and metallicity structure
of the Coma cluster early-type galaxy population. We therefore
measured a S/N at the central rest wavelength of \emph{each}\/ line index investigated in this study.
The line indices H$\beta_{\rm{G}}$ and $[$MgFe$]$
have a mean S/N for their combined exposures 
of 58.7 and 66.7 per \AA\, respectively if a minimum cut-off 
of S/N$\geq$35 per \AA\, is applied. This minimum S/N cut-off is chosen to keep the errors
of the line indices small in subsequent stellar population analyses.
Fig. \ref{fig:snplots} shows the distribution of S/N for our sample
as measured at the H$\beta_{\rm{G}}$ index.

\subsection{Line index velocity dispersion correction}
\label{sec:veldispcorrections}

The observed spectrum of a galaxy is the convolution of the integrated 
spectrum of its stellar population with the instrumental broadening and 
the distribution of line-of-sight velocities of the stars (parameterized 
by the velocity dispersion measurement). 
The broadening of the spectra generally causes the indices to appear
weaker than they intrinsically are.

To probe the stellar population of a galaxy it is necessary
to remove the effects of the instrumental and velocity dispersion 
broadening. This gives an index measurement corrected to zero 
velocity dispersion. This was done by using the spectra of standard stars that 
were observed during the run. These stellar spectra were convolved in logarithmic wavelength space with 
a Gaussian function of widths 0--460\,km\,s$^{-1}$ (in steps of 20\,km\,s$^{-1}$)
to simulate the velocity dispersion broadening within a galaxy.
They were then converted in linear wavelength
space to the Lick/IDS resolution using the method detailed in Section \ref{sec:lickids}.
Index strengths were measured for each spectrum. These values were compared to the
values measured from the zero velocity dispersion stellar spectra which
have also been transformed to the Lick/IDS resolution.
A correction function versus velocity dispersion for each line index 
was then computed by calculating the difference between
the broadened value and the zero velocity dispersion value and then dividing it by the zero
velocity dispersion value.
For each line index, a second order polynomial was fit to the correction function data from 
all the observed standard stars. This function
was then evaluated at the velocity dispersion of each observed galaxy and 
the measured line index value for that galaxy corrected to a zero velocity dispersion value.
After Kuntschner (2000), stars with low H$\beta$ ($<1.6$~\AA) which are 
unrepresentative of bright elliptical galaxies are excluded 
from the analysis.

\begin{figure}
\begin{tabular}{cc}
\psfig{file=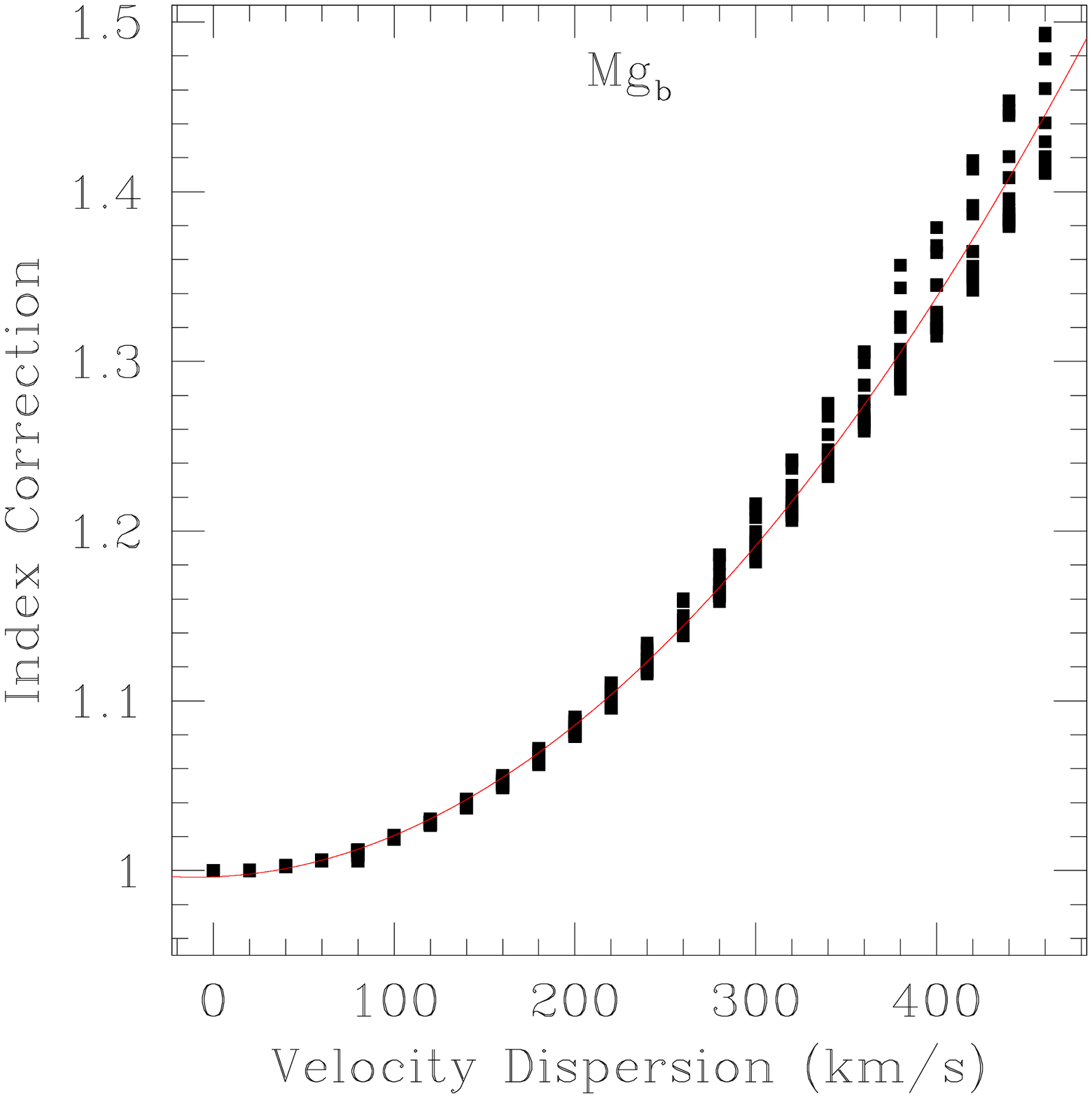,width=40mm} &
\psfig{file=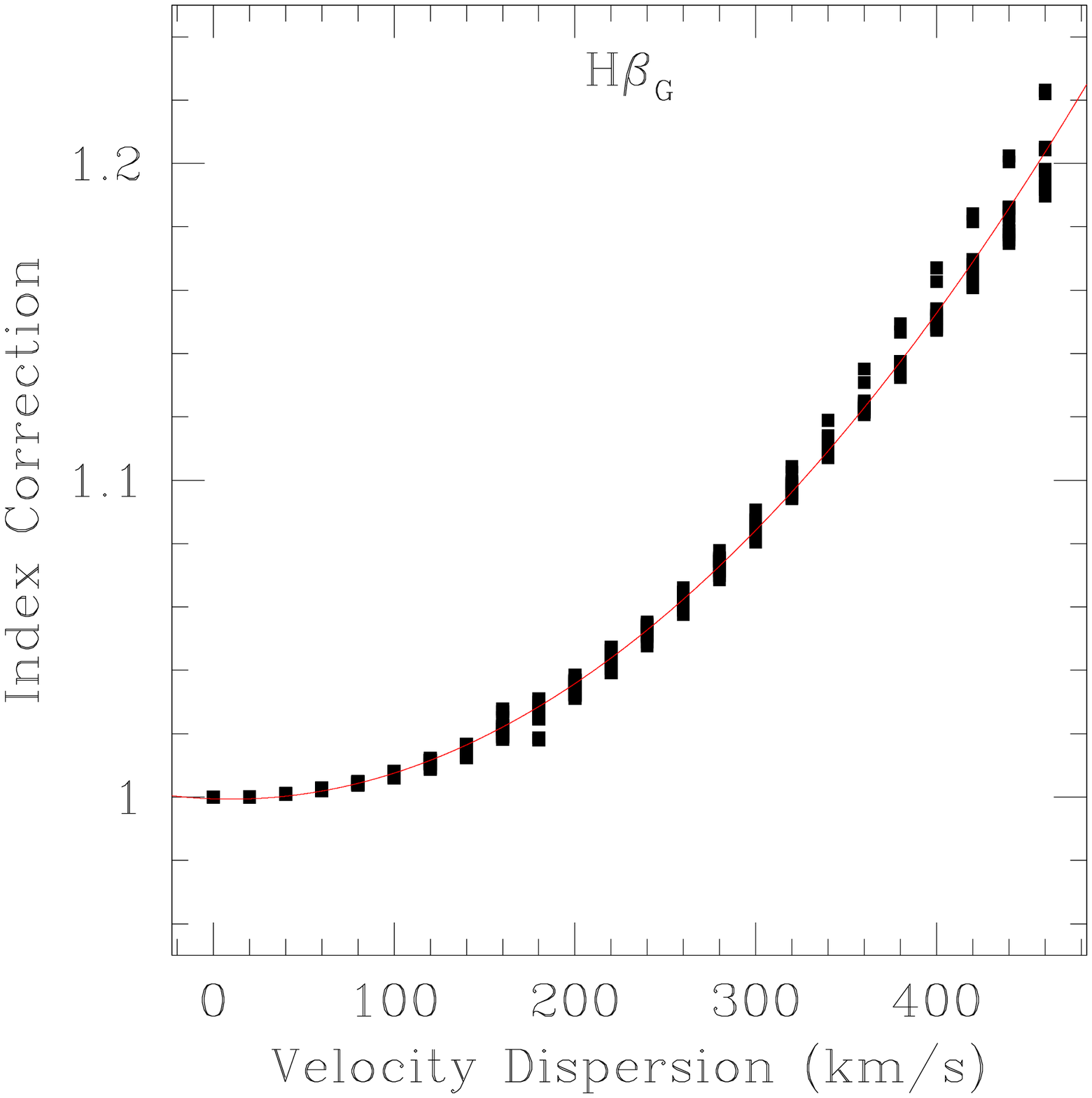,width=40mm} \\
\end{tabular}
\caption
{Velocity dispersion correction functions for the Mg$_{\rm{b}}$ and 
H$\beta_{\rm{G}}$ line indices. Each point corresponds to an index correction
calculated from broadening an observed star and comparing the index
measured from that spectrum to that from the unbroadened spectrum.
The second order polynomial fit to the data is superimposed on the plot.}
\label{fig:veldispcorr}
\end{figure}

The correction functions for the line indices Mg$_{\rm{b}}$
and H$\beta_{\rm{G}}$ are shown in Fig. \ref{fig:veldispcorr}. Table \ref{tab:veldispcorr}
gives the polynomial coefficients for each line index correction
function and the size of the correction for a galaxy with a 
velocity dispersion, $\sigma$ of 200\,km\,s$^{-1}$.
The corrections are multiplicative except for the Mg$_1$ and Mg$_2$
indices where the corrections are additive (since they are measured
in magnitudes rather than in equivalent widths).

\begin{table}
\caption{Velocity dispersion correction polynomial coefficients.}
\label{tab:veldispcorr}
\begin{tabular}{lrrrr}
\hline
\hline
\multicolumn{4}{c}{\ \ } & Correction \\
\ \   & \multicolumn{3}{c}{correction\,=\,a0\,+\,a1.$\sigma$\,+\,a2.$\sigma^2$} & \multicolumn{1}{c}{at $\sigma$ =} \\
Index & \multicolumn{1}{c}{a0} & \multicolumn{1}{c}{a1} & \multicolumn{1}{c}{a2} & \multicolumn{1}{c}{200\,km\,s$^{-1}$} \\
\hline
C4668               &   9.994e--01 & --5.779e--06 &   9.102e--07 & $\times$\/ 1.035 \\
Fe4930              &   9.945e--01 &   1.087e--04 &   4.743e--06 & $\times$\/ 1.206 \\
Fe5015              &   9.893e--01 &   3.144e--04 &   1.494e--06 & $\times$\/ 1.112 \\
Fe5270              &   9.914e--01 &   2.538e--04 &   1.553e--06 & $\times$\/ 1.104 \\
Fe5335              &   1.001e+00  & --7.494e--05 &   5.355e--06 & $\times$\/ 1.200 \\
Fe5406              &   1.007e+00  & --2.662e--04 &   5.744e--06 & $\times$\/ 1.184 \\
H$\beta$            &   1.003e+00  & --6.225e--05 &   7.326e--07 & $\times$\/ 1.020 \\
H$\beta_{\rm{G}}$ &   9.994e--01 & --1.935e--05 &   1.007e--06 & $\times$\/ 1.036 \\
Mg$_1$              & --4.380e--04 &   1.154e--05 &   2.233e--08 & $+$\/ 0.0028 \\
Mg$_2$              &   2.811e--05 &   1.888e--06 &   3.147e--08 & $+$\/ 0.0017 \\
Mg$_{\rm{b}}$     &   9.963e--01 &   4.038e--05 &   2.034e--06 & $\times$\/ 1.086 \\
\hline
\end{tabular}
\begin{minipage}{8.25cm}
{\em Notes:}\/ The final column 
gives the correction for a $\sigma=200$\,km\,s$^{-1}$ galaxy as an 
example of the scale of the correction necessary.
\end{minipage}
\end{table}

\subsection{Emission correction}
\label{sec:emission}

An important issue when estimating ages and metallicities from line 
strength indices is nebular emission. Elliptical galaxies 
normally contain much less dust and ionized gas than spirals, and 
were regarded as dust and gas free for a long time. Surveys of 
large samples of early type galaxies 
(Phillips et al. 1986; Caldwell 1984; Goudfrooij et al. 1994)
have revealed however that 50--60 per cent of the galaxies show weak optical 
emission lines. The measured emission line strengths of [O{\sc{ii}}], [H$\alpha$]
and [N{\sc{ii}}]$\lambda$6584 indicate a presence of only 10$^3$--10$^5{\rm{M}}_\odot$ of 
warm ionized gas in the galaxy centre. Additionally, HST images
of nearby bright early type galaxies revealed that approximately 70--80 per cent
show dust features in the nucleus (van Dokkum \& Franx 1995).
Stellar absorption line strength measurements can be severely affected if there 
is emission present in the galaxy 
(e.g. Gonz\'{a}lez 1993; Goudfrooij \& Emsellem 1996): nebular H$\beta$ emission on 
top of the integrated stellar H$\beta$ absorption weakens the H$\beta$
index and leads therefore to incorrectly older age estimates.

In the Gonz\'{a}lez (1993) study of bright elliptical galaxies in groups and clusters, 
he noted that [O{\sc{iii}}] 
emission at 4959~\AA\, and 5007~\AA\, are clearly detectable in about half of the 
nuclei in his sample and that most of these galaxies also have detectable H$\beta$
emission (see his Fig 4.10). For galaxies in his sample with 
strong emission, H$\beta$ is fairly tightly correlated with [O{\sc{iii}}] 
such that:

\begin{equation}
\frac{\rm{H}\beta \; \rm{emission}}{[\rm{O\sc{iii}}]} \sim 0.7
\label{eq:emissioncorr}
\end{equation}

\noindent
A statistical correction of:

\begin{equation}
\Delta\rm{H}\beta \, = \, 0.7 \, [\rm{O\sc{iii}}] 
\label{eq:emissioncorr2}
\end{equation}

\noindent
was therefore added to H$\beta$ to correct for this residual emission.

Trager et al. (2000a,b) re-examined the accuracy of this correction 
by studying H$\beta$/[O{\sc{iii}}] among the Gonz\'{a}lez (1993) galaxies, 
supplemented by additional early type galaxies from the emission line 
catalogue of Ho, Filippenko \& Sargent (1997). The sample was restricted 
to include only normal, non-AGN Hubble types E through to S0 and 
well measured objects with H$\alpha>1.0$~\AA. For 27 galaxies meeting 
these criteria, they found that H$\beta$/[O{\sc{iii}}] varies from 
0.33 to 1.25, with a median value of 0.60. They suggest that 
a better correction coefficient in Equation (\ref{eq:emissioncorr2})
is 0.6 rather than 0.7: 

\begin{equation}
\Delta\rm{H}\beta \, = \, 0.6 \, [\rm{O\sc{iii}}]
\label{eq:emissioncorr3}
\end{equation}

\noindent
implying that the average galaxy 
in the Gonz\'{a}lez (1993) sample is slightly over-corrected. 
For a median [O{\sc{iii}}] strength through the Gonz\'{a}lez (1993) 
$r_e/8$ aperture of 0.17~\AA, the error due to this correction 
difference would be 0.02~\AA\, or $\sim$3 per cent in age. This systematic 
error for a typical galaxy is negligible compared to other sources 
of error.

\begin{figure}
\psfig{file=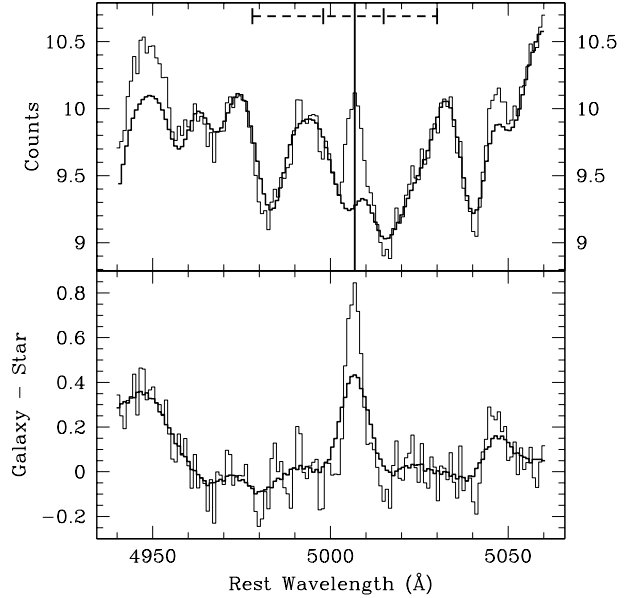,width=85mm}
\caption{Example of our [O{\sc{iii}}]$\lambda$5007~\AA\, emission line strength measurement. Galaxy 
NGC 4850, type E/S0, is shown. The top part of the figure shows the 
fluxed galaxy spectrum (arbitrary units) overlaid with a zero emission 
template (thick line) whose continuum has been matched to the galaxy 
by minimising the maximum absolute deviation between the two spectrum.
The vertical line marks the centre of the [O{\sc{iii}}]$\lambda$5007~\AA\, feature, whilst
the dashed horizontal bar at the very top marks the continuum side 
bandpasses and width of the feature.
The bottom part of the figure shows the difference between the galaxy 
and the zero emission template (a standard star) overlaid with the
difference broadened to the Lick resolution (thick line). The 
spectrum shows clear [O{\sc{iii}}]$\lambda$5007~\AA\, emission.}
\label{fig:emission_example}
\end{figure}

In this study we adopt the 0.6 multiplicative factor to correct the 
H$\beta$ index for nebular emission using the [O{\sc{iii}}]$\lambda$5007~\AA\, emission 
line strength. Whilst there is evidence that this correction factor is 
uncertain for individual galaxies (Mehlert et al. 2000), it is good 
in a statistical sense for the study sample.
After Kuntschner (2000), we adopt a slightly different 
definition of the [O{\sc{iii}}] emission line strength index bandpasses
which we have found better measures the true [O{\sc{iii}}] emission.
After the Lick/IDS system of measuring line indices, we define the 
feature bandpass to be 4998--5015~\AA\, and the continuum side bandpasses 
to be 4978--4998~\AA\, and 5015--5030~\AA. This new definition 
does not affect the conclusions of Trager et al. (2000a,b) nor 
Gonz\'{a}lez (1993) on the relationship of [O{\sc{iii}}] to H$\beta$ emission.
To further improve the measurement of the [O{\sc{iii}}] emission line 
strength in this study, we measure our best estimate of the emission by first subtracting 
a zero emission template from a galaxy spectrum and then measuring
the residual equivalent width. The zero emission template is simply 
a standard star. The process is repeated for a set of zero emission templates 
and an average [O{\sc{iii}}] emission line strength calculated. 
An example of this process is shown in Fig. \ref{fig:emission_example}.

\begin{figure}
\psfig{file=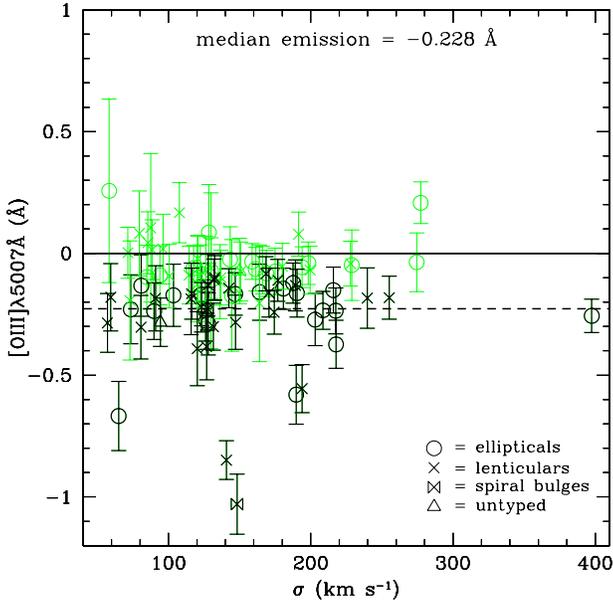,width=85mm}
\caption{Summary of [O{\sc{iii}}]$\lambda$5007~\AA\, emission line strength measurements.
A total of 50 galaxies (shown in a darker shade on the graph) were found to have 1 sigma evidence of 
[O{\sc{iii}}]$\lambda$5007~\AA\, emission, with a median emission of 0.228~\AA\, 
(marked as a bold dashed line on the graph)
giving a median H$\beta$ correction of 0.137~\AA. The H$\beta$ correction 
is calculated separately for each galaxy using Equation (\ref{eq:emissioncorr3})
and our best estimate of the  [O{\sc{iii}}]$\lambda$5007~\AA\, emission for that 
galaxy.}
\label{fig:OIII_emission_summary}
\end{figure}

A total of 50 galaxies were found to have 1 sigma evidence of 
[O{\sc{iii}}]$\lambda$5007~\AA\, emission, with a median emission of 0.228~\AA\, giving 
a median H$\beta$ correction of 0.137~\AA\, (see Fig. \ref{fig:OIII_emission_summary}). 
The H$\beta$ correction 
is calculated separately for each galaxy using Equation (\ref{eq:emissioncorr3})
and our best estimate of the [O{\sc{iii}}]$\lambda$5007~\AA\, emission for that 
galaxy.

\subsection{Line strength index errors}
\label{sec:errors}

The line index measurement errors were calculated by internal comparison 
during a night and between nights. With the large amount of multiple observations 
with different fibre configurations and
high S/N data this allows accurate mapping of the  
random and some of the systematic errors.

The method assumes that the errors have an underlying Gaussian 
nature and exploits the central limit theorem.
Firstly it is necessary to compute the difference between 
the multiple line index measurements 
taken during a night to the `true' line index measurement, 
taken to be the measurement from the combined exposure for that night.
To prevent any contaminating 
systematics, only exposures from a particular night were compared. 
In this way we mapped the random errors as a function of 
galaxy S/N up to 
a maximum S/N governed by the individual exposures.
To extend this random error mapping to a higher S/N limit, we used the 
fact that a number of galaxies were observed every night during the 
observing run and 
further compared the line index measurement from the 
combined exposure for a night to the mean line index measurement 
from all of the nights, taken here to be the `true' measurement 
as before. This mapping to higher S/Ns
is only done for galaxies observed on all nights (often with
different fibres due to the different field fibre configurations) to minimize 
any systematic error contamination of the random error mapping.

\begin{figure}
\begin{tabular}{cc}
\psfig{file=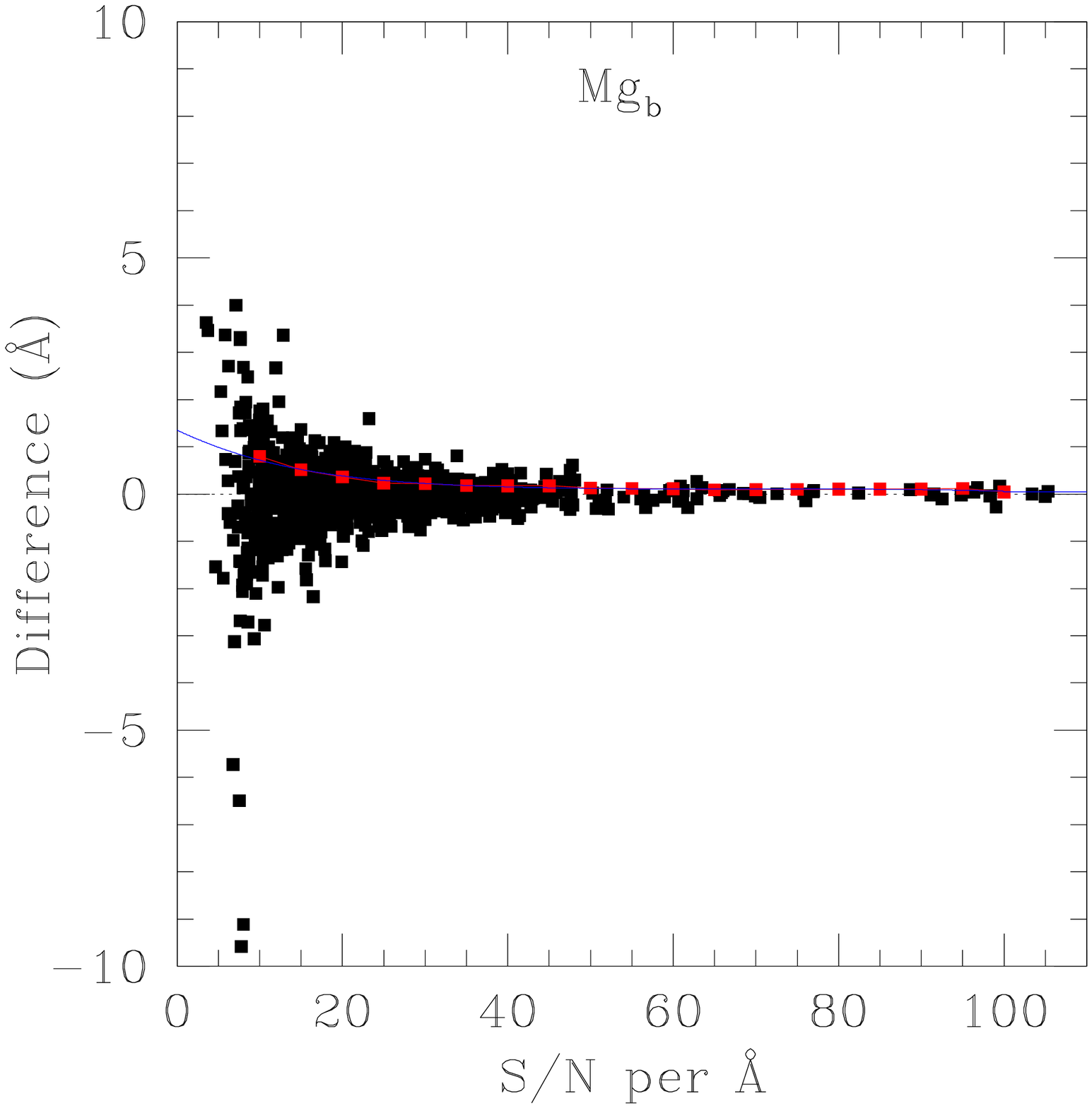,width=40mm} &
\psfig{file=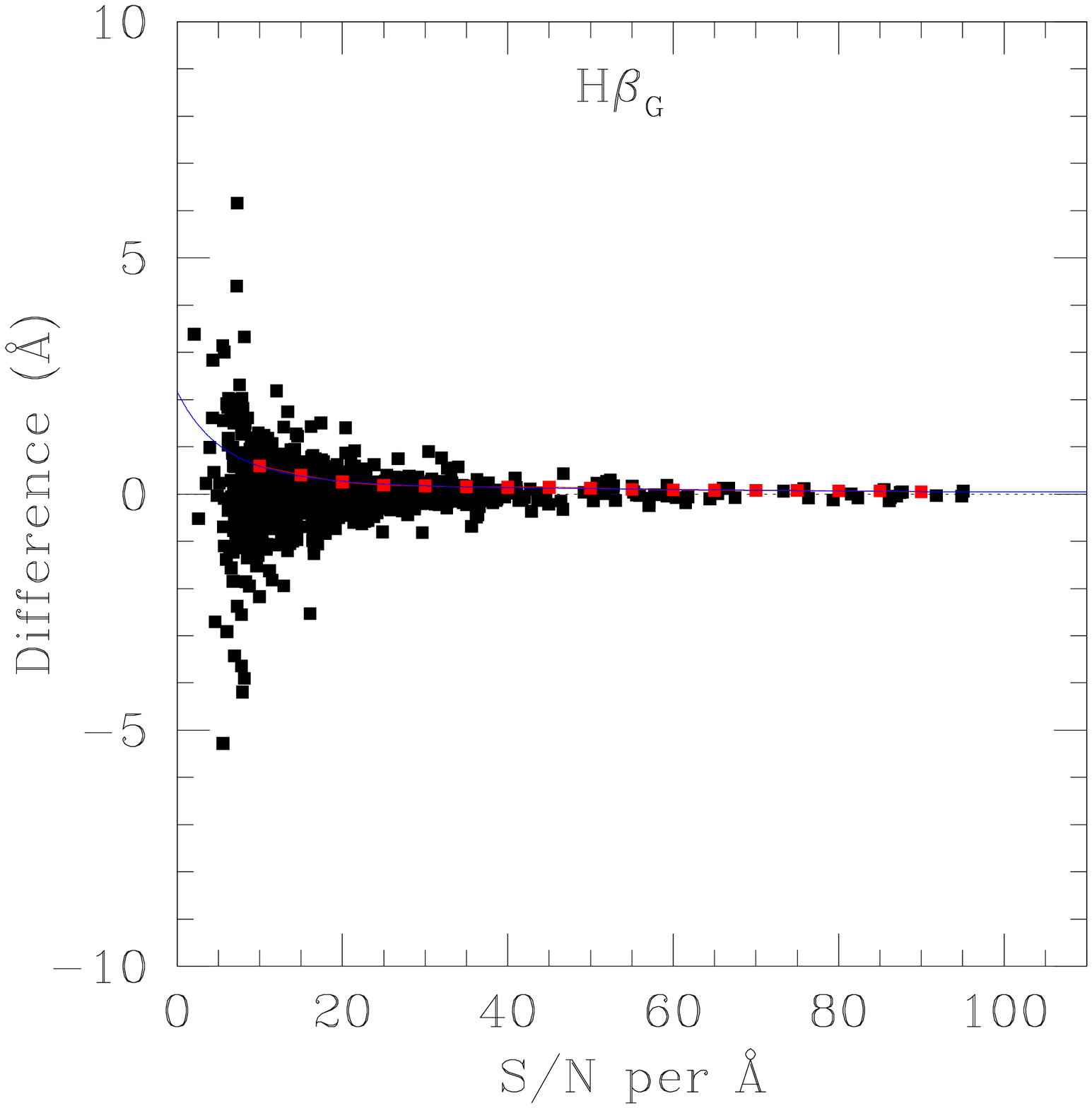,width=40mm} \\
\end{tabular}
\caption{Error calculation plots for the Mg$_{\rm{b}}$
and H$\beta_{\rm{G}}$ line indices.}
\label{fig:errorplots}
\end{figure}

Once we obtained the dependence of the random errors 
with S/N for a particular line index, we
deduced the error function for that index. The error function 
is calculated by binning the data by S/N 
from 5--35 S/N per \AA\, with bin widths of 3 S/N per \AA\, 
(the lower limit is to exclude very low S/N 
spectra which would contaminate the derivation of the 
error function). These bins were then analysed and a 
standard deviation computed for each bin. For 
spectra with a S/N greater than 35 per \AA\, 
binning is no longer used to prevent contamination by 
small number statistics. Instead a standard deviation 
was computed for the differences for all galaxies 
with a S/N greater than 35 per \AA\, and 
then this lower limit was incremented by the bin width and 
the standard deviation re-computed. This process was repeated 
up to a maximum S/N of 120 per \AA. 
This procedure results in a data set of standard deviation versus S/N. 
A 4th order polynomial was then fit to the natural log of the variation of 
standard deviation with S/N (the function is fit 
to the natural logarithm of the data to simply fit a smoother 
function to the data, without introducing any erroneous 
high order fluctuations).
Fig. \ref{fig:errorplots}
shows the error calculation plots for the Mg$_{\rm{b}}$
and H$\beta_{\rm{G}}$ line indices.

The computed error function versus S/N was 
subsequently used to calculate the errors for all of the
line index measurements.

\begin{table}
\caption{Summary of error calculation results.}
\label{tab:medianerrors}
\begin{tabular}{lcccc}
\hline
\hline
\multicolumn{2}{c}{\ \ }  & \multicolumn{1}{c}{Median}  & \multicolumn{1}{c}{Systematic}  & Scale test \\
Index                & N$_{\rm{gals}}$ & \multicolumn{1}{c}{error} & \multicolumn{1}{c}{error} & result \\
\hline
C4668                &   75 &  0.638~\AA & --0.007~\AA & 1.052 \\
Fe4930               &   97 &  0.160~\AA & --0.002~\AA & 1.047 \\
Fe5015               &  101 &  0.729~\AA & --0.001~\AA & 0.892 \\
Fe5270               &  109 &  0.136~\AA & --0.002~\AA & 1.062 \\
Fe5335               &  109 &  0.180~\AA & $\;\,$0.002~\AA & 1.007 \\
Fe5406               &   54 &  0.118~\AA & --0.000~\AA & 1.647 \\
H$\beta$             &   95 &  0.106~\AA & --0.002~\AA & 1.104 \\
H$\beta_{\rm{G}}$  &   95 &  0.103~\AA & --0.001~\AA & 1.042 \\
$\langle$Fe$\rangle$ &  109 &  0.114~\AA & $\;\,$0.002~\AA & 1.035 \\
Mg$_1$               &  103 & 0.0090~mag & --0.0002~mag & 0.904 \\
Mg$_2$               &  102 & 0.0066~mag & --0.0001~mag & 0.953 \\
Mg$_b$               &  103 &  0.123~\AA & $\;\,$0.000~\AA & 1.033 \\
$[$MgFe$]$           &  109 &  0.085~\AA & --0.006~\AA & 1.034 \\
$[$O{\sc{iii}}$]$$_1$ &   99 & 0.203~\AA & $\;\,$0.001~\AA & 0.979 \\
$[$O{\sc{iii}}$]$$_2$ &  101 & 0.122~\AA & $\;\,$0.003~\AA & 1.026 \\
$[$O{\sc{iii}}$]$$_{\rm{hk}}$ &   93 & 0.075~\AA & --0.002~\AA & 1.048 \\
\hline
\end{tabular} \\
\begin{minipage}{8.25cm}
{\em Notes:}\/ Median errors for all data with a S/N$\geq$35 per \AA\,
are shown. The results of an internal systematic error analysis and of
the scale test check are also included (see text). 
There are no internal systematic errors during a night nor between nights.
The only
significant scale test result is that for Fe5406; this result implies that the median 
error should be 0.194~\AA.
\end{minipage}
\end{table}

To test the correctness of the error determination the central 
limit theorem was exploited to perform a scale test on the data.
If the errors computed are appropriate then the following 
function will have a standard deviation equal to unity:

\begin{equation}
\begin{array}{c} 
\rm{scale \; test} \\
\rm{parameter}
\end{array}
\quad = \quad 
\frac{
\begin{array}{c} 
\rm{line \; index} \\
\rm{measurement}
\end{array}
\/ - \/ 
\begin{array}{c} 
\rm{true \; line} \\
\rm{index \; value}
\end{array}
}
{
\begin{array}{c} 
\rm{line \; index} \\
\rm{error}
\end{array}}
\label{eq:scaletest}
\end{equation}

\noindent
This scale test was performed on all data with a S/N
greater than 10 per \AA\, to prevent any contamination by very low 
S/N measurements. In our case the true line index value is equal to the mean 
line index value. It is therefore necessary to include the 
error on the mean in the line index error.
Table \ref{tab:medianerrors} includes the results of the
scale test for each line index measured.
The scale test parameter does indeed have a standard deviation 
approximately equal to unity (apart from the Fe5406 index), showing that the errors calculated are truly 
representative.
For the Fe5406 index, the scale test implies that the median error should be
0.194~\AA. A possible explanation for the difference between our error
estimate and the conclusion of the scale test is the proximity of the index to the end of 
the wavelength range. In a conservative approach we adopt a final error 
for Fe5406 scaled by a factor 1.647.

In addition to the scale test, we conducted an
internal systematic error analysis. 
A mean difference was calculated for data with a S/N$\geq$10 per \AA\,
(the same low S/N cut-off used in the scale test),
however only the central 68.3 per cent of this sample (i.e. 1 sigma clipping) were used 
so that effect of any rogue outliers in the sample distribution was minimized.
The conclusion of this analysis was that there are no internal systematic errors
either during a night or between nights.

\subsection{Lick/IDS index absorption line strength data}
\label{sec:finallinestrengths}

Table \ref{tab:linestrengths} lists 
the Lick/IDS index absorption line strength data for
132 galaxies observed in the Coma cluster in this study.
The three remaining galaxies (RB\,71, RB\,199, GMP\,4469) had insufficient S/N to permit
any line strength measurement.
Missing values in the table indicate either that the line strength measured
had a low S/N or that it could not be measured.
Where a galaxy was observed on multiple nights with the same wavelength range, 
the line strength measurements from each night were combined using the square of 
the S/N to weight the measurements. The H$\beta$ and H$\beta_{\rm{G}}$ 
line strengths given in the table have not been corrected for nebula emission.
The [O{\sc{iii}}]$\lambda$5007~\AA\, emission line strength measurement used 
for this correction is in the column $[$O{\sc{iii}}$]$$_{\rm{sm}}$.

\section{Comparison with previous data}
\label{sec:comparison}

Although the spectral resolution of the Lick/IDS system has been 
well matched, small systematic offsets in the indices introduced 
by continuum shape differences are generally present (note that the 
original Lick/IDS spectra are \emph{not}\/ flux calibrated). 
These offsets do not depend on the velocity dispersion of 
the galaxy itself.
To establish these offsets we compared our measurements with data from 
studies that have galaxies in common:

~\\
\noindent{(i)$\;\;$ Seven Samurai (Dressler et al. 1987; Faber et al. 1987);} \\
\noindent{(ii)$\;$  Lick/IDS database (Trager et al. 1998);} \\
\noindent{(iii)   J{\o}rgensen (1999);}\\
\noindent{(iv)$\,$  SMAC (Hudson et al. 2001);}\\
\noindent{(v)$\;\,$ Mehlert et al. (2000)\footnote{Mehlert et al. (2000) measured high 
S/N long-slit spatially resolved spectra, giving line strength 
measurements as a function of radius from the galaxy centre.
After J{\o}rgensen, Franx \& Kjaergaard (1995a,b) and Mehlert et al. (2000) we used
the equation 
$r_L = \left( r_A\over{1.025} \right)^2 \, {\pi\over{2b}}$
to convert the aperture radius used in this study ($r_A$) to a `slit-equivalent'
radius ($r_L$). $b$ is the width of the slit used by Mehlert et al. (2000). 
An index value at this slit-equivalent radius was then calculated.};} \\
\noindent{(vi)\,  Kuntschner et al. (2001).}

~\\
Whilst some of the above studies partly contain data from the same source,
for simplicity they are treated as independent studies.

This comparative analysis investigated the following hypothesis:
an offset could be present between the data in this study and 
the published data sets, but there should be no offset between
each of the comparison sets (since these have already been corrected to 
a common Lick/IDS system). This analysis also allowed a direct comparative
measure of the quality of the data in this study.

\subsection{Method of analysis}
\label{sec:comparisonmethod}

The following statistics were calculated for each sample 
of data from this study (with S/N$\geq$35 per \AA) that matches with that
of a previous study:

~\\
\noindent{(i)$\;\;$ mean offset to this study;} \\
\noindent{(ii)$\;$  root mean squared of sample differences (rms);} \\
\noindent{(iii)   intrinsic root mean squared of sample differences, 
taking into account the measurement errors (rms$_{intr}$).}

~\\
All offsets were calculated as:

\begin{equation}
\rm{offset}
\quad = \quad
\begin{array}{c} 
\rm{data \; from} \\
\rm{this \; study}
\end{array}
\quad - \quad
\begin{array}{c} 
\rm{data \; from \; a} \\
\rm{comparison \; study}
\end{array}
\end{equation}

The intrinsic root mean squared of sample differences is a test of the quality
of the data errors: if the quoted errors on the parameters are correct, then the 
intrinsic rms should be negligible (i.e. close to zero).

\subsection{Results of comparisons}
\label{sec:comparisonresults}

Table \ref{tab:comparisons} shows the results of the comparisons between
the data in this study with that in previous studies.
Intrinsic scatters are not calculated for the Seven Samurai data
(Dressler et al. 1987) because they do not quote individual measurement errors.

\begin{figure}
\begin{tabular}{c}
\psfig{file=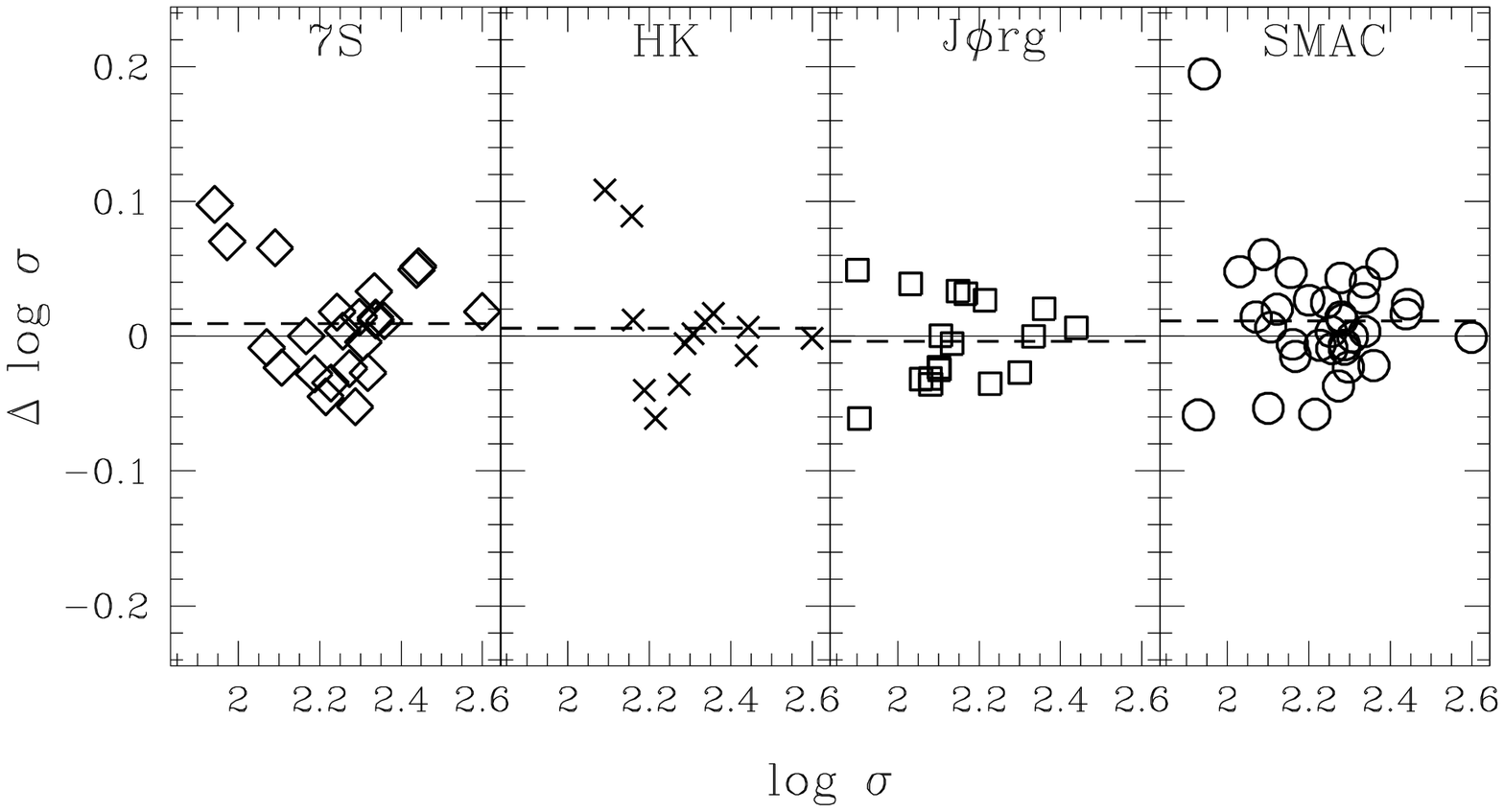,width=85mm,height=40mm,bbllx=18bp,bblly=387bp,bburx=566bp,bbury=694bp,clip=} \\
\psfig{file=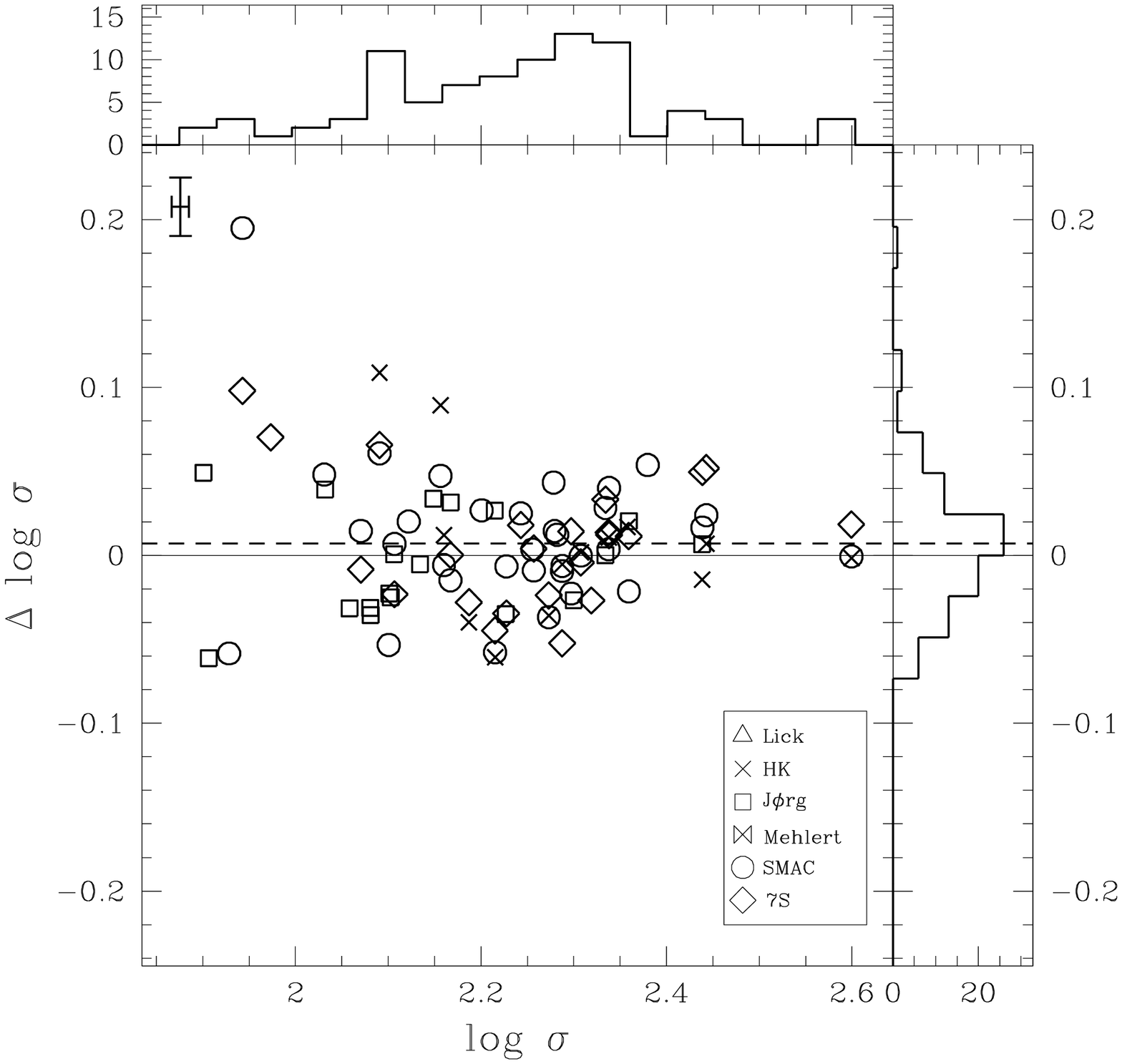,width=85mm} \\
\end{tabular}
\caption{Comparison between the velocity dispersions in 
this study and those from other studies of the Coma cluster.
The difference is calculated as the data from this study minus the data from a comparison study. 
The plots at the top of the figure break down the comparison, showing 
the matching data between this study and each of the matching studies. 
The plot at the bottom of the figure shows the total data set used
for the comparative analysis.
The horizontal dashed lines indicate the mean offset
and a median error bar is shown.}
\label{fig:compare_velocity_dispersions}
\end{figure}

Our velocity dispersion measurements show good agreement with 
published values, with the mean offsets all being less than 3 per cent
(see Fig. \ref{fig:compare_velocity_dispersions}). After allowing
for the quoted measurement errors, only a small intrinsic
scatter of 0.0216\,dex is seen. To remove this intrinsic scatter the
quoted measurement errors have to be increased by only $\sim15$ per cent.
These velocity dispersions will be used in Paper III with photometric data to
investigate various spectro-photometric relations, using the 
stellar population parameters to probe their fundamental nature 

\begin{table}
\caption{Comparison between this study and other studies of the Coma cluster.}
\label{tab:comparisons}
\begin{tabular}{llrrcr}
\hline \hline 
Data & S & N & \multicolumn{1}{c}{Offset} & rms & \multicolumn{1}{c}{rms$_{intr}$} \\
\hline
$\log\sigma$         & 7 &  23 &   0.0093$\pm$0.0081 & 0.0379 &   \ \ \\
$\log\sigma$         & S &  33 &   0.0115$\pm$0.0078 & 0.0443 &   0.0222$\pm$0.0039  \\ 
$\log\sigma$         & J &  18 & --0.0037$\pm$0.0075 & 0.0308 &   0.0158$\pm$0.0037 \\
$\log\sigma$         & H &  14 &   0.0059$\pm$0.0121 & 0.0437 &   0.0231$\pm$0.0062 \\
$\log\sigma$         & A &  88 &   0.0070$\pm$0.0043 & 0.0405 &   0.0216$\pm$0.0027$^\dag$ \\ 
\hline
C4668                & L &   9 &     0.783$\pm$0.272 &  0.770 &    0.383$\pm$0.128  \\
\hline
Fe5015               & L &  11 &     0.006$\pm$0.192 &  0.609 &    0.532$\pm$0.160 \\
\hline
Fe5270               & L &  11 &   --0.026$\pm$0.147 &  0.464 &    0.065$\pm$0.020  \\ 
\hline
Fe5335               & L &  10 &     0.329$\pm$0.135 &  0.405 &    0.129$\pm$0.041  \\
\hline
Fe5406               & L &   5 &     0.134$\pm$0.245 &  0.491 &    0.097$\pm$0.044  \\
\hline
$\langle$Fe$\rangle$ & J &  36 &   --0.121$\pm$0.050 &  0.294 &    0.046$\pm$0.008  \\
$\langle$Fe$\rangle$ & H &  14 &     0.121$\pm$0.084 &  0.304 &    0.090$\pm$0.024  \\
$\langle$Fe$\rangle$ & L &  10 &     0.204$\pm$0.050 &  0.150 &    0.183$\pm$0.058  \\
$\langle$Fe$\rangle$ & M &  18 &     0.157$\pm$0.039 &  0.159 &    0.050$\pm$0.012 \\
$\langle$Fe$\rangle$ & A &  78 &     0.028$\pm$0.033 &  0.291 &    0.048$\pm$0.005  \\
\hline
Mg$_1$               & J &  36 &   0.0203$\pm$0.0026 & 0.0152 &   0.0040$\pm$0.0007  \\
Mg$_1$               & L &  11 &   0.0208$\pm$0.0073 & 0.0219 &   0.0075$\pm$0.0024  \\
Mg$_1$               & A &  47 &   0.0204$\pm$0.0025 & 0.0169 &   0.0050$\pm$0.0007  \\
\hline
Mg$_2$               & 7 &  23 & --0.0063$\pm$0.0036 & 0.0169 &   \ \  \\
Mg$_2$               & S &  33 &   0.0051$\pm$0.0029 & 0.0164 &   0.0094$\pm$0.0017  \\
Mg$_2$               & J &  36 &   0.0024$\pm$0.0033 & 0.0198 &   0.0099$\pm$0.0017  \\
Mg$_2$               & H &  14 &   0.0157$\pm$0.0059 & 0.0204 &   0.0111$\pm$0.0031  \\
Mg$_2$               & L &  11 &   0.0221$\pm$0.0073 & 0.0232 &   0.0087$\pm$0.0026  \\
Mg$_2$               & A & 117 &   0.0048$\pm$0.0019 & 0.0205 &   0.0111$\pm$0.0012$^\dag$  \\
\hline
Mg$_b$               & J &  36 &     0.015$\pm$0.053 &  0.312 &    0.036$\pm$0.006  \\
Mg$_b$               & H &  14 &     0.180$\pm$0.062 &  0.222 &    0.012$\pm$0.003  \\
Mg$_b$               & L &  11 &     0.007$\pm$0.135 &  0.426 &    0.002$\pm$0.001  \\
Mg$_b$               & M &  18 &     0.132$\pm$0.068 &  0.279 &    0.034$\pm$0.008  \\
Mg$_b$               & A &  79 &     0.070$\pm$0.036 &  0.318 &    0.035$\pm$0.004  \\
\hline
$[$MgFe$]$           & J &  36 &   --0.143$\pm$0.039 &  0.229 &    0.044$\pm$0.007  \\
$[$MgFe$]$           & H &  14 &     0.148$\pm$0.067 &  0.240 &    0.081$\pm$0.022  \\
$[$MgFe$]$           & L &  10 &     0.158$\pm$0.080 &  0.238 &    0.031$\pm$0.010  \\
$[$MgFe$]$           & M &  18 &     0.152$\pm$0.040 &  0.165 &    0.004$\pm$0.001  \\
$[$MgFe$]$           & A &  78 &     0.016$\pm$0.030 &  0.264 &    0.078$\pm$0.009  \\
\hline
H$\beta$             & J &  35 &   --0.210$\pm$0.059 &  0.345 &    0.085$\pm$0.014  \\
H$\beta$             & H &  13 &     0.194$\pm$0.102 &  0.354 &    0.139$\pm$0.038  \\
H$\beta$             & L &  10 &   --0.005$\pm$0.148 &  0.443 &    0.091$\pm$0.029  \\
H$\beta$             & M &  18 &   --0.099$\pm$0.067 &  0.276 &    0.051$\pm$0.012  \\
H$\beta$             & A &  76 &   --0.088$\pm$0.043 &  0.376 &    0.120$\pm$0.014  \\ 
\hline
H$\beta_{\rm{G}}$  & J &  35 &   --0.106$\pm$0.055 &    0.321 &    0.129$\pm$0.022  \\
\hline
\end{tabular}
\begin{minipage}{8.25cm}
{\em Notes:}\/ The different sources, \emph{S}\/ used for comparison are given by the key below: \\
~\\
\begin{tabular}{ll}
7 = 7S & = Seven Samurai studies (Dressler et al. 1987) \\
S = SMAC & = Streaming Motions of Abell Clusters \\
\ \      & \,\, (Hudson et al. 2001) \\
J = J{\o}rg & = J{\o}rgensen (1999) \\
H = HK & = Kuntschner et al. (2001) \\
L = Lick & = Lick/IDS database (Trager et al. 1998) \\
M = Mehlert & = Mehlert et al. (2000) \\
A = All & = Total matching data set \\
\end{tabular} \\
~\\
$N$ is the number of matching data between the studies. \\
~\\
$^\dag$ this does not include the 7S data \\
\end{minipage}
\end{table}

As in our study of stellar population ages and metallicities (Paper II),
we focus our comparative analysis on the indices
H$\beta_{\rm{G}}$ (predominantly age dependent) and $[$MgFe$]$ (predominantly metallicity
dependent). 
We can see that the analysis gives an initial average offset over all studies
in H$\beta$ of --0.088\,$\pm$\,0.043~\AA\, and 
 in $[$MgFe$]$ of 0.016\,$\pm$\,0.030~\AA. 
This implies that the $[$MgFe$]$ values in this study require no correction
to place them fully on the Lick/IDS system (as the computed correction is not statistically
significant), but that the H$\beta$ do require a correction (though only at a 2.0 sigma level).
However a 
closer examination shows that there are systematic 
offsets between the comparison data sets. 
If the J{\o}rgensen (1999) data set 
(which has the largest offset from this study for the line indices H$\beta$ and $[$MgFe$]$)
is excluded from the comparative analysis, an average offset 
in H$\beta$ of 0.017\,$\pm$\,0.059~\AA\, and 
 in $[$MgFe$]$ of 0.152\,$\pm$\,0.033~\AA\, is found. This implies the reverse of the 
previous result, namely that the $[$MgFe$]$ values in this study 
\emph{do}\/ require a correction (at a 4.6 sigma level), whilst the H$\beta$ values do not.
This analysis highlights problems with the comparison data sets, indicating that
either there are underlying problems with the line index measurements or 
that the data sets have not been fully corrected to the Lick/IDS system.
This leads to the conclusion that any
systematic correction to the data set in this study would be
uncertain because of the discrepancies between published data sets,
therefore no corrections are applied.
The size of the correction would in
any case only be $\sim0.1$~\AA\, which corresponds to 
either a correction of $\sim0.05$ in stellar population mean metallicity, $[$Fe/H$]$ 
or $\sim2$\,Gyr in stellar population mean age, depending on where the data point is on a 
stellar population grid (e.g. the grids of Worthey 1994). Such corrections would be approximately
a systematic shift for the entire data set and would therefore not
significantly affect the analysis of distributions or relative trends within the 
Coma cluster bright early-type galaxy stellar populations.

Table \ref{tab:comparisons} also contains comparative analyses of
other parameters. 
No highly statistically significant
evidence for any offsets between the data from this study and
the comparison data are found, except for the Mg$_1$ index. This is found to have
a mean offset of 0.0204\,$\pm$\,0.0025\,mag. 
This offset was therefore removed from the Mg$_1$ line strengths published in this paper.

\subsection{Effect of aperture corrections}
\label{sec:apcorr}

\begin{figure*}
\begin{minipage}{170mm}
\begin{centering}
\begin{tabular}{cc}
\psfig{file=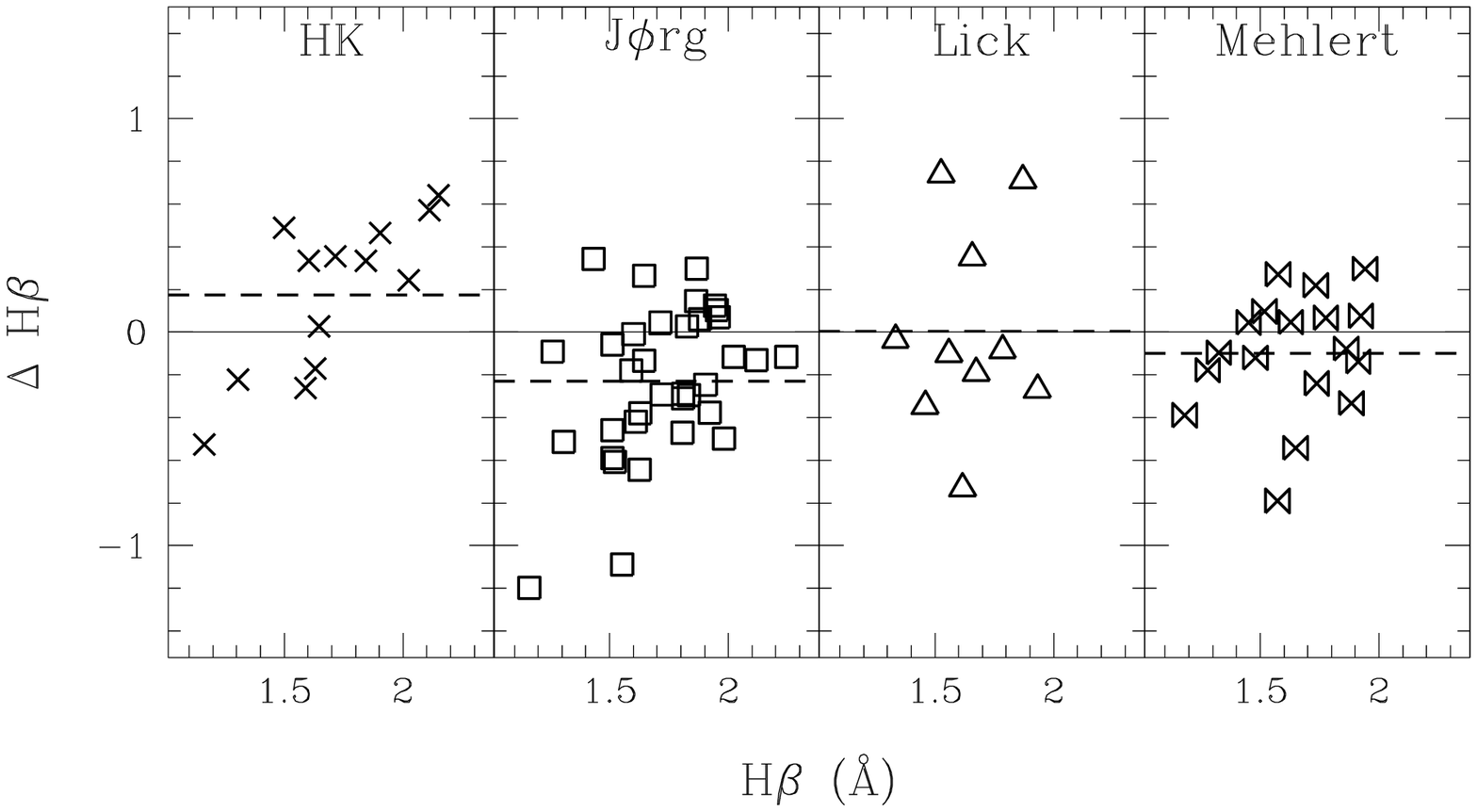,width=85mm,height=40mm,bbllx=18bp,bblly=387bp,bburx=566bp,bbury=694bp,clip=} &
\psfig{file=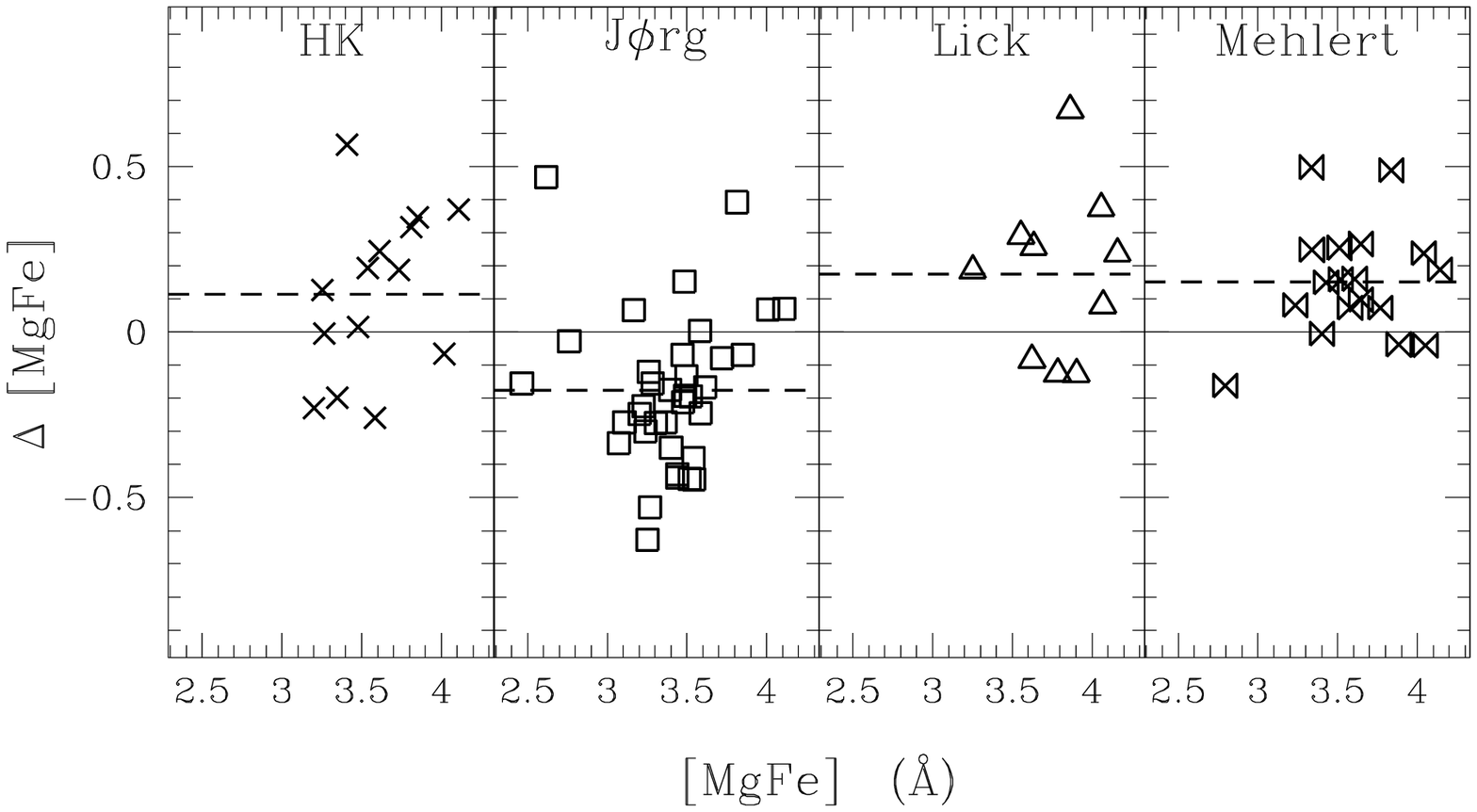,width=85mm,height=40mm,bbllx=18bp,bblly=387bp,bburx=566bp,bbury=694bp,clip=} \\
\psfig{file=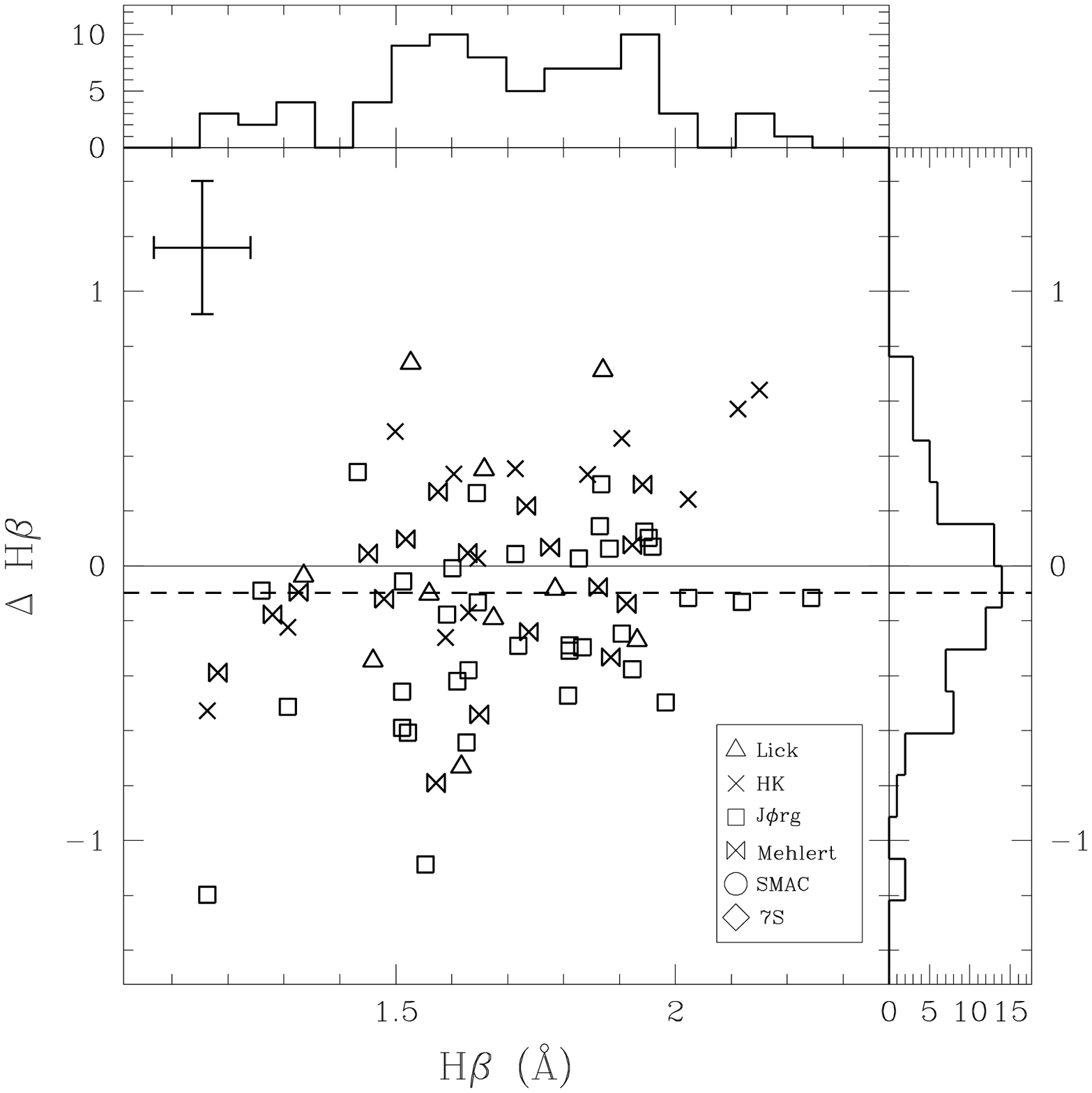,width=85mm} &
\psfig{file=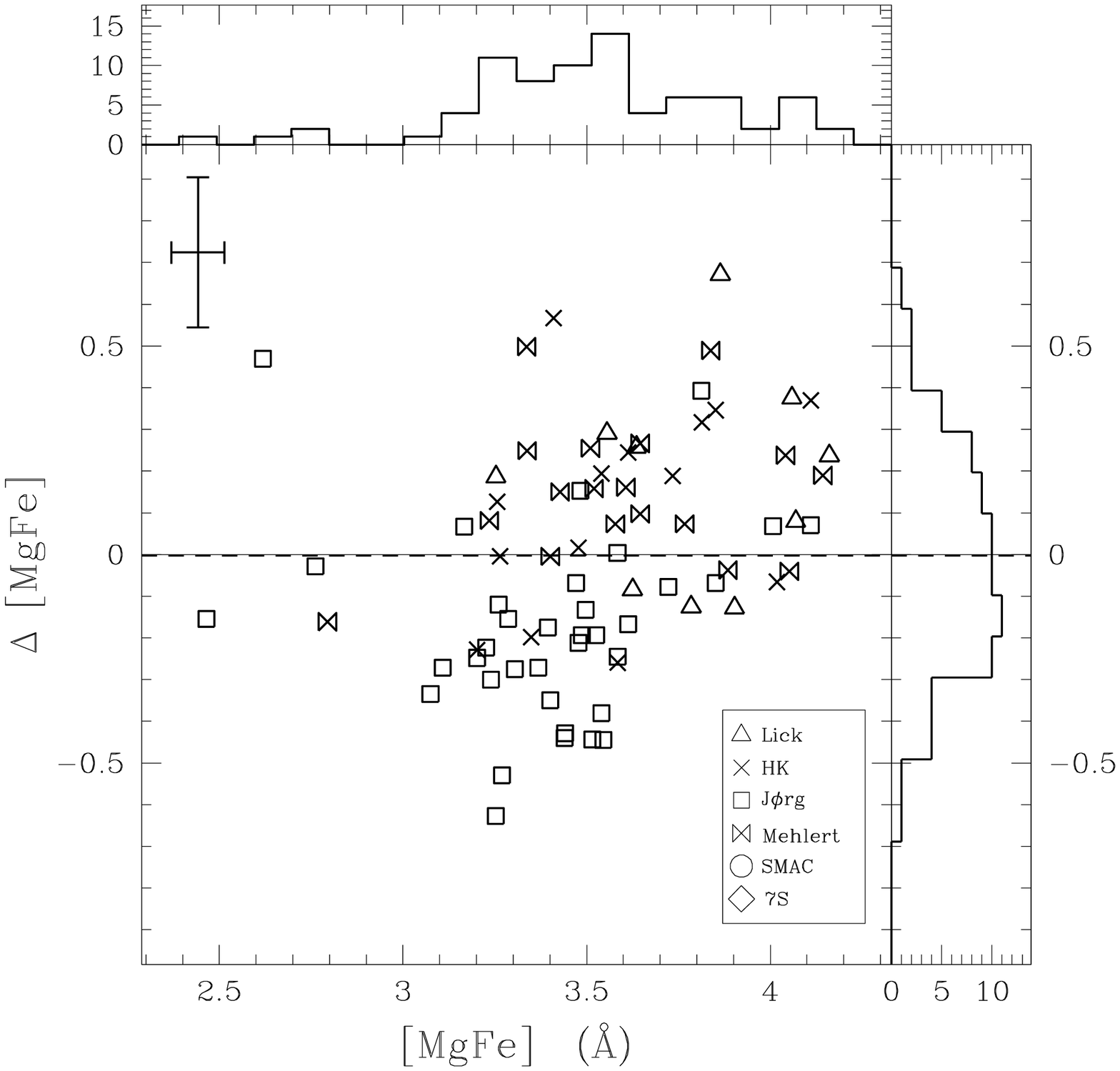,width=85mm} \\
\end{tabular}
\caption{Comparison between this study and other studies 
of the Coma cluster after aperture corrections
for the line indices H$\beta$ and $[$MgFe$]$.
The difference is calculated as the data from this study minus the data from a comparison study. 
The plots at the top of the figure break down the comparison, showing 
the matching data between this study and each of the matching studies. 
The plots at the bottom of the figure show the total data set used
for the comparative analysis.
The horizontal dashed lines indicate the mean offset 
and a median error bar is shown.}
\label{fig:aperture_corrections3}
\end{centering}
\end{minipage}
\end{figure*}

The above comparative analysis did not 
take into account the effect of different aperture sizes. 
Galaxies exhibit a radial dependence for line
strength measurements (see e.g. Mehlert et al. 2000) so it is necessary
to understand the offsets introduced when comparing data
from studies with different aperture dimensions.
Following J{\o}rgensen et al. (1995a,b) and Mehlert et al. (2000) we 
calculate `slit-equivalent' radii to match the aperture width 
of 2.7$''$ used in this study with the standard 3.4$''$ diameter aperture
and the long-slit of dimension 1.4$''$\,$\times$\,4$''$ (Trager et al. 1998)
used in the comparison studies. These radii are then used to
convert the long-slit absorption line strengths
of Mehlert et al. (2000) to equivalent values and a mean 
aperture correction factor calculated 
(Table \ref{tab:aperture_corrections}). 
Since Mehlert et al. (2000) only measured H$\beta$, Mg$_b$ and 
$\langle$Fe$\rangle$ (and therefore $[$MgFe$]$ as well as it is 
a composite index), this data 
can only be used to corrected these indices.
The calculated aperture corrections are small ($<\pm0.05$~\AA) in
comparison to the data errors ($\sim0.1$~\AA) and therefore do not affect the previous 
conclusion about the presence of a mean offset between the Lick/IDS 
calibrated data and the data in this study (the data in Fig. \ref{fig:aperture_corrections3}
and in Table \ref{tab:aperture_corrections3}
give a very similar average offset to Table \ref{tab:comparisons} for H$\beta$ and $[$MgFe$]$).

\begin{table}
\caption{Aperture corrections for the line indices H$\beta$, Mg$_b$, 
$\langle$Fe$\rangle$ and $[$MgFe$]$.}
\label{tab:aperture_corrections}
\begin{tabular}{lcc}
\hline
\hline
\ \                  & Mean aperture correction      & Mean aperture correction \\
Index                & 2.7$''$ -- 3.4$''$ & 2.7$''$ -- 1.4$''$\,$\times$\,4$''$ \\
\hline
H$\beta$             & 0.019\,$\pm$\,0.010~\AA & --0.009\,$\pm$\,0.007~\AA \\
$\langle$Fe$\rangle$ & 0.025\,$\pm$\,0.010~\AA & --0.012\,$\pm$\,0.009~\AA \\
Mg$_b$               & 0.042\,$\pm$\,0.011~\AA & --0.026\,$\pm$\,0.008~\AA \\
$[$MgFe$]$           & 0.033\,$\pm$\,0.009~\AA & --0.018\,$\pm$\,0.008~\AA \\
\hline
\end{tabular} \\
\begin{minipage}{8.25cm}
{\em Notes:}\/ All data is calculated from
the long slit data of Mehlert et al. (2000).
Subtract the mean correction factor from the
2.7$''$ line indices presented in this study
to convert them to line indices equivalent 
to 3.4$''$ diameter fibre measurements or long-slit data from a $1.4''\times4''$ slit.
\end{minipage}
\end{table}

\begin{table}
\caption{Comparison between this study and other studies of 
the Coma cluster after aperture corrections
for the line indices H$\beta$ and $[$MgFe$]$.}
\label{tab:aperture_corrections3}
\begin{centering}
\begin{tabular}{llrcl}
\hline
\hline
Parameter            & Source & \multicolumn{3}{c}{Offset to this study} \\
\hline 
H$\beta$             &   HK           &    0.175 & $\pm$ &    0.102~\AA \\
H$\beta$             &   J{\o}rg      &  --0.229 & $\pm$ &    0.059~\AA \\
H$\beta$             &   Lick         &    0.004 & $\pm$ &    0.148~\AA \\
H$\beta$             &   Mehlert      &  --0.099 & $\pm$ &    0.067~\AA \\
\emph{H$\beta$}      & \emph{ALL} & \emph{--0.099} & \emph{$\pm$} & \emph{0.044~\AA} \\
\hline
$[$MgFe$]$           &   HK           &    0.115 & $\pm$ &    0.067~\AA \\
$[$MgFe$]$           &   J{\o}rg      &  --0.176 & $\pm$ &    0.039~\AA \\
$[$MgFe$]$           &   Lick         &    0.176 & $\pm$ &    0.080~\AA \\
$[$MgFe$]$           &   Mehlert      &    0.152 & $\pm$ &    0.040~\AA \\
\emph{$[$MgFe$]$}    & \emph{ALL} & \emph{--0.003} & \emph{$\pm$} & \emph{0.031~\AA} \\
\hline
\end{tabular} \\
\end{centering}
\end{table}

\subsection{Implications for the errors within our study}
\label{sec:errimplications}

The intrinsic rms of the differences between this study and the comparative studies
for the indices H$\beta$ and $[$MgFe$]$ was non-zero in Table \ref{tab:comparisons}.
An intrinsic rms of 0.120\,$\pm$\,0.014~\AA\, and 0.078\,$\pm$\,0.009~\AA\,
was found respectively for the indices H$\beta$ and $[$MgFe$]$.
If there are no systematic differences between the comparison data 
sets and the data set in this study then the presence of an intrinsic rms implies that
the random errors of the data have been underestimated.
In Section \ref{sec:errors} the median error of the H$\beta$ measurements
in this study was found to be 0.106~\AA, whilst $[$MgFe$]$ had a median
error of 0.085~\AA. Since the error calculation method used in this
study is completely independent and truly statistical,
we believe that the
published errors of the comparison data sets are underestimated.
These published data sets rely greatly upon comparisons between 
each other to normalise their error estimations. 
We believe this approach has led to the underestimation of
the line strength index errors.
In the worst case scenario, if the errors 
in the comparison data sets are however perfect and it is the errors in this study
that are underestimated, this analysis implies that the median errors 
for the indices H$\beta$ and $[$MgFe$]$ should in fact be 0.177~\AA\, and 0.115~\AA\,
respectively. The true situation 
is likely to be somewhere in between, with both errors requiring some scale factor 
to be applied.
A scale factor is not applied to the errors in this study 
because of the large uncertainties 
of this scaling and the question of the validity of such a scaling to our independent
error estimates. However it does highlight the importance of rigorous error treatments
and of obtaining high-quality repeat observations to fully characterise 
both the random and systematic errors in a data set. Both of these approaches have
been taken in this study.

\section{Conclusions}
\label{sec:conclusions}

In this paper we have presented data from a new spectroscopic study on
 the central 1$^\circ$ 
of the nearby rich Coma cluster. This study aims to accurately measure the ages and metallicities 
of the bright ($b_j\leq18.0$, $B\la-17.7$) early-type galaxy population and 
to probe their distributions and the
impact upon spectro-photometric relations.
The data set described in this paper is
a homogeneous and high signal-to-noise 
set of Lick/IDS stellar population line indices
 and central velocity dispersions for 132 galaxies.
Our observations include 73 per cent (100 out of 137) 
of the total early-type galaxy population ($b_j\leq18.0$).

\begin{figure*}
\begin{minipage}{185mm}
\begin{tabular}{ccc}
\psfig{file=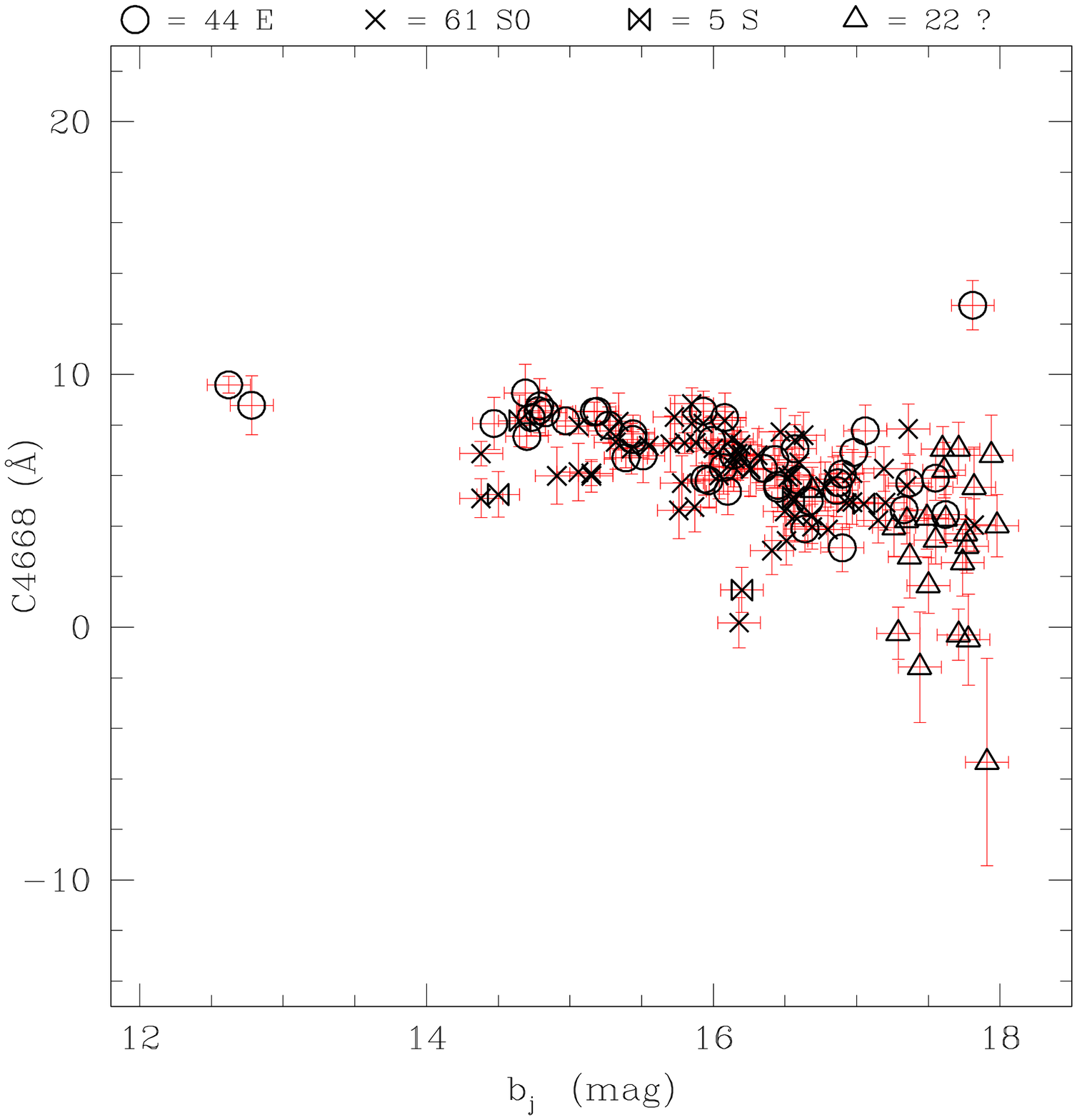,width=55mm} &
\psfig{file=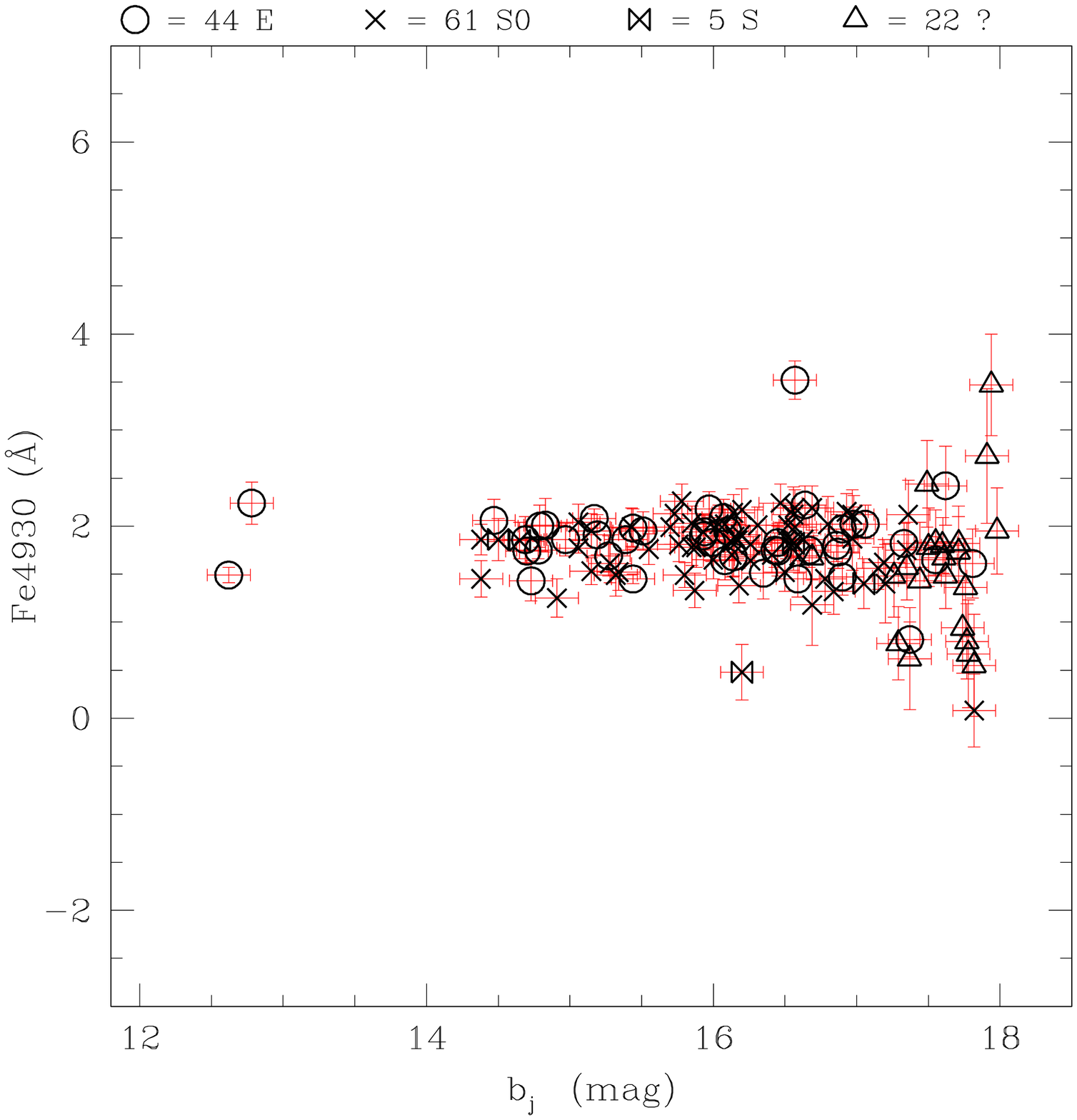,width=55mm} &
\psfig{file=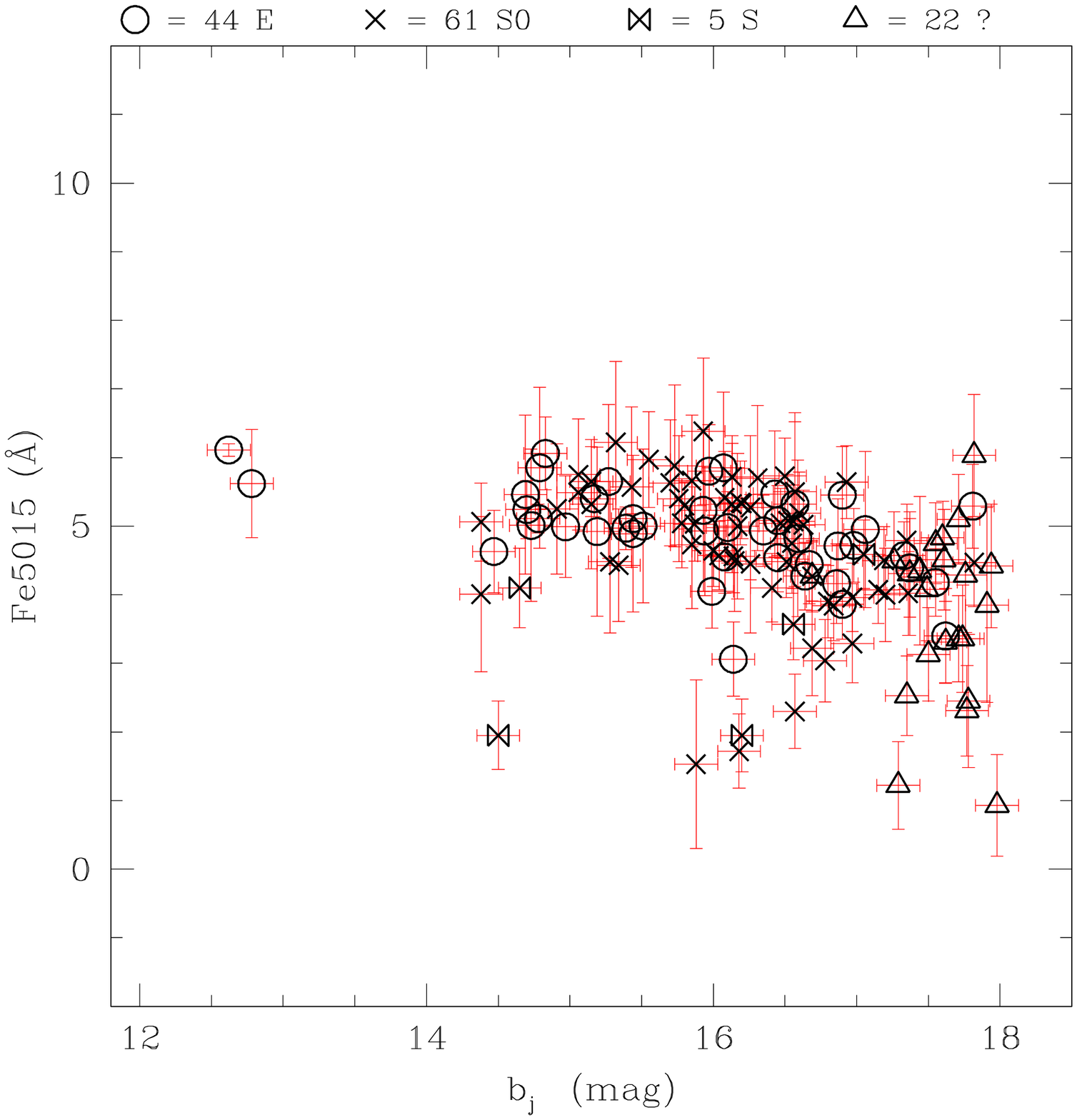,width=55mm} \\
\psfig{file=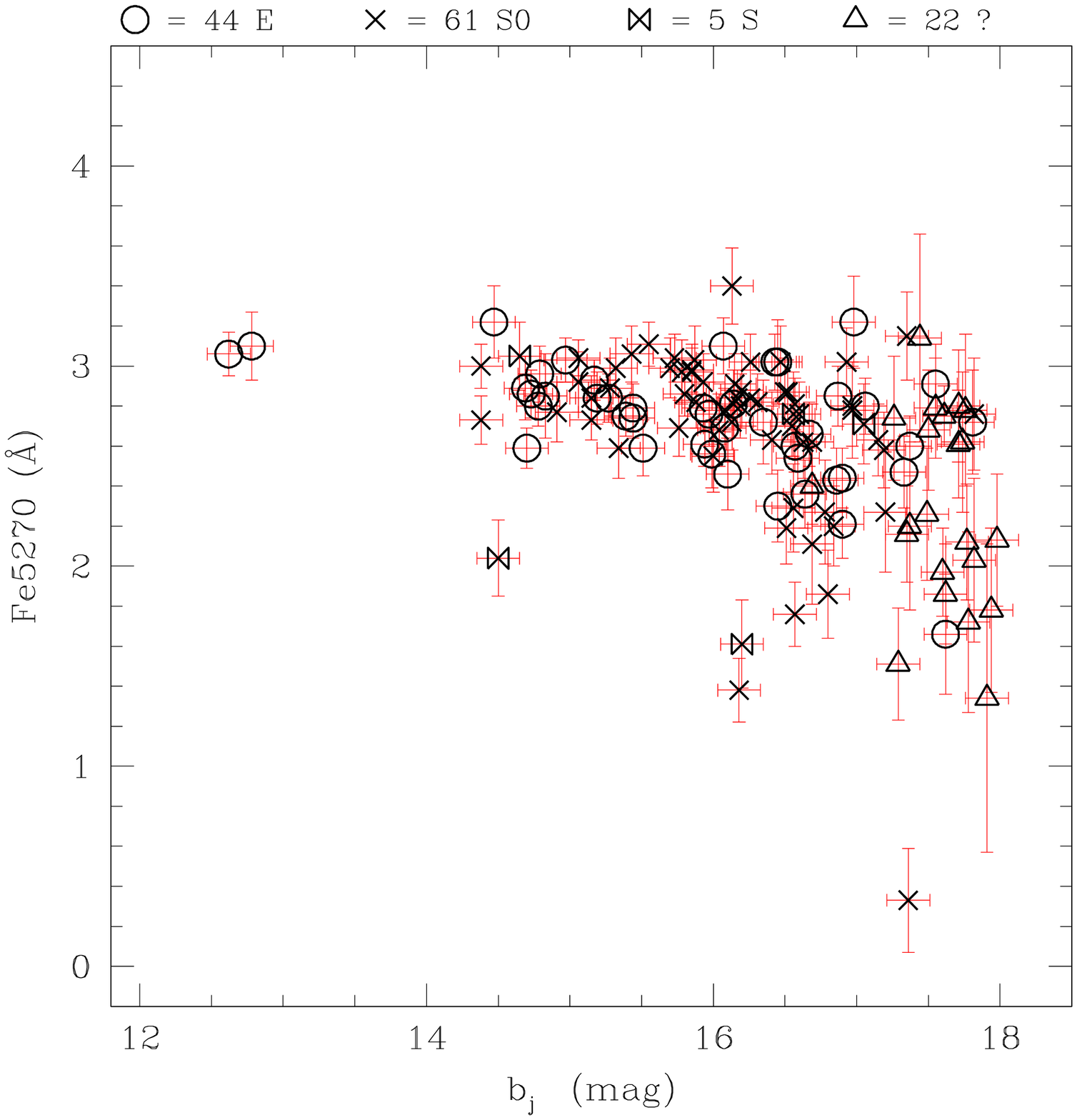,width=55mm} &
\psfig{file=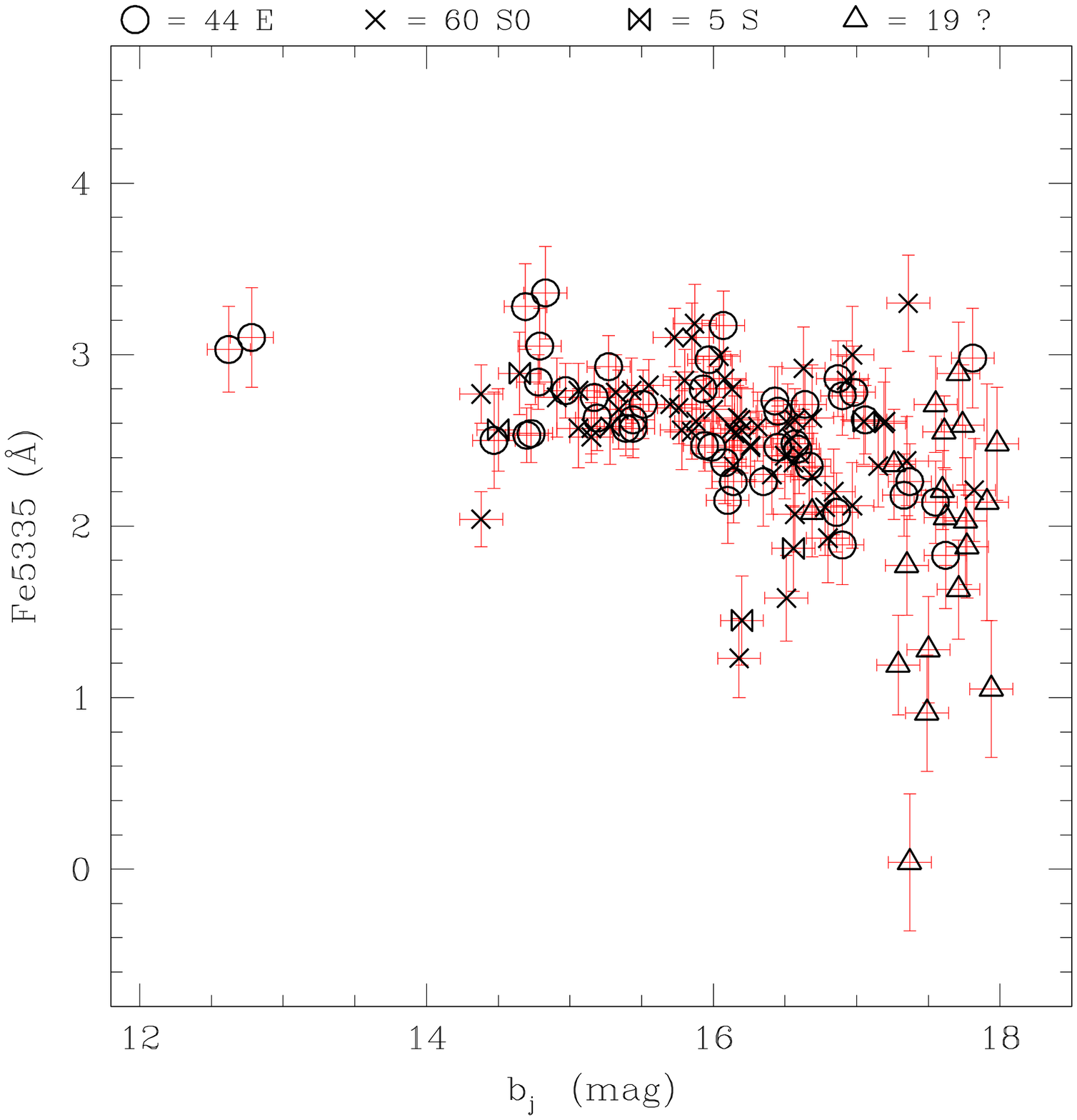,width=55mm} &
\psfig{file=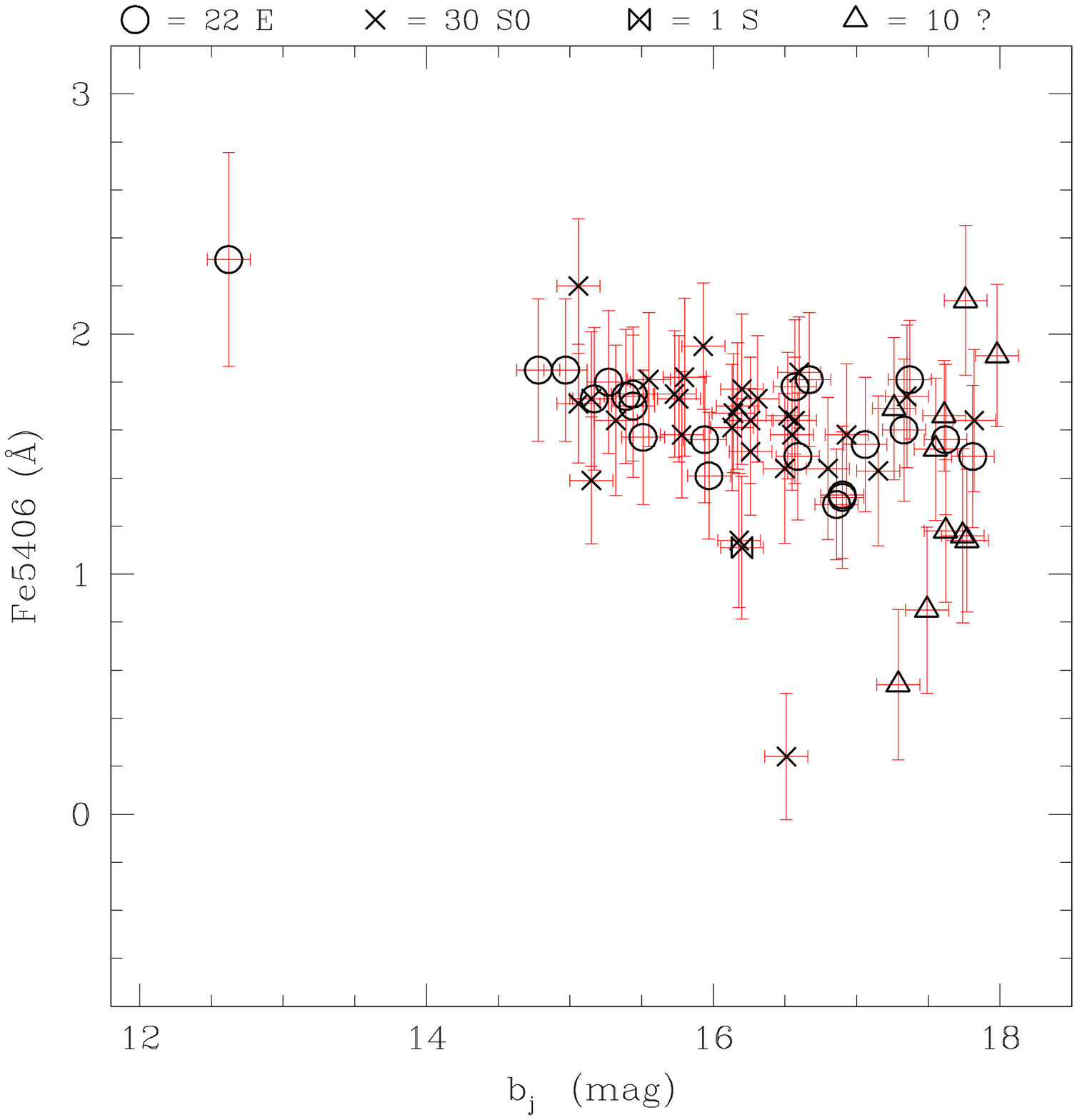,width=55mm} \\
\psfig{file=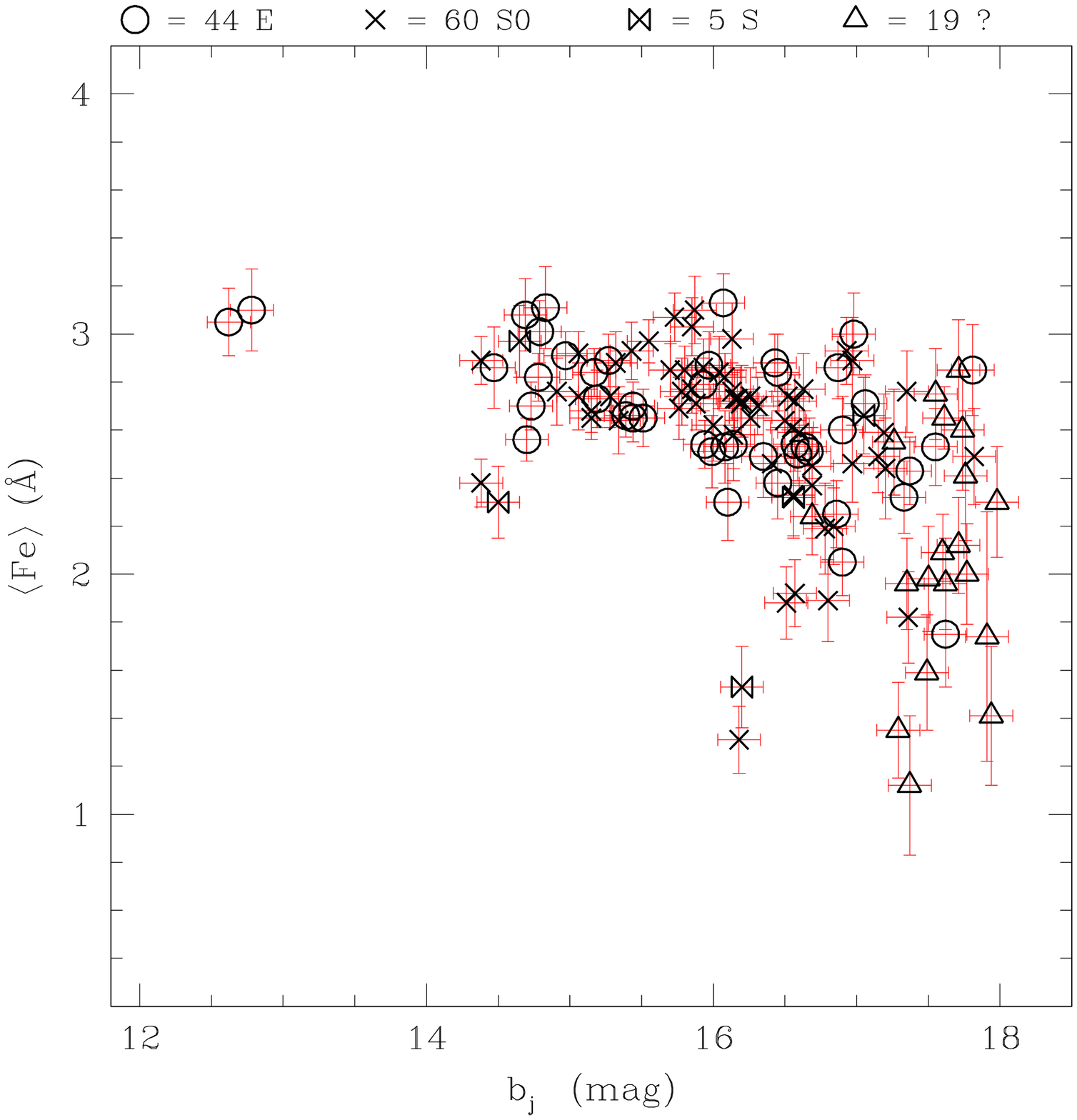,width=55mm} &
\psfig{file=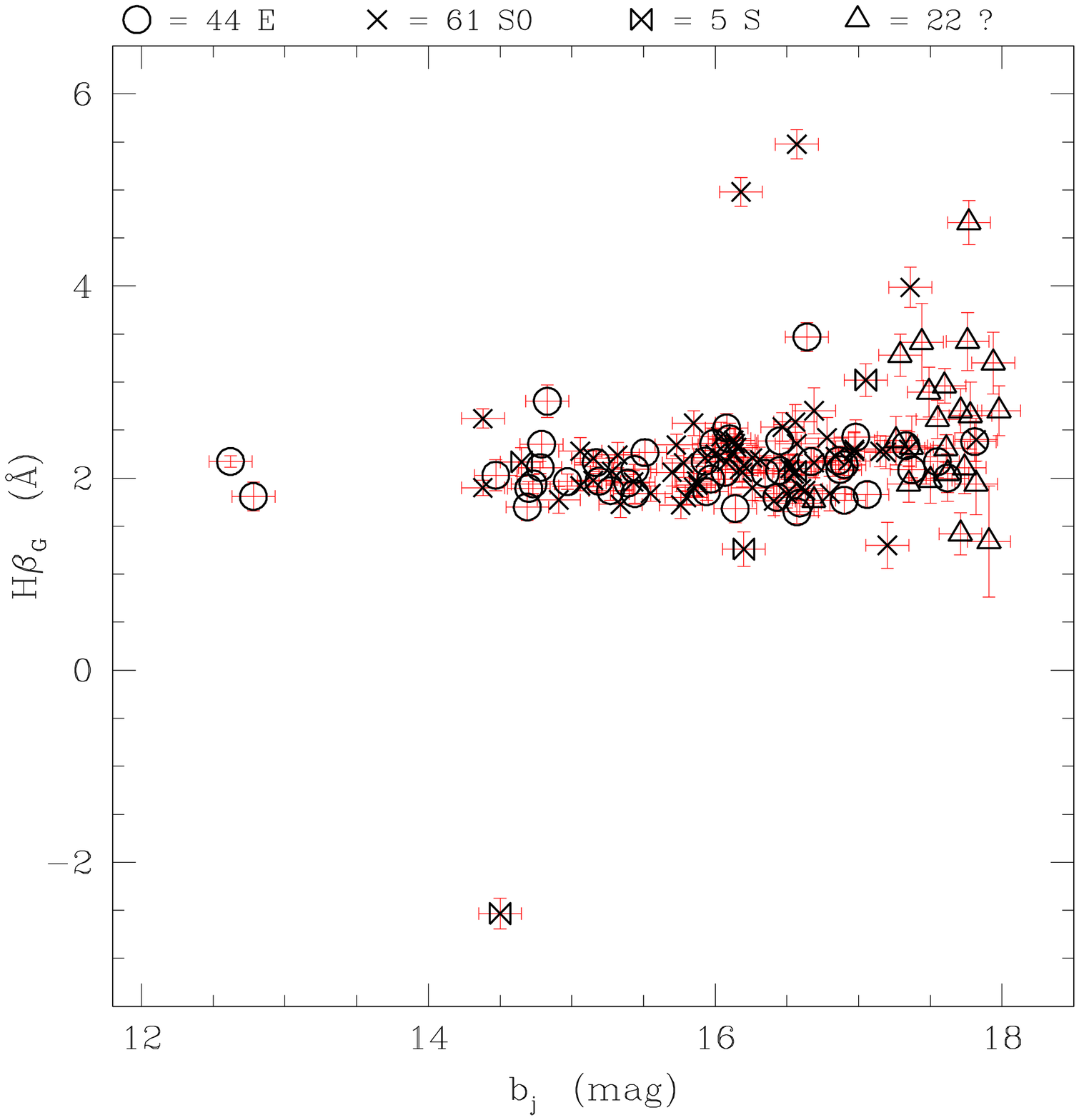,width=55mm} &
\psfig{file=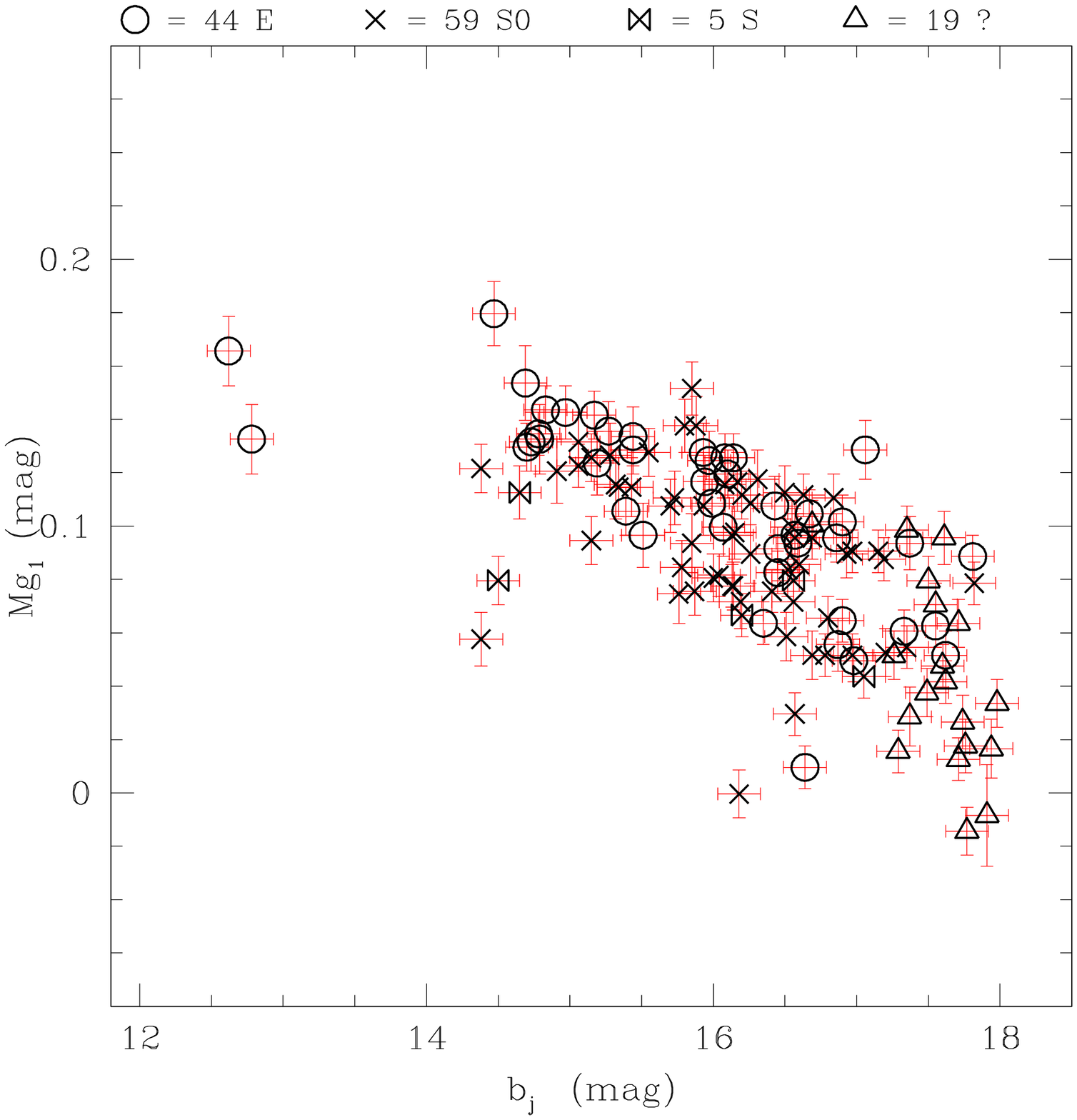,width=55mm} \\
\psfig{file=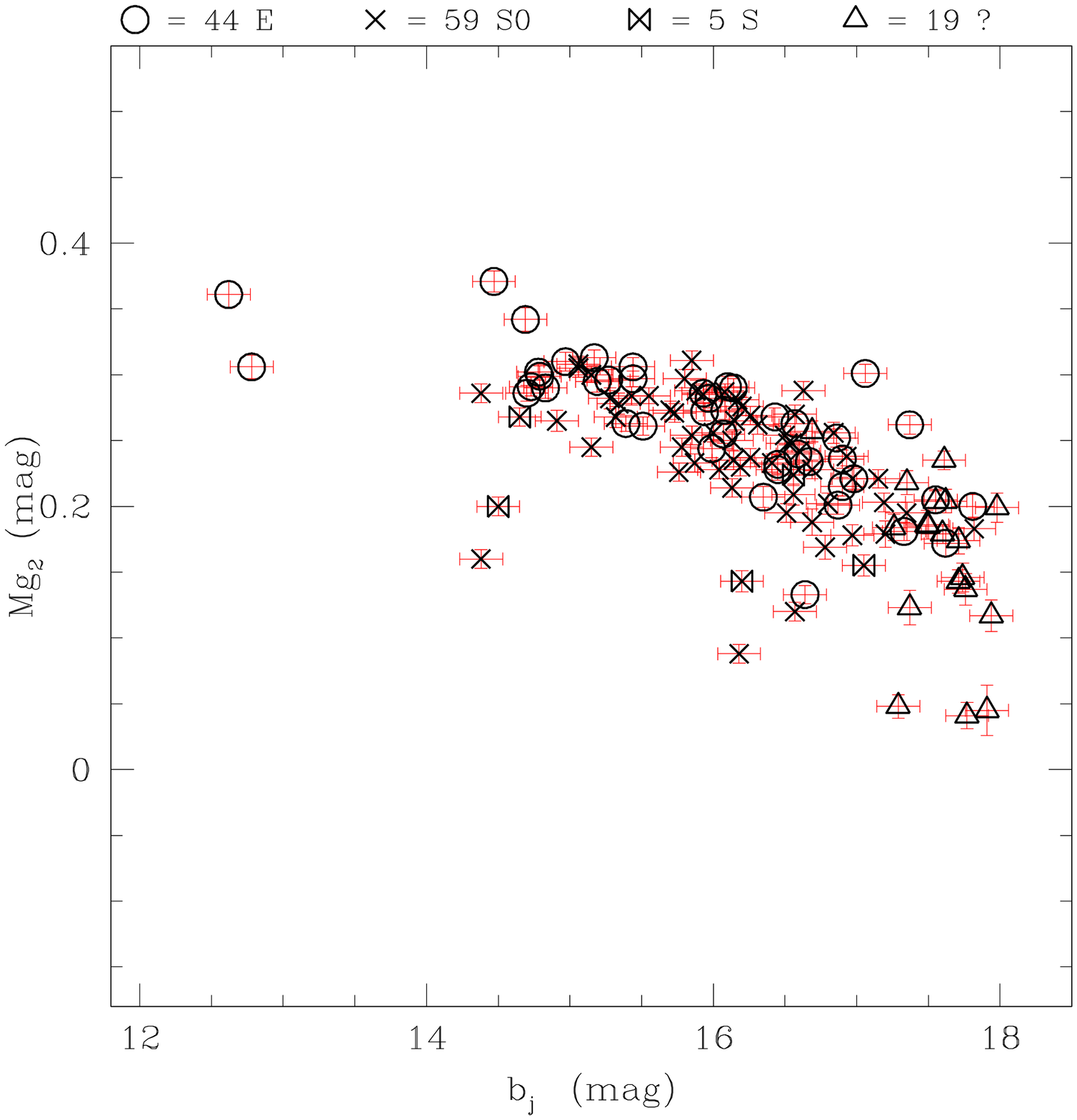,width=55mm} &
\psfig{file=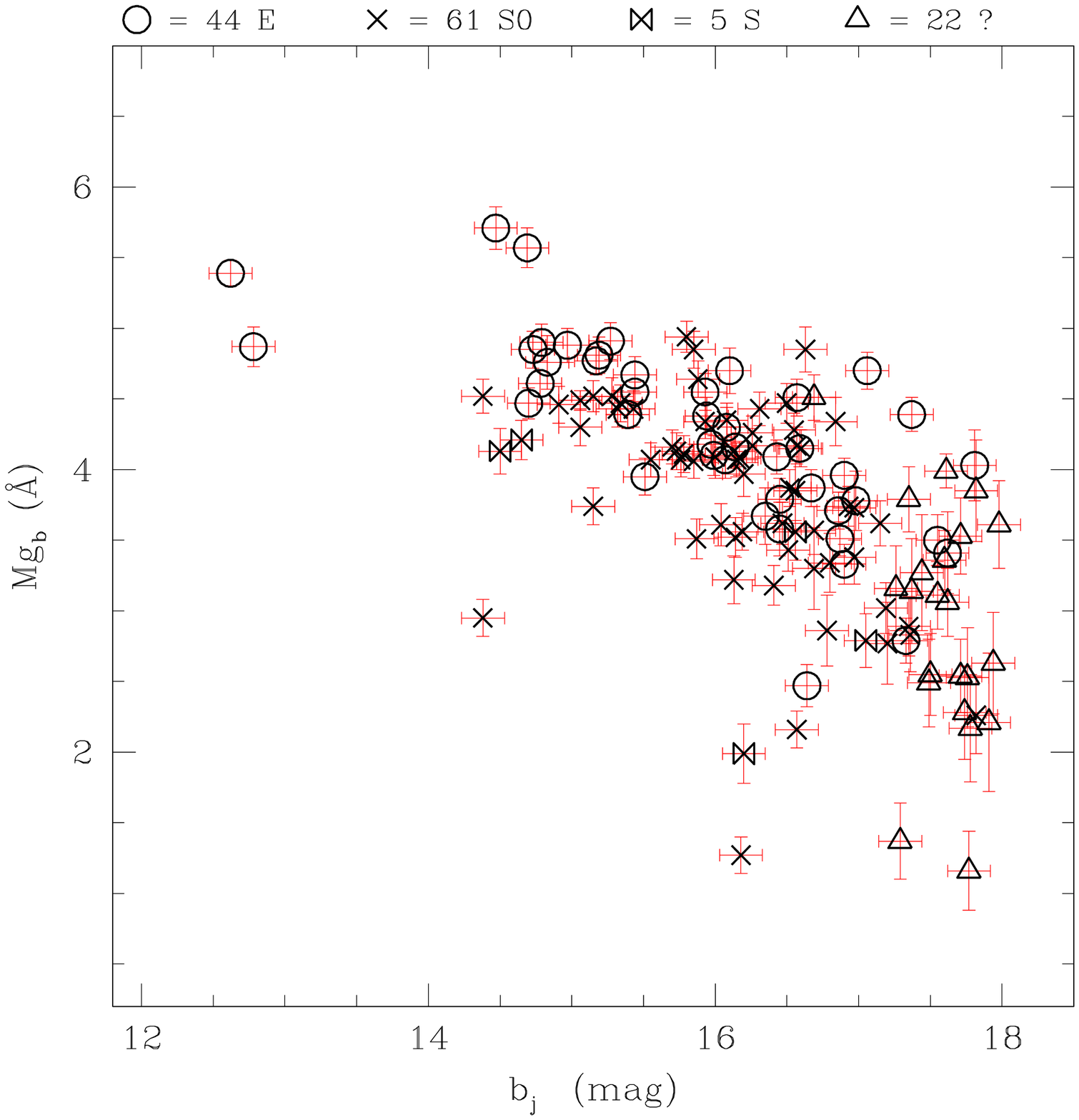,width=55mm} &
\psfig{file=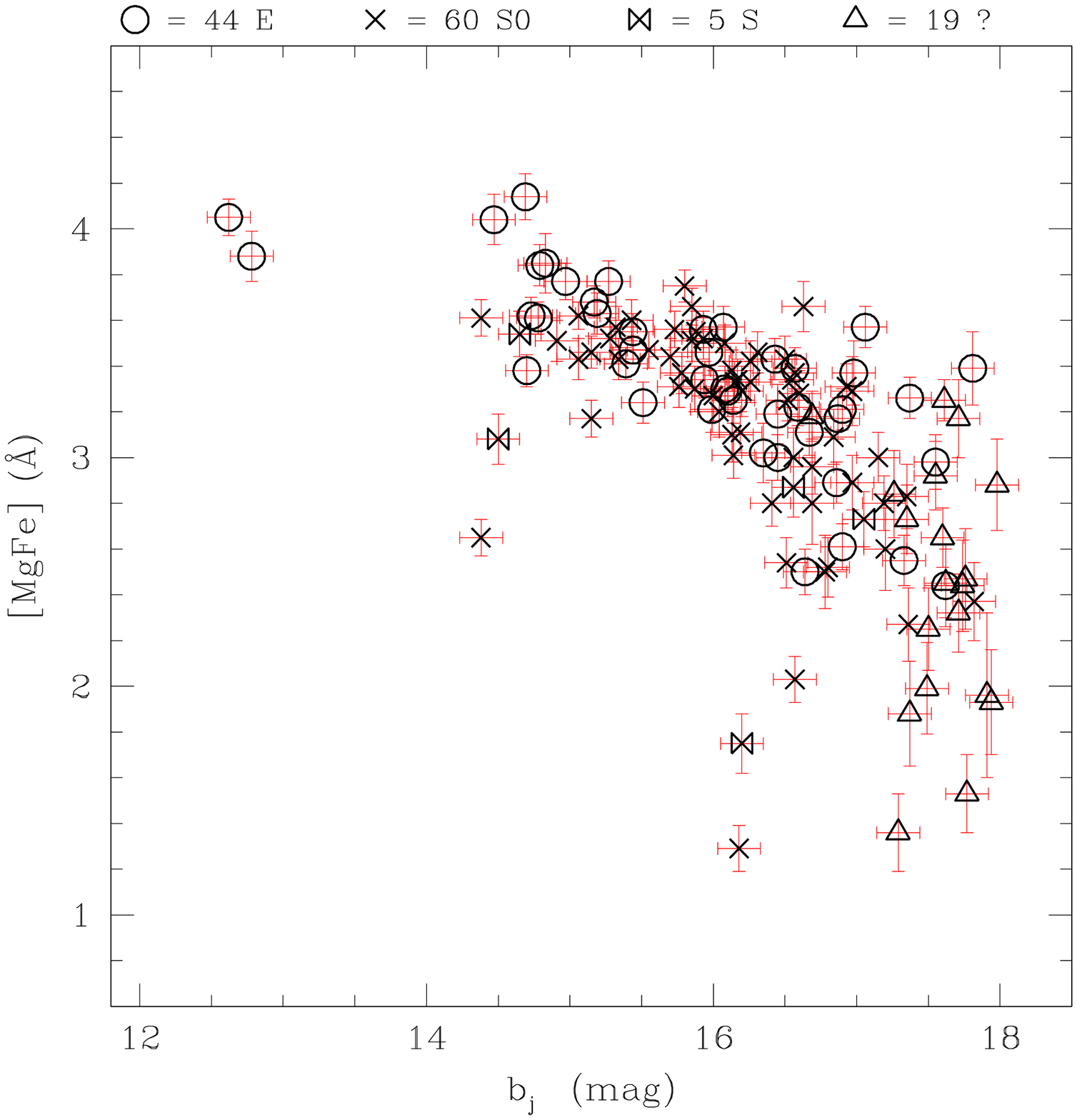,width=55mm} \\
\end{tabular}
\caption{Lick/IDS index absorption line strength data versus magnitude, $b_j$.
The key at the top of each plot gives the number of ellipticals, lenticulars,
spiral bulges and unknown morphological type galaxies. 
The Mg$_1$ line strengths have been corrected to the Lick/IDS system
by subtracting a mean offset of 0.0204\,$\pm$\,0.0025\,mag.
The Fe5406 line strength errors have been multiplied by 1.647 to correct for
an underestimation of the errors demonstrated in the scale test analysis.
}
\label{fig:datavsmag}
\end{minipage}
\end{figure*}

The data has been compared to previous studies and found to be of high quality. 
No significant systematic offsets are found between 
the study data and any previously published data, 
except for the Mg$_1$ index. A correction of 0.0204$\pm$0.0025\,mag
is applied to the Mg$_1$ index. 
The significance of any other offsets is at the level of $\sim0$--5 per cent.
The data is therefore shown to be fully on the Lick/IDS system.
This data can now be used to analyse the galaxy stellar populations and cluster properties
using models based upon the Lick/IDS system (e.g. the Worthey 1994 models).
Since the data set is homogeneous, there are in addition no inherent 
internal systematic errors clouding any analyses;
such errors are often introduced when multiple data sets are combined.
The data in this paper also has well characterised errors which have
been calculated in a rigorous and statistical way.
Fig. \ref{fig:datavsmag} shows the line strength data versus magnitude, $b_j$.
This data provides an important baseline at $z\sim0$ for studies of 
distant, high redshift clusters. It also expands the existing 
knowledge base of galaxy formation and evolution in rich clusters.

The companion papers in this study will use the data presented in this paper 
to measure stellar population mean ages and metallicities
to probe the evolutionary history of the
bright early-type galaxies within the core of the Coma cluster (Paper II). These
stellar populations act as fossil records of galaxy formation and evolution
and provide a powerful tool to constrain theoretical models.
They will also combine the data
with photometry to investigate in detail various spectro-photometric relations (Paper III).
These relations will be used to place further constraints on the models of formation 
and evolution.

\section*{Acknowledgments}
\label{sec:acknowledgments}

This work is based on observations made with the 4.2m William Herschel 
Telescope operated on the island of La
Palma by the Isaac Newton Group in the Spanish Observatorio del Roque de los
Muchachos of the Instituto de Astrofisica de Canarias.
This research has made use of the USNOFS Image and Catalogue Archive
operated by the United States Naval Observatory, Flagstaff Station
(www.nofs.navy.mil/data/fchpix/).
It has also used the UK Schmidt Telescope situated at Siding Spring Observatory, 
New South Wales, Australia which is operated as part of the Anglo-Australian Observatory.
Schmidt plates were scanned in using the advanced photographic plate
digitising machine SuperCOSMOS operated by the Royal Observatory Edinburgh.

\bsp

\begin{table*}
\begin{minipage}{185mm}
\caption{Lick/IDS index absorption line strength data.}
\label{tab:linestrengths}
\begin{tabular}{lcccccccccccccc}
\hline \hline 
name  & C4668 & Fe4930 & Fe5015 & Fe5270 & Fe5335 & Fe5406 & $\langle$Fe$\rangle$ & H$\beta$ & H$\beta_{\rm{G}}$ &   Mg$_1$ &   Mg$_2$ &   Mg$_b$ &   $[$MgFe$]$ & $[$O{\sc{iii}}$]$$_{\rm{sm}}$ \\ 
\hline
d26      &    0.18 &   1.38 &   1.72 &   1.39 &   1.22 &   1.14 &   1.31 &   5.16 &   4.98 &  0.000 &  0.088 &   1.27 &   1.29  & $\;\,$0.00 \\  
\ \      &    0.83 &   0.15 &   0.56 &   0.15 &   0.20 &   0.25 &   0.12 &   0.13 &   0.13 &  0.009 &  0.006 &   0.12 &   0.09  & $\;\,$0.10 \\  
\ \      &    41.0 &   47.3 &   50.3 &   53.5 &   53.3 &   51.4 &   53.4 &   44.0 &   43.9 &   52.0 &   52.0 &   52.0 &   52.7 &   46.5 \\
d27      &    5.60 &   1.74 &   5.08 &   3.02 &   2.67 &    \ \ &   2.84 &   2.15 &   2.28 &  0.092 &  0.228 &   3.58 &   3.19  & --0.18 \\  
\ \      &    0.81 &   0.23 &   0.51 &   0.19 &   0.24 &    \ \ &   0.15 &   0.18 &   0.16 &  0.008 &  0.008 &   0.17 &   0.11  & $\;\,$0.15 \\ 
\ \      &    28.2 &   35.4 &   37.0 &   41.3 &   41.8 &    \ \ &   41.5 &   33.3 &   33.3 &   38.2 &   38.1 &   38.1 &   39.8 &   33.8 \\
d28      &    4.55 &   1.79 &   4.55 &   2.66 &   2.51 &   1.76 &   2.58 &   1.86 &   2.06 &  0.104 &  0.235 &   3.91 &   3.18  & --0.17 \\
\hline
\end{tabular}
\end{minipage}
\end{table*}
\begin{table*}
\begin{minipage}{185mm}
\contcaption{}
\begin{tabular}{lcccccccccccccc}
\hline \hline 
name  & C4668 & Fe4930 & Fe5015 & Fe5270 & Fe5335 & Fe5406 & $\langle$Fe$\rangle$ & H$\beta$ & H$\beta_{\rm{G}}$ &   Mg$_1$ &   Mg$_2$ &   Mg$_b$ &   $[$MgFe$]$ & $[$O{\sc{iii}}$]$$_{\rm{sm}}$ \\ 
\hline
d75      &    7.63 &   2.38 &   5.67 &   2.99 &   2.31 &   1.61 &   2.65 &   1.57 &   1.68 &  0.071 &  0.201 &   2.95 &   2.79  & $\;\,$0.08 \\  
\ \      &    0.81 &   0.16 &   0.53 &   0.15 &   0.20 &   0.23 &   0.13 &   0.13 &   0.13 &  0.009 &  0.007 &   0.12 &   0.09  & $\;\,$0.17 \\  
\ \      &    36.2 &   45.0 &   46.9 &   52.2 &   53.3 &   58.5 &   52.7 &   42.9 &   42.9 &   49.1 &   49.1 &   49.1 &   50.9 &   36.5 \\
d81      &    5.38 &   1.95 &   4.98 &   2.46 &   2.15 &    \ \ &   2.30 &   2.13 &   2.41 &  0.120 &  0.291 &   4.70 &   3.29  & --0.03 \\  
\ \      &    0.81 &   0.19 &   0.52 &   0.17 &   0.22 &    \ \ &   0.14 &   0.15 &   0.14 &  0.009 &  0.007 &   0.14 &   0.10  & $\;\,$0.13 \\ 
\ \      &    33.1 &   40.8 &   42.4 &   48.7 &   49.6 &    \ \ &   49.2 &   38.2 &   38.2 &   44.1 &   43.8 &   43.9 &   46.4 &   40.4 \\
d83      &    7.84 &   2.12 &   4.02 &   0.33 &   3.30 &    \ \ &   1.82 &   4.12 &   3.81 &    \ \ &    \ \ &   2.83 &   2.27  & --0.28 \\
\hline
\end{tabular}
\end{minipage}
\end{table*}
\begin{table*}
\begin{minipage}{185mm}
\contcaption{}
\begin{tabular}{lcccccccccccccc}
\hline \hline 
name  & C4668 & Fe4930 & Fe5015 & Fe5270 & Fe5335 & Fe5406 & $\langle$Fe$\rangle$ & H$\beta$ & H$\beta_{\rm{G}}$ &   Mg$_1$ &   Mg$_2$ &   Mg$_b$ &   $[$MgFe$]$ & $[$O{\sc{iii}}$]$$_{\rm{sm}}$ \\ 
\hline
d147     &    7.22 &   1.91 &   5.35 &   2.81 &   2.62 &    \ \ &   2.72 &   1.97 &   2.06 &  0.072 &  0.230 &   3.56 &   3.11  & $\;\,$0.17 \\  
\ \      &    0.84 &   0.15 &   0.60 &   0.14 &   0.19 &    \ \ &   0.12 &   0.12 &   0.12 &  0.009 &  0.006 &   0.12 &   0.08  & $\;\,$0.12 \\ 
\ \      &    40.5 &   50.5 &   52.8 &   59.0 &   59.9 &    \ \ &   59.4 &   47.6 &   47.6 &   55.1 &   54.6 &   54.6 &   57.0 &   50.2 \\
d153     &    6.81 &   1.68 &   3.06 &   2.81 &   2.26 &    \ \ &   2.54 &   1.39 &   1.51 &  0.126 &  0.290 &   4.16 &   3.25  & --0.29 \\  
\ \      &    0.81 &   0.17 &   0.53 &   0.16 &   0.21 &    \ \ &   0.13 &   0.14 &   0.13 &  0.009 &  0.007 &   0.13 &   0.09  & $\;\,$0.13 \\ 
\ \      &    34.8 &   44.0 &   45.9 &   52.7 &   54.1 &    \ \ &   53.4 &   41.7 &   41.7 &   47.7 &   47.7 &   47.8 &   50.5 &   44.2 \\
d154     &    3.03 &   1.70 &   4.09 &   2.62 &   2.30 &    \ \ &   2.46 &   1.14 &   1.59 &  0.076 &  0.233 &   3.18 &   2.80  & --0.29 \\
\hline
\end{tabular}
\end{minipage}
\end{table*}
\begin{table*}
\begin{minipage}{185mm}
\contcaption{}
\begin{tabular}{lcccccccccccccc}
\hline \hline 
name  & C4668 & Fe4930 & Fe5015 & Fe5270 & Fe5335 & Fe5406 & $\langle$Fe$\rangle$ & H$\beta$ & H$\beta_{\rm{G}}$ &   Mg$_1$ &   Mg$_2$ &   Mg$_b$ &   $[$MgFe$]$ & $[$O{\sc{iii}}$]$$_{\rm{sm}}$ \\ 
\hline
d225     &    3.45 &   2.06 &   4.04 &   1.99 &   1.88 &   1.39 &   1.94 &   1.67 &   1.85 &  0.067 &  0.204 &   3.41 &   2.57  & $\;\,$0.23 \\  
\ \      &    0.80 &   0.23 &   0.51 &   0.20 &   0.24 &   0.26 &   0.15 &   0.18 &   0.16 &  0.008 &  0.007 &   0.17 &   0.11  & $\;\,$0.21 \\  
\ \      &    28.2 &   34.9 &   36.1 &   39.7 &   40.7 &   40.3 &   40.2 &   33.2 &   33.2 &   37.8 &   37.6 &   37.6 &   38.9 &   30.9 \\
d231     &    5.82 &   2.18 &   5.05 &   2.94 &   2.48 &   1.58 &   2.71 &   1.85 &   2.05 &  0.084 &  0.245 &   4.06 &   3.32  & --0.13 \\  
\ \      &    0.84 &   0.15 &   0.70 &   0.13 &   0.18 &   0.13 &   0.11 &   0.11 &   0.11 &  0.010 &  0.006 &   0.11 &   0.08  & $\;\,$0.13 \\  
\ \      &    42.6 &   54.9 &   57.6 &   65.0 &   66.4 &   76.1 &   65.7 &   51.5 &   51.5 &   60.3 &   60.0 &   60.0 &   62.8 &   52.0 \\
ic3943   &    7.18 &   1.76 &   5.97 &   3.11 &   2.82 &   1.81 &   2.97 &   1.53 &   1.79 &  0.128 &  0.284 &   4.07 &   3.47  & --0.08 \\
\hline
\end{tabular}
\end{minipage}
\end{table*}
\begin{table*}
\begin{minipage}{185mm}
\contcaption{}
\begin{tabular}{lcccccccccccccc}
\hline \hline 
name  & C4668 & Fe4930 & Fe5015 & Fe5270 & Fe5335 & Fe5406 & $\langle$Fe$\rangle$ & H$\beta$ & H$\beta_{\rm{G}}$ &   Mg$_1$ &   Mg$_2$ &   Mg$_b$ &   $[$MgFe$]$ & $[$O{\sc{iii}}$]$$_{\rm{sm}}$ \\ 
\hline
ngc4853  &    5.09 &   1.45 &   4.01 &   2.73 &   2.04 &    \ \ &   2.38 &   1.98 &   2.11 &  0.058 &  0.160 &   2.95 &   2.65  & --0.85 \\  
\ \      &    0.54 &   0.15 &   0.94 &   0.09 &   0.12 &    \ \ &   0.08 &   0.07 &   0.08 &  0.008 &  0.006 &   0.11 &   0.07  & $\;\,$0.08 \\ 
\ \      &    62.3 &   77.4 &   81.0 &   88.5 &   89.5 &    \ \ &   89.0 &   72.8 &   72.7 &   84.2 &   84.6 &   84.7 &   86.8 &   78.7 \\
ngc4860  &    9.27 &   1.86 &   5.46 &   2.89 &   3.28 &    \ \ &   3.08 &   1.33 &   1.70 &  0.154 &  0.342 &   5.57 &   4.14  & $\;\,$0.21 \\  
\ \      &    0.78 &   0.19 &   1.05 &   0.12 &   0.18 &    \ \ &   0.11 &   0.09 &   0.10 &  0.013 &  0.008 &   0.13 &   0.08  & $\;\,$0.09 \\ 
\ \      &    53.8 &   67.3 &   68.6 &   76.6 &   77.8 &    \ \ &   77.2 &   64.0 &   64.0 &   70.3 &   69.2 &   69.3 &   73.1 &   65.7 \\
ngc4864  &    7.55 &   1.74 &   5.25 &   2.59 &   2.53 &    \ \ &   2.56 &   1.61 &   1.82 &  0.130 &  0.286 &   4.47 &   3.38  & --0.12 \\
\hline
\end{tabular}
\end{minipage}
\end{table*}
\begin{table*}
\begin{minipage}{185mm}
\contcaption{}
\begin{tabular}{lcccccccccccccc}
\hline \hline 
name  & C4668 & Fe4930 & Fe5015 & Fe5270 & Fe5335 & Fe5406 & $\langle$Fe$\rangle$ & H$\beta$ & H$\beta_{\rm{G}}$ &   Mg$_1$ &   Mg$_2$ &   Mg$_b$ &   $[$MgFe$]$ & $[$O{\sc{iii}}$]$$_{\rm{sm}}$ \\ 
\hline
ngc4923  &    8.56 &   1.77 &   5.13 &   2.79 &   2.84 &   1.85 &   2.82 &   1.78 &   2.11 &  0.135 &  0.303 &   4.62 &   3.61  & --0.04 \\  
\ \      &    0.31 &   0.09 &   0.35 &   0.10 &   0.18 &   0.15 &   0.10 &   0.07 &   0.06 &  0.006 &  0.007 &   0.07 &   0.06  & $\;\,$0.09 \\  
\ \      &    72.2 &   90.9 &   94.5 &  109.5 &  112.8 &  117.9 &  111.2 &   85.8 &   85.7 &   98.5 &   98.6 &   98.7 &  104.7 &   95.5 \\
rb58     &    5.53 &   0.55 &   6.02 &   2.03 &    \ \ &    \ \ &    \ \ &   1.35 &   1.94 &    \ \ &    \ \ &   3.85 &    \ \  & $\;\,$0.36 \\ 
\ \      &    1.36 &   0.51 &   0.83 &   0.40 &    \ \ &    \ \ &    \ \ &   0.38 &   0.30 &    \ \ &    \ \ &   0.35 &    \ \  & $\;\,$0.27 \\ 
\ \      &    15.7 &   19.6 &   20.7 &   22.6 &    \ \ &    \ \ &    \ \ &   18.5 &   18.5 &    \ \ &    \ \ &   21.4 &    \ \  &   17.3 \\
rb60     &    3.00 &   1.89 &   4.50 &   2.79 &   2.78 &   1.52 &   2.79 &   2.34 &   2.44 &  0.072 &  0.207 &   3.13 &   2.95  & --0.05 \\
\hline
\end{tabular}
\end{minipage}
\end{table*}
\begin{table*}
\begin{minipage}{185mm}
\contcaption{}
\begin{tabular}{lcccccccccccccc}
\hline \hline 
name  & C4668 & Fe4930 & Fe5015 & Fe5270 & Fe5335 & Fe5406 & $\langle$Fe$\rangle$ & H$\beta$ & H$\beta_{\rm{G}}$ &   Mg$_1$ &   Mg$_2$ &   Mg$_b$ &   $[$MgFe$]$ & $[$O{\sc{iii}}$]$$_{\rm{sm}}$ \\ 
\hline
gmp3298  &    4.15 &   1.58 &   4.69 &   2.61 &   2.38 &   1.69 &   2.50 &   2.13 &   2.35 &  0.048 &  0.180 &   3.16 &   2.81  & --0.34 \\  
\ \      &    0.99 &   0.37 &   0.64 &   0.28 &   0.30 &   0.30 &   0.20 &   0.28 &   0.23 &  0.009 &  0.009 &   0.25 &   0.16  & $\;\,$0.40 \\  
\ \      &    20.2 &   25.1 &   26.3 &   29.4 &   30.0 &   33.2 &   29.7 &   23.5 &   23.5 &   27.8 &   27.8 &   27.8 &   28.7 &   21.9 \\
gmp3585  &  --0.23 &   0.79 &   1.17 &   1.51 &   1.18 &   0.54 &   1.34 &   2.07 &   2.31 &  0.016 &  0.048 &   1.37 &   1.36  & --1.63 \\  
\ \      &    0.91 &   0.36 &   0.61 &   0.28 &   0.30 &   0.35 &   0.21 &   0.27 &   0.22 &  0.009 &  0.009 &   0.24 &   0.16  & $\;\,$0.21 \\  
\ \      &    22.0 &   25.7 &   27.3 &   29.5 &   29.7 &   29.1 &   29.6 &   24.1 &   24.1 &   28.4 &   29.1 &   29.0 &   29.3 &   24.4 \\
gmp3588  &    2.66 &   1.12 &   4.26 &   2.92 &   1.89 &   2.13 &   2.40 &   3.28 &   3.19 &  0.022 &  0.138 &   2.42 &   2.41  & --0.32 \\
\hline
\end{tabular}
\end{minipage}
\begin{minipage}{185mm}
{\em Notes:}\/ There are three lines for each galaxy in this table.
The first line gives the absorption line strength data,
whilst errors are given on the second line
and signal-to-noise values for each index are given on the third.
Missing values in the table indicate either that the line strength measured
had a low S/N or that it could not be measured.
The H$\beta$ and H$\beta_{\rm{G}}$ 
line strengths given in the table have not been corrected for nebula emission. 
The Mg$_1$ line strengths have been corrected to the Lick/IDS system
by subtracting a mean offset of 0.0204\,$\pm$\,0.0025\,mag.
The Fe5406 line strength errors have been multiplied by 1.647 to correct for
an underestimation of the errors demonstrated in the scale test analysis.
There are a total of 132 galaxies in this data table.
\label{lastpage}
\end{minipage}
\end{table*}

\end{document}